\documentclass[11pt,english]{article}
\usepackage[T1]{fontenc}
\usepackage[latin9]{inputenc}
\usepackage{geometry}
\usepackage{amsmath,amsfonts,epsfig}
\usepackage{amssymb}
\usepackage{jheppub}
\usepackage{mathrsfs}
\usepackage{hyperref}
\usepackage{hyperref}
\newcommand{\CP}{\mathop{\textrm{CP}}}
\newcommand{\mm}{{\mathfrak{m}}}
\newcommand{\scL}{{\cal L}}
\newcommand{\scG}{{\cal G}}
\newcommand{\al}{\alpha}

\newcommand{\tom}{\tilde{\omega}}
\newcommand{\om}{{\omega}}
\newcommand{\de}{{\delta}}
\newcommand{\ltl}{{\log\tfrac{2}{\ell}}}

\newcommand{\intinf}{\mathop{\int\limits_{-\infty}^\infty}}

\newcommand{\intli}{\mathop{\int\limits_{-\log\tfrac{2}{\ell}}^\infty}}

\newcommand{\intil}{\mathop{\int\limits_{-\infty}^{\log\tfrac{2}{\ell}}}}

\begin{document}
\numberwithin{equation}{section}
\setlength{\unitlength}{.8mm}

%%%%%%%%%%%%%%%%%%%%%

%\begin{document}

\title{Integrable formulation of the OPE coefficients in the UV limit of the sine-Gordon model}

\author{\'Arp\'ad Heged\H us}

\affiliation{
HUN-REN Wigner Research Centre for Physics,\\
H-1525 Budapest 114, P.O.B. 49, Hungary}

\emailAdd{hegedus.arpad@wigner.hun-ren.hu}

\abstract{
In the repulsive regime of the sine-Gordon model, we work out a method, that enables one to formulate the UV limit 
of finite volume expectation values in terms of the integrable description of the UV limit of the corresponding 
spectral problem. Since these expectation values are related to 3-point couplings containing at least two identical operators, 
our computations provide with an integrable formulation to the 3-point couplings of the  UV CFT. 
Our approach is based on the fermionic description of operators in the sine-Gordon theory, in which these 
expectation values are expressed in terms of a finite number of elements of an infinite matrix. % $\om$. % $\om.$ 
Focusing on vacuum expectation values, in this paper, the first two nontrivial coefficients in the UV 
series representation of this  matrix are expressed in terms of integrability data. 
This allows one to get an integrable description to some specific ratios of vacuum expectation values 
in the complex Liouville CFT. }

\maketitle

%%%%%%%%%%%%%%%%%%%%%%%%%%%%%%%%%%%%%%%%%%%%%%%%%%%%%%%%%

\section{Introduction} \label{intro}

Solving conformal field theories (CFT) is an important problem in various areas of theoretical physics, 
from string-theory to statistical physics. As a consequence of the restrictions imposed by the conformal 
symmetry, knowing the CFT data, namely the scaling dimensions and the 3-point couplings, all the correlation 
functions of the theory can in principle be determined.

In general, the determination of the CFT data is a very hard and upto now unsolved problem. 
In 2-dimensions, the existence of the infinite dimensional Virasoro algebra \cite{Belavin:1984vu} 
allows one to 
construct the CFT data analytically, but in higher dimensions one can rely only on the recently 
elaborated conformal bootstrap method. Both its analytical \cite{Poland:2022qrs} 
and numerical versions \cite{Poland:2018epd} can 
provide important information on the CFT data. There are a few higher dimensional models,  
where a 2-dimensional integrable scattering description of the CFT is available. Important examples of this 
class are the models corresponding to the planar limit of various $AdS/CFT$ dualities \cite{Maldacena:1997re}. 
Among these the most elaborated one is the $AdS_5/CFT_4$ correspondence: % \cite{Maldacena:1997re}. 
a duality  between $N = 4$ supersymmetric 
Yang-Mills theory in 3+1 dimensions and type IIB superstring 
theory on $AdS_5  \times S^5$.

In this model the integrable description allows one to describe the spectrum in terms of the 
quantum spectral curve method (QSC) \cite{Gromov:2013pga,Gromov:2014caa} 
and the so-called hexagon approach \cite{Basso:2015zoa} 
has been worked out to determine the 3-point couplings of the theory. While the QSC method proved to be 
very efficient in determining the spectrum, the hexagon approach can be mostly used at weak coupling, 
since the treatment of all wrapping corrections remained a great challenge in this approach.  
 
In such CFT models having an integrable formulation, one might hope, that there exists a relatively 
nice description of the 3-point couplings in terms of the integrable description of the spectral 
problem of the theory. To solve the spectral problem of the $AdS_5/CFT_4$ duality, the formerly elaborated theoretical 
methods of 2-dimensional integrable quantum field theories gave a useful guideline.
We think, that the mathematical techniques worked out to compute 3-point functions in 2-dimensional CFTs 
in terms of integrability data, might facilitate the solution of the same problem in the $AdS/CFT$ dualities, as well. 

Thus, our goal is to give such a description of the 3-point couplings of 2-dimensional CFTs, in which 
they are expressed in terms of functions coming from the integrable formulation of the spectrum 
of the theory. 

This approach was initiated in \cite{Bajnok:2022ucr}, where the Lee-Yang and 3-state Potts CFTs were discussed. 
In this paper, such a special set of 3-point couplings were investigated, in which at least two of the three 
operators were identical and the 3rd one corresponded to the perturbing operator of a massive integrable 
perturbation of the CFT. Namely,  3-point couplings of the type $C_{{\cal O} \Phi {\cal O}}$  
were considered in \cite{Bajnok:2022ucr}, where ${\cal O}$ stands for the 
identical operators, while $\Phi$ stands for the 3rd (perturbing) operator.\footnote{The ${\cal O}=\Phi$ 
choice was also allowed.}.

Thus, the 3-point couplings accessible with the method of \cite{Bajnok:2022ucr} remained  in general very limited. 
 
In \cite{Bajnok:2022ucr}, 
the 3-point couplings were computed from finite volume expectation values of local operators, 
such that the sandwiching states correspond to the two identical operators ${\cal O}$ 
in the 3-point coupling, while the sandwiched operator corresponds to the perturbing one $\Phi.$  

The limitation of the sandwiched operator to be the perturbing operator leading to a massive integrable 
quantum field theory, came from the fact, that these special 3-point couplings could be extracted 
from  the ultraviolet (UV) limit of the finite volume energy of the sandwiching state in the 
perturbed massive integrable theory. 

In this work, we would like to extend the results of \cite{Bajnok:2022ucr} 
to a larger set of sandwiched operators 
(i.e. 3-point couplings), but in another interesting family of models where former results available in the 
literature makes this possible. Namely, the model we are investigating in this paper, is the UV CFT limit of 
the sine-Gordon model.  

This extension is possible due to the results of \cite{Jimbo:2010jv}, where using the fermionic basis \cite{Boos:2007wv,Boos:2006mq,Boos:2008rh,Jimbo:2008kn,Boos:2010qii}, 
in principle all finite volume expectation values of the theory can be computed from the integrable 
description of the spectral problem of the theory.  
The formulas in \cite{Jimbo:2010jv} and \cite{Hegedus:2019rju} 
give the volume dependence of the massive theory, but to get the 
3-point couplings in the limiting CFT, the ultraviolet limit of the formulas should be computed in terms of 
integrability data. 

The key result in \cite{Jimbo:2010jv} is that in the integrable description, all expectation values are expressed 
in terms of a finite number of elements of an infinite matrix $\om_{2k-1,2j-1},$ where $j,k \in \mathbb{Z}.$ 

In this paper, we compute the first two nontrivial coefficients in the UV 
series representation of this $\om$ matrix, which allows us to get the integrable formulation 
of some specific ratios of vacuum expectation values in the complex Liouville CFT.

In principle relying on \cite{Hegedus:2019rju}, the evaluation of our results to any sandwiching state would be 
possible, too. Nevertheless, for the sake of simplicity and to avoid proliferating details in the 
computations, we perform the actual computations for the vacuum expectation values. 
As a result, the ratios of 3-point couplings of the UV limiting CFT are expressed in terms of 
the integrable formulation of the spectral problem of the corresponding CFT. 

The outline of the paper is as follows: 

In section \ref{2}, we recall how the expectation values are related to 3-point couplings in a CFT 
and the most important results on the integrable description of 
finite volume expectation values of local operators in the sine-Gordon theory.  
In section \ref{3}, we discuss the UV behaviour of the solution of the nonlinear-integral equation 
(NLIE) governing the groundstate energy in finite volume.  
In section \ref{4}, we summarize, the predictions to the UV behaviour of the expectation values 
coming from field theoretical considerations. In section \ref{5}, we present our final integrability based 
expressions for  the leading order in the UV coefficients of the matrix elements $\om_{2k-1,1-2j},$ with $j,k \in \mathbb{N}.$ 
Section \ref{6}, contains the analog result, but for the next to leading order coefficients.  % at $\al=0.$  
In section \ref{7}, contains some useful technical details about the practical evaluation 
of the integrals entering our UV formulas. 
In the short section \ref{7b}, for completeness 
the leading order UV coefficients, for $\om_{j,k}(\al)$ with subscripts having the same sign are discussed.
In section \ref{8}, we share a few important technical details on the numerical computations, and 
shortly summarize the outcome of our numerical checks.   
The main part of the paper is closed by a short discussion of the results in section \ref{9}.

The details of the bulky computations of the derivation of our main results 
are relegated into appendix \ref{appA}. In appendix \ref{appB} we summarized 
the large argument series representation of the most important integral kernel of the problem. 
Finally, in appendix \ref{appC}, we provide with some tables of numerical data.

\section{Preliminaries} \label{2}

The basic tool of our approach in investigating the conformal data of a CFT, is to put it into a finite volume, 
 or equivalently define it on a cylinder of circumference $L.$  
The energy spectrum of the CFT in finite volume is 
fully determined by the scaling dimensions and the central charge of the model. In CFTs, due to the state-operator 
correspondence, the energy levels can be labeled by the operators of the theory. The volume dependence of an energy level 
corresponding to an operator ${\cal O},$ is completely determined by the scaling dimensions of ${\cal O}:$
\begin{equation} \label{EcftO}
\begin{split}
E_{\cal O}(L)=\frac{2 \, \pi}{L} \left( \Delta_+^{\cal O} +\Delta_-^{\cal O} -\frac{c}{12} \right),
\end{split}
\end{equation}
where $\Delta_\pm^{\cal O}$ denote the left and right conformal dimensions of the operator ${\cal O},$ and $c$ is the central charge of the CFT. 

In finite volume, the matrix elements of the fields give access to the  
  3-point couplings of the model through the relation as follows:
\begin{equation} \label{cftexpL}
\begin{split}
\langle {\cal O}_1 | {\cal O}  | {\cal O}_2 \rangle=\left( \frac{2 \, \pi}{L} \right)^{\Delta_+^{\cal O} +\Delta_-^{\cal O}}\! \! \! C_{{\cal O}_1 {\cal O} {\cal O}_2}.
\end{split}
\end{equation}

In this paper, we would like to give an integrability based description to the spectrum $\Delta_\pm^{\cal O},$ and to the 3-point couplings 
of the UV limit of the sine-Gordon theory. Due to limitations in the available mathematical apparatus, we restrict ourselves 
to 3-point couplings, which contain at least 2 identical operators (${\cal O}_1={\cal O}_2$ in (\ref{cftexpL})) 
and so they correspond to diagonal matrix elements in the finite volume description.

\subsection{The model}

In this paper we study the sine-Gordon model described by its Euclidean action as follows:. 
\begin{equation} \label{LSG}
\begin{split}
{\cal A}_{SG}=\int \bigg\{ \frac{1}{4 \pi} \partial_z \varphi(z,\bar{z}) \, \partial_{\bar{z}} \varphi(z,\bar{z})
-\frac{2 {\bf \mu}^2}{\sin \pi \beta^2} \cos(\beta \, \varphi(z,\bar{z})) \bigg\} \, \frac{i \, dz \wedge \, d\bar{z}}{2},
\end{split}
\end{equation}
where $z=x+i\, y$ and $\bar{z}=x-i \, y,$ with $x,y$ being the coordinates of the Euclidean space-time.

In accordance with \cite{Jimbo:2010jv}, first we consider the model as a perturbed complex Liouville CFT: 
\begin{equation} \label{PCLT}
\begin{split}
{\cal A}_{SG}={\cal A}_{L}- \frac{{\bf \mu}^2}{\sin \pi \beta^2} \int e^{-i \, \beta \, \varphi(z,\bar{z})} \frac{i \, dz \wedge \, d\bar{z}}{2},\quad 
\end{split}
\end{equation}
where ${\cal A}_L$ denotes the action of the unperturbed CFT:
\begin{equation} \label{AL}
\begin{split}
{\cal A}_{L}=\int \bigg\{ \frac{1}{4 \pi} \partial_z \varphi(z,\bar{z}) \, \partial_{\bar{z}} \varphi(z,\bar{z})
-\frac{ {\bf \mu}^2}{\sin \pi \beta^2} e^{i \, \beta \, \varphi(z,\bar{z})} \bigg\} \, \frac{i \, dz \wedge \, d\bar{z}}{2}.
\end{split}
\end{equation}
The central charge of this CFT is given by:
\begin{equation} \label{centLiou}
\begin{split}
c_L=1-6 \frac{\nu^2}{1-\nu}, \qquad \nu=1-\beta^2. 
\end{split}
\end{equation}
The primary fields are labeled by the real continuous parameter $\alpha$:
\begin{equation} \label{primLiou}
\begin{split}
\Phi_\alpha(z,\bar{z})=e^{\tfrac{i\, \alpha \beta \nu}{2 (1-\nu)}  \, \varphi(z,\bar{z})},
\end{split}
\end{equation}
and have scaling dimensions $2 \Delta_\alpha$ with:
\begin{equation} \label{Dalfa}
\begin{split}
\Delta_\alpha=\frac{\nu^2}{4(1-\nu)} \alpha \, (\alpha-2). 
\end{split}
\end{equation}
In this paper, similarly to \cite{Jimbo:2010jv}, we consider the range $ 0\leq \al <2.$ A wider range for $\al$ can be obtained by an 
appropriate analytical continuation. % out of this regime. 

Primary fields (\ref{primLiou}) and their descendants span the basis in the 
space of operators of the theory.
Using the fermionic basis, in \cite{Jimbo:2010jv} it has been shown, 
that specific ratios of vacuum expectation values of the fields of the theory 
can be computed in terms of a single function $\om(\lambda,\mu|\alpha).$ To be more precise, 
the expectation values can be formulated in terms of a finite number of  
coefficients of its expansions around the $\lambda,\mu \to \pm \infty$ points: 
\begin{equation} \label{ommatr}
\begin{split}
\omega(\mu,\lambda|\alpha)\!=\!\!\!\sum\limits_{j,k=1}^{\infty} \!\!e^{-\epsilon_1 \mu \tfrac{2j-1}{\nu}} e^{-\epsilon_2 \lambda \tfrac{2k-1}{\nu}} 
\omega_{\epsilon_1 (2j-1),\epsilon_2(2k-1)}(\alpha),
\quad \! \epsilon_{1,2}\!=\!\pm, \quad \epsilon_1 \, \mu \!\to \!\infty, \quad \!
\epsilon_2 \, \lambda \! \to \! \infty.
\end{split}
\end{equation}
The coefficients of these expansions define an infinite matrix 
$\omega_{2j-1,2k-1}(\alpha), \, j,k\in \mathbb{Z},$ the UV behaviour of which is the primary 
interest of this paper.

In the rest of the paper, we will focus on computing the first two terms in the UV series 
representation of the matrix $\om_{2k-1,1-2j}(\alpha),$ for $j,k=1,...,\infty,  $ in terms of 
an integrable description of the spectral problem of the UV CFT limit of the model.

The importance of this matrix is that, the ratios of vacuum expectation values of the primary fields 
can be expressed in terms of its matrix elements, in the following way  
 \cite{Jimbo:2010jv}: 
\begin{equation}  \label{primVEV}
\begin{split}
\frac{\langle  \Phi_{\alpha+2 \,m\, \tfrac{1-\nu}{\nu}}(0) \rangle}{\langle  \Phi_{\alpha}(0) \rangle}=
\mu^{2 \, m \alpha-2 \, m^2\,(1-\tfrac{1}{\nu})}\, C_m(\alpha) \, \underset{1 \leq j,k \leq m }{\mbox{det}} \, \Omega_{kj}(\alpha), \qquad m=1,2,...
\end{split}
\end{equation}
where $C_m(\alpha)$ is a constant given by the formula for $m>0,$ as follows:
\begin{equation} \label{Cm}
\begin{split}
C_m(\alpha)=\prod\limits_{j=0}^{m-1} C_1(\alpha+2j\tfrac{1-\nu}{\nu}),
\end{split}
\end{equation}
with
\begin{equation} \label{C1}
\begin{split}
&C_1(\alpha)=i \, \nu \Gamma(\nu)^{4 x(\alpha)}\, \frac{\Gamma(-2 \nu x(\alpha))}{\Gamma(2 \nu x(\alpha))}\, \frac{\Gamma(x(\alpha))}{\Gamma(x(\alpha)+1/2)} \,
\frac{\Gamma(-x(\alpha)+1/2)}{\Gamma(-x(\alpha))} \cot(\pi x(\alpha)), \\
&\text{and} \qquad x(\alpha)=\frac{\alpha}{2}+\frac{1-\nu}{2 \nu}.
\end{split}
\end{equation}
The matrix elements of $\Omega_{j,k}(\alpha)$ are expressed in terms of $\om_{2k-1,1-2j}(\alpha)$ 
 as follows: 
\begin{equation}  \label{OM}
\begin{split}
\Omega_{j,k}(\alpha)=\om_{2k-1,1-2j}(\alpha)+\frac{i}{\nu} \, \delta_{j,k}\, 
\coth\left[ \frac{\pi}{2 \nu} (2k-1+\nu \, \alpha)\right], \qquad j,k=1,...m.
\end{split}
\end{equation}
Finally, $\mu$ in (\ref{primVEV}), is nothing, but the coupling constant in the action (\ref{LSG}) of the sine-Gordon model. It is related to the soliton mass ${\cal M},$ by the formula \cite{Zamolodchikov:1995xk}:
\begin{equation} \label{muM}
\begin{split}
\mu={\cal M}^\nu \, \Pi(\nu)^\nu, \qquad \text{where} \qquad 
\Pi(\nu)=\frac{\sqrt{\pi}}{2} \frac{\Gamma\left(\tfrac{1}{2 \nu}\right)}{\Gamma\left(\tfrac{1-\nu}{2 \nu}\right)} \, \Gamma\left( \nu \right)^{-\tfrac{1}{\nu}}.
\end{split}
\end{equation}
So, far we have given the definitions of all quantities in (\ref{primVEV}), apart from those of 
$\om_{2k-1,1-2j}(\alpha).$ 
Now, we fill this gap, too. 
%%%%%%%%%%%%%%%%%%%%%%%%%%%%%%%%%%%%%%%%%%%%%%%%%%%%%%%%%%%%%%%%%%%%%%%%%%%%
First, we eliminate some trivial constant factors from the original definition of \cite{Jimbo:2010jv}:
\begin{equation}  \label{omtom}
\begin{split}
\om_{2k-1,1-2j}(\alpha)=\frac{i}{\pi \, \nu}\, \tom_{2k-1,1-2j}(\alpha),
\end{split}
\end{equation}
where the nontrivial part of this definition is involved in $\tilde{\omega}_{2k-1,1-2j}(\alpha).$ 
It is defined by an integral containing functions related to the integrable description of the 
finite volume groundstate energy\footnote{This definition implies the symmetry property:  $\tom_{1-2j,2k-1}(\al)=\tom_{2k-1,1-2j}(-\al).$}:
\begin{equation} \label{omegatilde}
\begin{split}
\tilde{\omega}_{2k-1,1-2j}(\alpha)=\int\limits_{-\infty}^\infty dx \, e^{(2k-1)x} \, {\mathfrak m}(x) \, {\cal G}^{(\alpha)}_{1-2j}(x), 
\qquad j,k=1,2,...
\end{split}
\end{equation}
Here ${\mathfrak m}(x)$ is a measure made out of the counting-function $Z(x)$ of the nonlinear integral  equation (NLIE) 
(\ref{DDV}) describing 
the finite volume ground state energy of the model:
\begin{equation} \label{m}
\begin{split}
{\mathfrak m}(x)\equiv{\mathfrak m}[Z(x)]={\cal L}_+[Z(x+i\, 0)]+{\cal L}_-[Z(x-i\, 0)], \qquad \mbox{with} \qquad
{\cal L}_\pm[Z]=\frac{e^{\pm i \, Z}}{1+e^{\pm i \, Z}},
\end{split}
\end{equation}
and ${\cal G}_{1-2j}^{(\alpha)}(x)$ satisfies the linear integral equation as follows\footnote{The definitions, (\ref{omtom}), (\ref{omegatilde}) and (\ref{G1m2j}) are also valid for 
arbitrary $j,k \in \mathbb{Z}.$}:
\begin{equation} \label{G1m2j}
\begin{split}
{\cal G}^{(\alpha)}_{1-2j}(x)-\int\limits_{-\infty}^\infty dy \, G_{\alpha}(x-y) \, {\mathfrak m}(y) \, {\cal G}^{(\alpha)}_{1-2j}(y)=e^{(1-2j)x},
\end{split}
\end{equation}
with $G_{\alpha}(x)$ being the deformed kernel of the NLIE\footnote{The relation of 
$G_{\alpha}(x)$ to the function $R(\lambda|\alpha)$ in \cite{Jimbo:2010jv}, is as follows: 
$G_{\alpha}(x)=-\nu \, R(\nu \, x|\alpha)$ with $\nu=1/(p+1).$}:
\begin{equation} \label{Galpha}
\begin{split}
G_{\alpha}(x)=\int\limits_{-\infty}^\infty \! \! \frac{d\omega}{2 \pi} \, \tilde{G}_{\alpha}(\omega) \, e^{i \, \omega \, x}, \qquad 
\mbox{with} \qquad 
\tilde{G}_{\alpha}(\omega)=\frac{ \sinh\left(\tfrac{\pi \omega (p-1)}{2}+i \tfrac{\pi \, \alpha}{2}\right) }
{2 \, \cosh\left(\tfrac{\pi \omega}{2}\right) \, \sinh\left(\tfrac{\pi \omega p}{2}+i \tfrac{\pi \, \alpha}{2}\right)},
\end{split}
\end{equation}
where, in order to be more conform with the literature of the NLIE \cite{Feverati:1998dt,Feverati:1998uz,Feverati:1999sr,Feverati:2000xa}, 
we introduced $p$ as an  alternative reparameterization of $\nu:$
\begin{equation}  \label{pnu}
\begin{split}
\nu=\frac{1}{1+p}, \qquad p=1-\frac{1}{\nu}.
\end{split}
\end{equation}
In this parameter the $0<p<1$ and $1<p<\infty$ regimes correspond to the attractive and repulsive regimes 
of the theory, respectively.

The parameter $\alpha$ in (\ref{Galpha}), is a kind of deformation parameter, such that $G_{\alpha}(x)$ at $\alpha=0$ corresponds to the  kernel of the NLIE (\ref{DDV}), which we denote simply by $G(x).$ 

The counting-function $Z(x)$ is an $i \pi\, (p+1)$ periodic function. 
In the fundamental regime: $|\text{Im}(x)|<\text{min}(\pi,p \,\pi ),$ it satisfies the nonlinear 
integral equation (NLIE) as follows \cite{Destri:1992qk,Destri:1994bv}:
\begin{equation} \label{DDV}
\begin{split}
Z(x)=\ell \, \sinh x+\alpha_z+\frac{1}{i} \!\!\! \int\limits_{-\infty}^\infty dy  \left\{ G(x-y-i \, 0) \, L_+(y+i \, 0)-  G(x-y+i \, 0) \, L_-(y-i \, 0) \right\},
\end{split}
\end{equation}
where for short we introduced the notations:
\begin{equation} \label{Lpmdef}
\begin{split}
L_\pm(x)=\log\left(1+e^{\pm i \, Z(x)} \right), \qquad \ell={\cal M} \, L,
\end{split}
\end{equation}
with ${\cal M}$ and $L$ being the infinite volume soliton mass and the finite volume, respectively.

The counting-function in (\ref{DDV}), is an important function in the integrable formulation of the spectral problem, 
since it allows one to determine the finite volume groundstate energy of the sine-Gordon model by the formula as follows\footnote{For the extension to the full spectrum, 
see \cite{Fioravanti:1996rz,Feverati:1998dt,Feverati:1998uz,Feverati:2000xa}. 
}:
\begin{equation}
E_0(L)=-\frac{\cal M}{2 \, \pi \, i}\, \int\limits_{-\infty}^\infty \!\! dx \, \sinh x \, \log \frac{1+e^{i \, Z(x+i\, 0)}}{1+e^{-i \, Z(x-i\, 0)}}.
\end{equation}

It is important to note, that we let the existence of a twist parameter $\alpha_z$ in the NLIE, in order 
to be able to discuss the vacuum states of twisted sine-Gordon theories \cite{Fioravanti:1996rz,Feverati:1999sr,Feverati:2000xa}. 
This parameter becomes important, 
when either minimal models and their integrable perturbations  or states in the underlying complex Liouville 
CFT are intended to be described with its help. 
Nevertheless, we pay the readers attention to keep in mind the difference between the two, so far introduced 
deformation parameters. The parameter $\alpha$ characterizes the operator, whose vacuum expectation 
value is investigated, while $\alpha_z$ introduces a twist in the NLIE describing the finite volume energy of the sandwiching state.

As we will see in the later sections, our quantity (\ref{omegatilde}) evolves the UV expansion as follows: 
\begin{equation} \label{UVexpansion}
\begin{split}
\tilde{\omega}_{2k-1,1-2j}(\alpha)\!\!=\!\!\left(\frac{2}{\ell}\right)^{2(k+j-1)+\tfrac{2 (\alpha-2)}{p+1}} \! \! \!
\left( \tilde{\omega}_{2k-1,1-2j|1}^{(\alpha)}+  \left(\tfrac{\ell}{2}\right)^{\tfrac{4}{p+1}} \tilde{\omega}_{2k-1,1-2j|2}^{(\alpha)}+O(\ell^{\tfrac{8}{p+1}}) \right).
\end{split}
\end{equation}
The main goal of this paper is to determine the first two coefficients of this expansion
%(\ref{UVexpansion}) 
as functionals built out of functions characterizing the UV limit of the model.

To do so, first we have to discuss in more detail the behaviour of $Z(x)$ in the UV limit.

\section{The UV behaviour of the counting-function} \label{3}

As we mentioned the counting-function $Z(x)$ is the solution of the NLIE (\ref{DDV}). The $\ell$ 
dependence of the quantities of our interest, arises mostly through the $\ell-$dependence of the 
counting-function. The reason for that, is that the measure is defined from $Z(x)$ by (\ref{m}), and 
this measure is present in both the direct definition (\ref{omegatilde}) of $\tom_{2k-1,1-2j}(\alpha)$ and 
in the linear problem (\ref{G1m2j}), the solution of which also enters  (\ref{omegatilde}). 

Consequently, the description the UV behaviour of $\tom_{2k-1,1-2j}(\alpha),$ requires 
 the knowledge of the UV behaviour of the counting-function, too. In this section, we 
discuss this problem. 

The qualitative UV behaviour of $Z(x)$ is well known from the literature \cite{Destri:1994bv,Destri:1997yz,Feverati:1998dt,Feverati:1998uz}. 
The $\ell \to 0$ behaviour of the source term $\ell \sinh x$ in (\ref{DDV}), gives the 
basic indications. Basically, it has three different types of behaviour, corresponding to 
three different regimes in $x:$
\begin{equation}  \label{lshUV}
\begin{split}
\ell \sinh \, x \simeq \left\{\begin{array}{ll} 
e^{x-\ltl} & \mbox{ $\ltl \lesssim x,$ } \\
0 & \mbox{ $-\ltl \lesssim x \lesssim \ltl,$ } \\
-e^{-(x+\ltl)} & \mbox{ $ x \lesssim-\ltl.$ }
\end{array} \right.
\end{split}
\end{equation}
Corresponding to the three different regimes, three different functions describe 
the solution of the NLIE in the UV limit. In the middle regime, which becomes infinitely 
large when $\ell \to 0,$ the solution of the NLIE tends to a constant, $z_0.$ It is called the 
plateau value. 
In the other two regimes, the $\pm$ kink functions can be defined, as appropriate limits of $Z(x)$: 
 \begin{equation}  \label{Zpmdef}
\begin{split}
Z_\pm(x)=\lim\limits_{\ell \to 0} Z(x\pm \ltl).
\end{split}
\end{equation}
They are important functions in the integrable description of the CFT describing the UV limit of the sine-Gordon model, 
since the spectral data of the CFT $(c,\Delta_\pm)$ corresponding to the groundstate can be expressed in terms of them as follows:
\begin{equation}  \label{EpmCFT}
\begin{split}
E_0^{CFT}(L)&=E_0^+(L)+E_0^-(L), \\
E_0^\pm(L)&=\frac{2 \pi}{L} \, \left(\Delta_\pm-\frac{c}{24} \right)=\mp \frac{1}{ L} \intinf \!\! \frac{dx}{4 \pi i } \, e^{\pm x} 
\log \frac{1+e^{i \, Z_\pm(x+i \, 0)}}{1+e^{-i \, Z_\pm(x-i \, 0)}}.
\end{split}
\end{equation}
Applying the standard UV computation technique \cite{Destri:1994bv,Destri:1997yz,Fioravanti:1996rz,Feverati:1998dt}, to the equations satisfied by the 
kinks (\ref{kinksol}), the following values can be obtained for the "chiral" energies:
\begin{equation}  \label{EpmCFTcomputed}
\begin{split}
E_0^\pm(L)&=-\frac{\pi}{12 \, L} \, \left(1-6\, (1-\nu)\, \left(\frac{\al_z}{\pi}\right)^2 \right).
\end{split}
\end{equation}
In the complex Liouville CFT formulation it implies the conformal dimensions as follows:
\begin{equation}  \label{DLCFT}
\begin{split}
\Delta_\pm \to &\Delta_{1\pm \kappa_z}, \qquad \text{with} \quad \kappa_z=\frac{\al_z}{\pi} \, \frac{1-\nu}{\nu},
\end{split}
\end{equation}
where $\Delta_{1\pm \kappa_z}$ should be computed from (\ref{Dalfa}).

From (\ref{DDV}) it can be shown, that $Z_\pm(x)$ satisfy the nonlinear integral equations, as follows:
\begin{equation} \label{kinksol}
\begin{split}
Z_\pm(x)=\pm e^{\pm x}+\alpha_z+\frac{1}{i} \!\!\! \int\limits_{-\infty}^\infty dy  \left\{ G(x-y-i \, 0) \, L^{(\pm)}_+(y+i \, 0)
-  G(x-y+i \, 0) \, L^{(\pm)}_-(y-i \, 0) \right\},
\end{split}
\end{equation}
where for short we introduced the notation:
\begin{equation} \label{Lpmdef}
\begin{split}
L^{(\sigma)}_\pm(x)=\log\left(1+e^{\pm i \, Z_\sigma(x)} \right), \qquad \sigma \in \{\pm\}.
\end{split}
\end{equation}
From these equations it can be shown that the plateau value $z_0,$ is nothing but certain 
limiting value of the functions $Z_{\pm}(x):$
\begin{equation}  \label{Zpm0}
\begin{split}
Z_{\pm}(\mp \infty)=z_0.
\end{split}
\end{equation}
It satisfies the so-called plateau equation \cite{Destri:1997yz}:
\begin{equation}  \label{plateq}
\begin{split}
z_0=\al_z+\frac{1}{i} \left( \intinf \! dx \, G(x) \right) \,
\left( \log\left( 1+e^{i  \, z_0} \right)-\log\left( 1+e^{-i \, z_0} \right) \right).
\end{split}
\end{equation}
Using naively the identity 
$\log\left( 1+e^{i \, z_0} \right)-\log\left( 1+e^{-i \, z_0} \right) =i \,z_0,$ 
in most of the cases the solution of (\ref{plateq}) reduces to solve a simple linear equation for $z_0.$  
Nevertheless, in certain regimes of the parameters $(p,\al_z)$ 
the existence of the branch cut of the $\log$ function can make the solution more cumbersome \cite{Destri:1997yz,Feverati:1998dt,Feverati:1998uz,Feverati:1999sr,Feverati:2000xa}.

All these facts imply the following qualitative leading order UV behaviour of the counting-function. 
Corresponding to the  three different regimes in $x,$ it has three different types of behaviour, as follows:  
\begin{equation}  \label{ZUV}
\begin{split}
Z(x) \simeq \left\{\begin{array}{ll} 
Z_+(x-\ltl) & \mbox{ $\ltl \lesssim x,$ } \\
z_0 & \mbox{ $-\ltl \lesssim x \lesssim \ltl,$ } \\
Z_-(x+\ltl) & \mbox{ $ x \lesssim-\ltl.$ }
\end{array} \right.
\end{split}
\end{equation}

Following the idea of \cite{Bajnok:2022ucr}, one can define an asymptotic solution of (\ref{DDV}), 
which shares the same qualitative leading UV behaviour with $Z(x).$
Following from (\ref{ZUV}), it is made out of the kinks and the plateau value 
and it accounts for the leading UV behaviour of the problem. 
The definition is as follows \cite{Bajnok:2022ucr}:
\begin{equation}  \label{Zasdef}
\begin{split}
Z_{as}(x)=Z_+(x-\ltl)+Z_-(x+\ltl)-z_0.
\end{split}
\end{equation}
Based on  perturbed conformal field theory (PCFT) considerations 
and on our high precision numerical experience in solving (\ref{DDV}), 
we assume that  the $\ell$ dependency structure of (\ref{Zasdef}) 
holds in higher UV corrections, as well.
This led us to the following 
Ansatz for describing the small $\ell$ series representation of $Z(x):$ 
\begin{equation}  \label{ZUVansatz}
\begin{split}
Z(x)&=Z_{as}(x)+\de Z(x), \\
\de Z(x)&= \sum\limits_{n=1}^{\infty} \left(\frac{2}{\ell}\right)^{\tfrac{4 \, n}{p+1}} \, 
\left( \de Z^{(n)}_{+}(x-\ltl)+\de Z^{(n)}_{-}(x+\ltl)-\de Z^{(n)}_0 \right),
\end{split}
\end{equation}
where $\de Z^{(n)}_{\pm}(x)$ denotes the $n$th UV correction to the kinks  
and $\de Z^{(n)}_0$ means the connecting constant plateau value for the kink corrections:
\begin{equation}  \label{dZn0}
\begin{split}
\de Z^{(n)}_0=\de Z^{(n)}_{\pm}(\mp \infty).
\end{split}
\end{equation}
Here, $\de Z_\pm(x)$ and $\de Z^{(n)}_0,$ are supposed to be $\ell-$independent quantities.

To summarize, it is assumed that corrections to the asymptotic solution, go with positive integer powers 
of $\ell^{\tfrac{4}{p+1}},$ such that the kink-plateau structure and so the qualitative 
$\ell-$dependence of the multiplying correction functions 
is the same as that of the asymptotic solution.

With the help of the NLIE (\ref{DDV}) and the kink- (\ref{kinksol}) and plateau-equations (\ref{plateq}), 
a linear equation can be derived for: 
\begin{equation}  \label{dZ1}
\begin{split}
\de Z^{(1)}(x)=\left( \frac{\ell}{2}\right)^{\tfrac{4}{p+1}}
\left(\de Z^{(1)}_+(x-\ltl)+\de Z^{(1)}_-(x+\ltl) -\de Z^{(1)}_0 \right) +O(\ell^{\tfrac{8}{p+1}}),
\end{split}
\end{equation}
which is the leading order in $\ell$ correction to the asymptotic solution.

The above mentioned linear equation becomes important, 
when computing the 2nd order UV terms to the quantities $\om_{2k-1,1-2j}(0).$  
Thus, the equation is presented in appendix \ref{appA} in formulas (\ref{dZdef})-(\ref{dZk}), when the 
derivation of the coefficients in the UV series representation for $\om_{1,-1}(0)$ is derived.

\section{UV predictions from field theory} \label{4}

Considering the sine-Gordon model as a perturbed complex Liouville theory, 
the Liouville theory's 3-point functions \cite{Dotsenko:1984nm,Zamolodchikov:1995aa,Teschner:1995yf}, 
give predictions to the 
leading UV behaviour of the ratios of expectation values of the primaries (\ref{primLiou}). 
Applying it to the quantity of our interest (\ref{primVEV}), the following analytical 
result is expected in the UV \cite{Jimbo:2010jv}:
\begin{equation} \label{VEVcft}
\begin{split}
\frac{\langle  \Phi_{\alpha+2 \,m\, \tfrac{1-\nu}{\nu}}(0) \rangle}{\langle  \Phi_{\alpha}(0) \rangle}\simeq 
\left(\frac{2 \, \pi}{ L} \right)^{2(\Delta_{\al+2 m (1-\nu)/\nu}-\Delta_{\al})} \, 
\prod\limits_{j=0}^{m-1}\, {\cal V}(\alpha+ 2\,j \, \tfrac{1-\nu}{\nu},\kappa_\Delta),
\end{split}
\end{equation}
where $\Delta$ denotes the conformal dimension of the sandwiching primary operator in the Liouville CFT
and $\kappa_\Delta$ is defined from it as follows:
\begin{equation}  \label{kappaDelta}
\begin{split}
\kappa_\Delta=\sqrt{1-\frac{\Delta}{\Delta_1}}, \qquad \text{with} \quad \Delta_1=-\frac{\nu^2}{4(1-\nu)}.
\end{split}
\end{equation}
We note, that to get the vacuum expectation values: the formula (\ref{VEVcft}) should be taken at the 
$\Delta \to \Delta_{1\pm \kappa_z}$  points, such that $\kappa_z$ is given in (\ref{DLCFT}).

The definitions of the functions entering (\ref{VEVcft}) are as follows \cite{Jimbo:2010jv}: 
\begin{equation} \label{prim2CFT}
\begin{split}
{\cal V}(\alpha,\kappa)=\mu^2 \, \Gamma(\nu)^2 \, Y(x(\alpha)) \, W(\alpha,\kappa) \, W(\alpha,-\kappa),
\end{split}
\end{equation}
where $\mu$ and $x(\alpha)$ are defined in (\ref{muM}) and (\ref{C1}) respectively, and 
\begin{equation} \label{Yx}
\begin{split}
Y(x)&=-2 \, \nu\, x \cdot \frac{\Gamma^2(\nu x+1/2-\nu/2) \, \Gamma(\nu-2 \nu x)}{\Gamma^2(1/2+\nu/2-\nu x) \, \Gamma(2 \nu x+1-\nu)}\cdot 
\frac{\Gamma(-2 \nu x)}{\Gamma(2 \nu x)}, \\
W(\alpha,\kappa)&=\frac{\Gamma(\alpha\nu/2-\nu+1+\kappa \nu)}{\Gamma(-\alpha \nu/2+\nu+\kappa\nu)}.
\end{split}
\end{equation}
Using the concrete expressions for the dimensions, the leading volume dependence of the expectation values
(\ref{primVEV}) take the form: 
\begin{equation} \label{simVEV}
\begin{split}
\frac{\langle  \Phi_{\alpha+2 \,m\, \tfrac{1-\nu}{\nu}}(0) \rangle}{\langle  \Phi_{\alpha}(0) \rangle}\sim
%\frac{1}{L^{2\, m \left(m-\tfrac{m+1}{p+1} \right)+\frac{2\, m \al}{p+1}}}
L^{-2\, m \left(m-(m+1)\nu \right)-2\, m \al \, \nu}, \qquad m=1,2,...
\end{split}
\end{equation}
Looking at the formula (\ref{primVEV}), one can conclude, that the analytical prediction for the 
ratios of expectation values of primaries, is not enough to extract the leading order in the UV limit expressions 
for the quantities $\om_{2k-1,1-2 j}(\al).$ Only the leading UV behaviour of their nonlinear determinant 
like combinations can be determined from comparing (\ref{primVEV}) and (\ref{VEVcft}).  

The only exception is the case\footnote{We note, that the leading UV behaviour of some other matrix elements of $\om,$ 
(e.g. $\om_{3,-1}(\al)$) can also be 
determined, if the descendant fields are also considered in this formalism. See \cite{Jimbo:2010jv,Hegedus:2019rju}. } of $m=1,$ when the UV behaviour of $\om_{1,-1}(\al)$ 
is completely determined by (\ref{primVEV}) and (\ref{VEVcft}) . 
This is because, in this special case, one is left with a determinant of a $1 \times 1$ matrix. Thus comparing
(\ref{primVEV}) to (\ref{VEVcft}) directly leads to a leading order prediction for $\om_{1,-1}(\al).$  
%in the repulsive regime. 
The predicted volume dependence in the repulsive regime takes the form:
\begin{equation} \label{om1m1pre}
\begin{split}
\om_{1,-1}(\al)\sim L^{-2+\tfrac{2\al-4}{p+1}}, %\quad \mbox{with} \quad \nu=\frac{1}{p+1}, 
\qquad \mbox{for} \quad 1<p, \, \quad 0<\al<2.
\end{split}
\end{equation}
We note, that we obtained the same result in appendix \ref{appA}.

\subsection{ The structure of the UV series from field theory, in the $\al=0$ case }

If one is interested in the ratios of expectation values given in (\ref{primVEV}), when $\al=0,$ then the 
perturbed compactified boson CFT formulation of the sine-Gordon model proves to be useful. 

In this description the cosine term in  
 (\ref{LSG}) plays the role of the perturbation:
\begin{equation} \label{PBform}
\begin{split}
{\cal A}_{SG}={\cal A}_{B}- \frac{ 2 {\bf \mu}^2}{\sin \pi \beta^2} \int \cos(\beta \, \varphi(z,\bar{z})) \frac{i \, dz \wedge \, d\bar{z}}{2},\quad  
\end{split}
\end{equation}
where ${\cal A}_B$ denotes the action of the $c=1$ free boson compactified on a circle of radius $R=\tfrac{1}{\beta}:$
\begin{equation} \label{Abe}
\begin{split}
{\cal A}_{B}=\int  \frac{1}{4 \pi} \partial_z \varphi(z,\bar{z}) \, \partial_{\bar{z}} \varphi(z,\bar{z})
 \, \frac{i \, dz \wedge \, d\bar{z}}{2}.
\end{split}
\end{equation}
The primary states in the unperturbed CFT correspond to vertex operators $V_{n,m}(z,\bar{z}).$ 
They are labeled by two quantum numbers: $n \in {\mathbb R},$ is called the momentum quantum number, 
and $m \in {\mathbb Z}$ is the winding number or topological charge.
The conformal dimensions of the primary states are given by \cite{Klassen:1992eq}:% (Ginsparg CFT):
\begin{equation} \label{Deltapm}
\begin{split}
\Delta^{\pm}_{n,m}=\left(\frac{n}{R} \pm \frac{1}{4} m R \right)^2.
\end{split}
\end{equation}
In \cite{Klassen:1992eq}, it has been shown, that the requirement of locality of the operator product 
algebra of the CFT,  
further restricts the allowed values of the quantum numbers $(n,m).$ 
It turns out, that the UV limit of the sine-Gordon model is described by a modular 
invariant compactified boson CFT, 
where the quantum numbers of the vertex operators can take only integer values: 
 $\{n \in {\mathbb Z}, m \in {\mathbb Z}\}.$

In this formulation the perturbing term in the action (\ref{PBform}) is given in terms of the vertex operators $V_{\pm 1,0}(z,\bar{z})$:
\begin{equation} \label{Apert}
\begin{split}
{\cal A}_{pert}=-\frac{\mu^2}{\sin \pi \beta^2}\int \, \frac{i \, dz \wedge \, d\bar{z}}{2}  \left( V_{1,0}(z,\bar{z})+V_{-1,0}(z,\bar{z})\right).
\end{split}
\end{equation}
In this paper our main interest is to study the UV limit of ratios of vacuum expectation values of primaries given in (\ref{primVEV}). 
To consider these quantities, it is enough to focus on the the zero winding number sector. 
It is easy to show, that the vertex operators $V_{m,0}(z,\bar{z})$  in the $c=1$ PCFT formulation, correspond to the primaries 
$\Phi_{2 \, m \, \tfrac{1-\nu}{\nu}}(z,\bar{z})$ in the perturbed complex Liouville formulation. At the level of vacuum expectation values, it reads:
\begin{equation} \label{VEVrel0}
\begin{split}
\lim\limits_{\al \to 0} \frac{\langle  \Phi_{\alpha+2 \,m\, \tfrac{1-\nu}{\nu}}(0) \rangle}{\langle  \Phi_{\alpha}(0) \rangle} \to 
\frac{\langle  V_{m,0}(0) \rangle}{\langle 1 \rangle}.
\end{split}
\end{equation}
To compute these vacuum expectation values in the framework of perturbed conformal field theory (PCFT), 
multi-point correction functions of the vertex operators of the $c=1$ CFT will enter the procedure \cite{Zamolodchikov:1990bk}. 
What is important for us, is that such a correlator is non-zero only, if the sum of momentum quantum numbers of the constituent vertex operators are zero:
\begin{equation} \label{corrFrule}
\begin{split}
\langle V_{m_1,0}(z_1,\bar{z}_1)  \, V_{m_2,0}(z_2,\bar{z}_2)...V_{m_k,0}(z_k,\bar{z}_k) \rangle_{CFT} \neq 0, 
\qquad \mbox{if :} \quad m_1+m_2+...+m_k=0.
\end{split}
\end{equation}
From (\ref{Apert}), one can see, that the dimensionful coupling constant of the perturbing term is effectively $\lambda \sim \mu^2.$ Following the usual steps  
of perturbation theory in the perturbed $c=1$ CFT formulation, one can conclude, that:
\begin{itemize}
\item{ the expectation value $\langle V_{m,0}(0)\rangle$ starts at the $m$th order in the perturbation theory. Namely, its leading order behaviour is given by: 
$\sim \lambda^m\sim \mu^{2 \, m}.$}
\item{The corrections to the leading order in $\lambda$ behaviour come in even powers of $\lambda \sim \mu^2.$}
\end{itemize}
Relying on this structure, some dimensional analysis supplemented with the  mass coupling relation (\ref{muM}), leads to the following formal small volume 
dependence of the expectation values of the vertex operators:
\begin{equation} \label{VEVell}
\begin{split}
\langle V_{m,0}(0) \rangle=K^{(m)}_0 \, L^{-2 \Delta_{m,0}+2 \, m \, \nu} \, \left( 1+\sum\limits_{s=1}^\infty  k^{(m)}_s \, \ell^{\, 4 \, s\, \nu}, %\ell^{\tfrac{4 s}{p+1}}
\right), \quad \mbox{with} \quad 
\ell= {\cal M} L.
\end{split}
\end{equation}
Namely, the leading order scaling in $L$ is governed by the sum of the conformal dimension $\Delta_{m,0}=m^2 \, (1-\nu)$ 
of the vertex operator in the $c=1$ CFT and the dimension of the $m$-th power of the coupling constant. 
Furthermore, the corrections go as positive integer powers of $\ell^{4 \nu}=\ell^{\, \tfrac{4}{p+1}}.$

To give a clearer vision to the readers on the volume dependence of the quantities $\om_{2k-1,1-2j}(\al),$ 
we show how the UV series representation of $\om_{1,-1}(0)$ is related to that of the ground state energy.
We do this, since the finite volume ground state energy was investigated in a wide range of integrable models, 
and we think that the presentation of this relation can be useful for those, who are more comfortable 
with the spectral problem. 

The relation is founded by the fact, that the vacuum expectation value of the trace of the stress-energy 
tensor denoted here by $\Theta,$ can be expressed either in terms of the ground state energy or in terms of $\om_{1,-1}(0).$  The two different 
formulations connect the UV series of the two different quantities. 

It is well known from PCFT considerations,  %\cite{}, 
that the ground state energy has the UV series representation as follows:
\begin{equation} \label{E0ser}
\begin{split}
E_0(L)=E_b \, L-\frac{\pi}{6 \, L} \, c_{}\, \left( 1+\sum\limits_{k=1}^\infty  \epsilon_k \, \ell^{\, 4 \, k\, \nu} \right), \qquad \ell={\cal M}\, L,
\end{split}
\end{equation}
where $c_{}=1$ is the  central charge of the unperturbed CFT, $E_b$ is the bulk energy constant:
\begin{equation} \label{Eb}
\begin{split}
E_b=\frac{{\cal M}}{4}\, \cot \frac{\pi}{2 \, \nu}.
\end{split}
\end{equation}
In principle, the coefficients $\epsilon_k$ can be computed as integrals of multi-point correlators 
of vertex operators in the unperturbed CFT. 

In the normalization used in \cite{Jimbo:2010jv}, the vacuum expectation value of the trace of the stress-energy tensor can be expressed in terms of the primary field $\Phi_{2 \tfrac{1-\nu}{\nu}}$ as follows:
\begin{equation} \label{THprim}
\begin{split}
\Theta=-\frac{2 \, \pi \, \nu}{\sin(\pi \, \nu)}\, \mu^2 \,\, \Phi_{2 \tfrac{1-\nu}{\nu}}.
\end{split}
\end{equation}
Then, its vacuum expectation value is related to the groundstate energy by the formula:
\begin{equation} \label{THdefs}
\begin{split}
\langle \Theta \rangle=\frac{\pi}{2} {\cal M} \, \left( \frac{1}{\ell}+\frac{d}{d \ell} \right) \, E_0(\ell),
\end{split}
\end{equation}
which together with (\ref{E0ser}), implies the following UV series representation:
\begin{equation} \label{THser}
\begin{split}
\langle \Theta \rangle={\cal M}^2 \, \pi \,  \left(  \frac{E_b}{{\cal M}}-\frac{\pi \, c \, \nu}{3} \sum\limits_{k=1}^{\infty} \, k \, \epsilon_k \, \ell^{-2+4 \, k \, \nu}  \right).
\end{split}
\end{equation}
On the other hand, from (\ref{primVEV})-(\ref{muM}) and (\ref{THprim}), $\langle \Theta \rangle$ can be expressed in terms of $\om_{1,-1}(0)$ \cite{Jimbo:2010jv}:
\begin{equation} \label{THom}
\begin{split}
\langle \Theta \rangle=-{\cal M}^2 \, \frac{2 \, \pi \, \nu}{\sin(\pi \nu)} \, C_1(0) \, \Pi(\nu)^2 \, \left(\om_{1,-1}(0)+\frac{i}{\nu} \, \cot \frac{\pi}{2 \nu} \right).
\end{split}
\end{equation}
Comparing (\ref{THser}) and (\ref{THom}), one can recognize, that on the one hand, 
the term being not proportional to $\om_{1,-1}(0)$ in (\ref{THom}) corresponds to the 
$\sim E_b$ bulk energy contribution in (\ref{THser}). 
On the other hand, the rest yields the UV series representation of $\om_{1,-1}(0):$ 
\begin{equation} \label{om1m10ser}
\begin{split}
\om_{1,-1}(0)=-i\,  \frac{4\, \pi \, c}{3} \sum\limits_{k=1}^{\infty} k \, \epsilon_k \, 
\ell^{-2+\tfrac{4 \, k}{p+1}}.
%\om_{1,-1}(0)=\frac{\sin (\pi \nu)}{2 \, \pi \, \nu \, C_1(0) \, \Pi(\nu)^2} \, \frac{\pi \, c}{12} \sum\limits_{k=1}^{\infty} \frac{4 \, k}{p+1} \, c_k \, \ell^{-2+\tfrac{4 \, k}{p+1}}
\end{split}
\end{equation}
Here, the coefficients of the UV series are expressed in terms of those of the ground state energy (\ref{E0ser}).
What is important for our later considerations is that the series goes in powers $\ell^{-2+\tfrac{4 \, k}{p+1}}$ with $k \in \mathbb{N}.$ 
We just notice, that in the final formula (\ref{om1m10ser}), we substituted $\nu \to \tfrac{1}{p+1},$ since 
in the subsequent integrable description based computations, we use this parameterization of the problem.

Similar structure holds for the other $\om_{2k-1,1-2j}(0)$ quantities. For example, from \cite{Jimbo:2010jv} 
and \cite{Hegedus:2019rju} it is known, that:
\begin{equation} \label{lm2PHI}
\begin{split}
\langle {\bf l}_{-2} \Phi_{2 \tfrac{1-\nu}{\nu}}\rangle \sim
%=\frac{C_1(0) \, \tfrac{1}{i \, \nu}\, \cot \tfrac{\pi}{2 \nu} \, \Pi(\nu)^{4-2\nu}  }{D_1\left(2 \tfrac{1-\nu}{\nu}\right) \,D_1\left(2-2 \tfrac{1-\nu}{\nu}\right)} \, 
\omega_{3,-1}(0), 
\end{split}
\end{equation}
which implies, that the UV series of $\omega_{3,-1}(0)$ goes as $\ell^{-4+\tfrac{2 \, k}{p+1}}$ with $k \in \mathbb{N}.$ 
Namely, only the global order independent integer part of the exponent, changes as the subscripts of $\om$ changes, but the structure of the 
corrections multiplying it, remains the same.

At this point, we should also discuss, how the UV series of $\langle \Phi_{2 \, m\, \tfrac{1-\nu}{\nu}}\rangle$ is related to those of $\om_{2k-1,1-2j}(0).$ Namely, we would like 
to answer the question, as follows. How many nontrivial coefficients of the UV series of 
$\langle \Phi_{2 \, m\, \tfrac{1-\nu}{\nu}}\rangle$ can\footnote{In order to fix 
notational normalization ambiguities,
we just note, that here and in the 
rest of the paper 
$\langle \Phi_{2 \, m\, \tfrac{1-\nu}{\nu}}\rangle$ stands for $\langle \Phi_{2 \, m\, \tfrac{1-\nu}{\nu}}\rangle/\langle {\bf 1} \rangle.$}
 be determined from the
knowledge of the first two nonzero coefficients in the UV series of $\om_{2k-1,1-2j}(0)$?   

As we have shown in (\ref{simVEV}), perturbed CFT considerations imply the leading UV series for the expectation 
value at $\al=0$ as follows:
\begin{equation} \label{phiexp}
\begin{split}
\langle \Phi_{2 \, m\, \tfrac{1-\nu}{\nu}}\rangle \sim \ell^{-2 \, m^2+\tfrac{4}{p+1} \tfrac{m(m+1)}{2}}(1+O(\ell^{\tfrac{4}{p+1}})).
\end{split}
\end{equation}
Later in this paper, in (\ref{UVcoeff1}) we have shown, 
that the leading UV behaviour of $\om_{2k-1,1-2j}(0),$ behaves as follows:
\begin{equation} \label{omexp}
\begin{split}
\om_{2k-1,1-2j}(0)\sim \ell^{-2(k+j-1)+\tfrac{4}{p+1}}.
\end{split}
\end{equation}
Based on the formula (\ref{primVEV}), one can show, that:
\begin{itemize}
\item{For $1<p,$ the leading UV contribution comes from that part of the determinant, which contains 
only the $\om$s. Namely, as if the $\de_{jk}$ term in (\ref{OM}) was deleted and only the   
  $\underset{1 \leq j,k \leq m }{\mbox{det}} \, \omega_{2k-1,1-2j}(\alpha)$ part gives  
 the relevant contribution. }
\item{ Assuming the leading UV behaviour (\ref{omexp}) for the $\om$s and that there are no special 
relations among the coefficients of the expansions of the $\om$s, (\ref{primVEV}) would suggest 
the following naive prediction for the UV behaviour of $\langle \Phi_{2 \, m\, \tfrac{1-\nu}{\nu}}\rangle:$ 
\begin{equation} \label{phiexpwrong}
\begin{split}
\langle \Phi_{2 \, m\, \tfrac{1-\nu}{\nu}}\rangle \sim \ell^{-2 \, m^2+\tfrac{4 \, m}{p+1} }
\left(\varphi_{0}^{(m)}  +\sum\limits_{k=1}^\infty \varphi_{k}^{(m)} \, \ell^{\tfrac{4 \, k}{p+1}} \right).
%O(\ell^{\tfrac{4}{p+1}})).
\end{split}
\end{equation}
This contradicts to field theory prediction (\ref{phiexp}), apart from the $m=1$ special case. 
The two formulas can be reconciled, if one  assumes, that the first few 
$\varphi_{k}^{(m)}$ coefficients in (\ref{phiexpwrong}) are zero. 
This can be easily fulfilled, if there are some nontrivial 
relations among the coefficients of the UV series representations of the different $\om$s. 
}

\item{Assuming such relations, the  computation of the leading order coefficient in the series of 
  $\langle \Phi_{2 \, m\, \tfrac{1-\nu}{\nu}}\rangle,$  requires the knowledge of
the UV series of the constituent $\om$s upto $m(m-1)/2+1$ order.  
This means, that our 2nd order computations allows one to get the leading UV coefficients of the  expectation 
values: 
$\langle \Phi_{2 \, \tfrac{1-\nu}{\nu}}\rangle,$ and $\langle \Phi_{4 \, \tfrac{1-\nu}{\nu}}\rangle$. }

\item{ Our leading order UV result for $\om_{2k-1,1-2j}(0)$ given in (\ref{UVcoeff1}), implies a 
simple separated product dependence on the subscripts:  
\begin{equation} \label{UV1subs}
\begin{split}
\om_{2k-1,1-2 j}(0) \sim f_j \, \tilde{f}_k \, \ell^{-2(j+k-1)+\tfrac{4}{p+1}},
\end{split}
\end{equation}
which substituted into (\ref{primVEV}) immediately, improves the naive prediction with one order in $\ell^{\tfrac{4}{p+1}}$:
\begin{equation} \label{phiexpwrong2}
\begin{split}
\langle \Phi_{2 \, m\, \tfrac{1-\nu}{\nu}}\rangle \sim \ell^{-2 \, m^2+\tfrac{4 \, (m+1)}{p+1}. } \bigg(\varphi_{1}^{(m)}+O(\ell^{\tfrac{4}{p+1}})\bigg), 
\qquad m\geq 2.
\end{split}
\end{equation}
This 1st order improvement allows one to compute the expectation value corresponding to the $m=2$ 
case, provided the 2nd order corrections to the $\om$s are known. This is what we computed in this paper 
with the tools of integrability. 
 }

\end{itemize}

\section{The leading order UV term in $\om_{2k-1,1-2j}(\al)$ with $j,k \in \mathbb{N}$} \label{5}

In this and  the next sections, we summarize the formulas for the first two coefficients of (\ref{UVexpansion}) 
as functionals built out of functions characterizing the UV limit of the model. 

This section is devoted to the discussion of the leading order coefficient, 
while the next to leading order results are summarized in the next section.

In order to be able to present the formulas, one needs to define several necessary quantities.

 Following from the definitions (\ref{Zpmdef}), the kink functions share the same 
periodicity property with the counting-function. Namely, they are periodic with period $i \,\pi\, (p+1).$
As a consequence, they admit the large argument expansion in the plateau regime as follows:
\begin{equation} \label{kinkplat}
\begin{split}
Z_-(x)&\stackrel{x\to +\infty}{=}z_0+c_1 \, e^{-\tfrac{2 \, x}{p+1}}+c_2 \, e^{-\tfrac{4 \, x}{p+1}}+O(e^{-\tfrac{6 \, x}{p+1}}),  \\
Z_+(x)&\stackrel{x \to -\infty}{=}z_0+\tilde{c}_1 \, e^{\tfrac{2 \, x}{p+1}}+\tilde{c}_2 \, e^{\tfrac{4 \, x}{p+1}}+O(e^{\tfrac{6 \, x}{p+1}}).
\end{split}
\end{equation}
Here, the coefficients $c_n$ and $\tilde{c}_n$  are related to the $n$th non-local conserved charge of the theory \cite{Bazhanov:1996dr,Bazhanov:1996aq}.

To formulate the leading order term in (\ref{UVexpansion}), solutions of linear problems defined on "kink" backgrounds are necessary. So, we define the ${\cal G}_{-(2j-1)}^{(\alpha)(-)}(x)$ and 
${\cal G}_{+(2k-1)}^{(-\alpha)(+)}(x)$ functions as solutions of the following linear equations:
\begin{equation} \label{kinkmalpha}
\begin{split}
{\cal G}_{-(2j-1)}^{(\alpha)(-)}(x)-\int\limits_{-\infty}^\infty \! \!dy \, G_{\alpha}(x-y) \, {\mathfrak m}_-(y) \, {\cal G}_{-(2j-1)}^{(\alpha)(-)}(y)
=e^{(1-2j)x}, \qquad j=1,2,...
\end{split}
\end{equation}
\begin{equation} \label{kinkpalpha}
\begin{split}
{\cal G}_{+(2k-1)}^{(-\alpha)(+)}(x)-\int\limits_{-\infty}^\infty \! \!dy \, G_{-\alpha}(x-y) \, {\mathfrak m}_+(y) \, {\cal G}_{+(2k-1)}^{(-\alpha)(+)}(y)
=e^{(2k-1)x}, \qquad k=1,2,...
\end{split}
\end{equation}
where for the sake of simplicity we introduced the short notation for the kink measures: 
${\mathfrak m}_\pm(x)={\mathfrak m}[Z_\pm(x)]$.

It is important to know the large $x$ asymptotics of these functions. They take the forms as follows:
\begin{equation} \label{scGmasy}
\begin{split}
{\cal G}_{-(2j-1)}^{(\alpha)(-)}(x)\,\, & \stackrel{x \to +\infty}{=}e^{\tfrac{\alpha \,x}{p+1}}\, \sum\limits_{n=1}^\infty c_n^{(-(2j-1),\alpha)} \, e^{-\tfrac{2 n x}{p+1}}, \\
%\qquad \mbox{as} \qquad x \to \infty \\
{\cal G}_{-(2j-1)}^{(\alpha)(-)}(x)-e^{-(2j-1)\, x} \,\, &\stackrel{x \to -\infty}{=} \sum\limits_{n=1-\Theta(\alpha)}^\infty a_n^{(-(2j-1),\alpha)} \, e^{\tfrac{\alpha+2 n}{p} x}+
 \sum\limits_{n=0}^\infty b_n^{(-(2j-1),\alpha)} \, e^{(1+2 n)\,x},
\end{split}
\end{equation}
where $\Theta(\alpha)$ is the unit-step function, thus $1-\Theta(\alpha)=1$ if $\alpha=0,$ and it is zero if $\alpha>0.$
Completely analogous formulas are valid in the "+"-kink regime. 
They take the forms as follows:
\begin{equation} \label{scGpasy}
\begin{split}
{\cal G}_{+(2k-1)}^{(-\alpha)(+)}(x)\,\, & \stackrel{x \to -\infty}{=}e^{-\tfrac{\alpha \,x}{p+1}}\, \sum\limits_{n=1}^\infty c_n^{(+(2k-1),-\alpha)} \, e^{\tfrac{2 n x}{p+1}}, \\
%\qquad \mbox{as} \qquad x \to \infty \\
{\cal G}_{+(2k-1)}^{(-\alpha)(+)}(x)-e^{(2k-1)\, x} \,\, &\stackrel{x \to \infty}{=} \! \! \! \! \! \sum\limits_{n=1-\Theta(\alpha)}^\infty \!\!\!\! a_n^{(+(2k-1),-\alpha)} \, e^{-\tfrac{\alpha+2 n}{p} x}+
 \sum\limits_{n=0}^\infty b_n^{(+(2k-1),-\alpha)} \, e^{-(1+2 n)\,x}.
\end{split}
\end{equation}
Here, we note, that the first formulas in (\ref{scGmasy}) and (\ref{scGpasy}) can be justified from the 
high precision numerical solutions of (\ref{kinkmalpha}) and (\ref{kinkpalpha}). On the other hand the 
second formulas in (\ref{scGmasy}) and (\ref{scGpasy}) can be derived directly from the large argument 
series representation of the kernel $G_{\al}(x)$ given in (\ref{Gsoral}).

Now, we are in the position to present the formula for the leading order coefficient in the UV expansion (\ref{UVexpansion}):
\begin{equation} \label{UVcoeff1}
\begin{split}
\tilde{\omega}_{2k-1,1-2j|1}^{(\alpha)}&=  c_1^{(-(2j-1),\alpha)} \! \! \! \int\limits_{-\infty}^{\infty} \! \! \! dx  
\,\, e^{\tfrac{\alpha-2}{p+1} x} \,
\bigg\{ e^{(2k-1) x}\, {\mathfrak m}_+(x)\\
&+({\mathfrak m}_+(x)-{\mathfrak m}[z_0])\, \left[{\cal G}_{+(2k-1)}^{(-\alpha)(+)}(x)-e^{(2k-1)\, x} \right]
\bigg\}.
\end{split}
\end{equation}
It is easy to show that in the repulsive regime $p>1$ for $0<\alpha<2$, the integrals in (\ref{UVcoeff1}) are convergent.
We note, that it follows from (\ref{m}), that  ${\mathfrak m}[z_0]\equiv {\mathfrak m}_0=1,$ independently of the actual value of the plateau.

An important remark concerning the leading order result (\ref{UVcoeff1}) is in order. 
It is easy to see, that the dependence of the result, on the subscripts $j$ and $k$ of $\om$ is 
multiplicative. Namely, the result factorizes into a product of a purely $k$ and purely $j$ dependent terms. 
With the help of this simple recognition, it is easy to see, that the coefficient of the naive 
leading order in $\ell$ term (\ref{phiexpwrong}) predicted by (\ref{primVEV}) and (\ref{UVcoeff1}) is 
zero\footnote{I.e. $\varphi_0^{(m)}=0$ in (\ref{phiexpwrong}), for $m\geq 2.$}, 
if $m \geq 2$. Thus, for $m \geq 2,$ the determination of the leading order coefficients for 
$\langle \Phi_{2 \, m \, \tfrac{1-\nu}{\nu}} \rangle$ requires the knowledge of the UV coefficients 
of $\om$s at least upto 2nd order.  

The leading order result (\ref{UVcoeff1}) gives an integrable description based formula to the ratio  
$\langle \Phi_{\al+2 \, \tfrac{1-\nu}{\nu}}\rangle/\langle \Phi_\al \rangle$, and  
to the expectation value $\langle \Phi_{2 \, \tfrac{1-\nu}{\nu}} \rangle$ 
in the complex Liouville CFT.

\section{The next to leading order UV term in $\om_{2k-1,1-2j}(\al)$ with $j,k \in \mathbb{N}$} \label{6}

For the sake of simplicity and necessity, we present the next to leading order correction only in the $\alpha=0$ undeformed case. 
To formulate this correction, one has to solve a series of linear problems, which describe the perturbations of the kinks (\ref{kinksol}) and 
the solutions of the linear problems on kink backgrounds (\ref{kinkmalpha}),(\ref{kinkpalpha}).
%\newline

To avoid carrying unnecessary amount of indexes, we rewrite (\ref{kinkmalpha}),(\ref{kinkpalpha}) again at $\alpha\!\!=\!\!0:\!$
\begin{equation} \label{kinkmalpha0}
\begin{split}
{\cal G}_{-(2j-1)}(x)-\int\limits_{-\infty}^\infty \! \!dy \, G(x-y) \, {\mathfrak m}_-(y) \, {\cal G}_{-(2j-1)}(y)
=e^{(1-2j)x}, \qquad j=1,2,...
\end{split}
\end{equation}
\begin{equation} \label{kinkpalpha0}
\begin{split}
{\cal G}_{+(2k-1)}(x)-\int\limits_{-\infty}^\infty \! \!dy \, G(x-y) \, {\mathfrak m}_+(y) \, {\cal G}_{+(2k-1)}(y)
=e^{(2k-1)x}, \qquad k=1,2,...
\end{split}
\end{equation}
where we deleted the superscripts from the solutions of (\ref{kinkmalpha}),(\ref{kinkpalpha}). Namely, 
${\cal G}_{-(2j-1)}(x)={\cal G}_{-(2j-1)}^{(\alpha=0)(-)}(x)$ and ${\cal G}_{+(2k-1)}(x)={\cal G}_{+(2k-1)}^{(\alpha=0)(+)}(x)$.

To get simpler looking formulas, using (\ref{scGmasy}) and (\ref{scGpasy}) we rewrite their large $x$ expansions, as well:
\begin{equation} \label{scGmasy0}
\begin{split}
{\cal G}_{-(2j-1)}(x)\,\, & \stackrel{x \to +\infty}{=} \sum\limits_{s=1}^\infty c_s^{(-(2j-1))} \, e^{-\tfrac{2 s x}{p+1}}, \\
%\qquad \mbox{as} \qquad x \to \infty \\
{\cal G}_{-(2j-1)}(x)-e^{-(2j-1)\, x} \,\, &\stackrel{x \to -\infty}{=} \sum\limits_{s=1}^\infty a_s^{(-(2j-1))} \, e^{\tfrac{2 s}{p} x}+
 \sum\limits_{s=0}^\infty b_s^{(-(2j-1))} \, e^{(1+2 s)\,x},
\end{split}
\end{equation}
\begin{equation} \label{scGpasy0}
\begin{split}
{\cal G}_{+(2k-1)}(x)\,\, & \stackrel{x \to -\infty}{=} \sum\limits_{s=1}^\infty c_s^{(+(2k-1))} \, e^{\tfrac{2 s x}{p+1}}, \\
%\qquad \mbox{as} \qquad x \to \infty \\
{\cal G}_{+(2k-1)}(x)-e^{(2k-1)\, x} \,\, &\stackrel{x \to \infty}{=}   \sum\limits_{s=1}^\infty \! a_s^{(+(2k-1))} 
\, e^{-\tfrac{2 s}{p} x}+
 \sum\limits_{s=0}^\infty b_s^{(+(2k-1))} \, e^{-(1+2 s)\,x}.
\end{split}
\end{equation}
First one must derive the linear equations for the kink corrections $\de Z_\pm(x)=\de Z_\pm^{(1)}(x)$ 
entering (\ref{ZUVansatz}). They take the form:
\begin{equation} \label{dZpm}
\begin{split}
\delta Z_\pm(x)-\int\limits_{-\infty}^\infty dy \, G(x-y) \, {\mathfrak m}_\pm(y) \, \delta Z_\pm(y)=
(G \star \delta f_\pm)(x),
\end{split}
\end{equation}
where 
\begin{equation} \label{dfpm}
\begin{split}
\delta f_+(x)&=c_1 \, e^{-\tfrac{2\,x}{p+1}} \, \left( {\mathfrak m}_+(x)-{\mathfrak m}_0 \right), \\
\delta f_-(x)&=\tilde{c}_1 \, e^{\tfrac{2\,x}{p+1}} \, \left( {\mathfrak m}_-(x)-{\mathfrak m}_0 \right), 
\qquad {\mathfrak m}_0 \equiv {\mathfrak m}[z_0],
\end{split}
\end{equation}
and $\star$ stands for convolution:
$ (f \star g)(x)=\int\limits_{-\infty}^\infty dy \, f(x-y) \, g(y).$

We also need to solve linear equations for the corrections to the solutions of (\ref{kinkmalpha0}),(\ref{kinkpalpha0}). 
They take the forms as follows:
\begin{equation} \label{kinkpalpha0corr}
\begin{split}
\delta {\cal G}_{+(2k-1)}^{(\sigma)}(x)-\int\limits_{-\infty}^\infty \! \!dy \, G(x-y) \, {\mathfrak m}_\sigma(y) 
\, \delta {\cal G}_{+(2k-1)}^{(\sigma)}(y)=(G \star \delta F_{+(2k-1)}^{(\sigma)})(x), \qquad \sigma=\{ \pm\}, 
\end{split}
\end{equation}
where 
\begin{equation} \label{dF2km1}
\begin{split}
\delta F_{+(2k-1)}^{(+)}(x)&=c_1 \, e^{-\tfrac{2 \, x}{p+1}} \,  {\cal G}_{+(2k-1)}(x) \, {\mathfrak m}'[Z_+(x)], \\ 
\delta F_{+(2k-1)}^{(-)}(x)&=c^{(+(2k-1))}_1 \, e^{\tfrac{2 \, x}{p+1}} \,  \left( {\mathfrak m}_-(x)-{\mathfrak m}_0 \right),
\end{split}
\end{equation}
with ${\mathfrak m}'[Z(x)]=\frac{d}{d \, Z(x)} {\mathfrak m}[Z(x)] .$

\begin{equation} \label{kinkmalpha0corr}
\begin{split}
\delta {\cal G}_{-(2j-1)}^{(\sigma)}(x)-\int\limits_{-\infty}^\infty \! \!dy \, G(x-y) \, {\mathfrak m}_\sigma(y) 
\, \delta {\cal G}_{-(2j-1)}^{(\sigma)}(y)=(G \star \delta F_{-(2j-1)}^{(\sigma)})(x), \qquad \sigma=\{ \pm\}, 
\end{split}
\end{equation}
where 
\begin{equation} \label{dFm2jm1}
\begin{split}
\delta F_{-(2j-1)}^{(+)}(x)&=  c^{(-(2j-1))}_1 \, e^{-\tfrac{2 \, x}{p+1}} \,  \left( {\mathfrak m}_+(x)-{\mathfrak m}_0 \right), \\
\delta F_{-(2j-1)}^{(-)}(x)&=  \tilde{c}_1 \, e^{\tfrac{2 \, x}{p+1}} \,  {\cal G}_{-(2j-1)}(x) \, {\mathfrak m}'[Z_-(x)].
\end{split}
\end{equation}
Each correction function has $\sim e^{-\tfrac{2|x|}{p+1}}$ behaviour at infinity.

To formulate the next to leading order correction, we introduce some new notations for the asymptotic 
behaviour of the measure and its derivatives on kink backgrounds.

\begin{equation} \label{masympts}
\begin{split}
{\mathfrak m}_\pm(x)={\mathfrak m}[Z_\pm(x)] &\stackrel{x \to \mp \infty}{=} 
{\mathfrak m}_0+\mm_1^{\pm} \, e^{\pm \tfrac{2 \, x}{p+1}}+\mm_2^{\pm} \, e^{\pm \tfrac{4 \, x}{p+1}}+..., \\
{\mathfrak m}'_\pm(x)={\mathfrak m}'[Z_\pm(x)] &\stackrel{x \to \mp \infty}{=} 
{\mathfrak m}'_0+\mm_1^{'\pm} \, e^{\pm \tfrac{2 \, x}{p+1}}+\mm_2^{'\pm} \, e^{\pm \tfrac{4 \, x}{p+1}}+..., \\
{\mathfrak m}''_\pm(x)={\mathfrak m}''[Z_\pm(x)] &\stackrel{x \to \mp \infty}{=} 
{\mathfrak m}''_0+\mm_1^{''\pm} \, e^{\pm \tfrac{2 \, x}{p+1}}+\mm_2^{''\pm} \, e^{\pm \tfrac{4 \, x}{p+1}}+.... 
\end{split}
\end{equation}
Defining the above asymptotic series seem to be superfluous, since from (\ref{m}), it is easy to see, that at the infinity, where the contours can be pushed into the real axis: ${\mathfrak m}[Z(x)]=1.$ Thus, it follows that 
${\mathfrak m}_0=1$ and all the other coefficients are zero. Namely: 
\begin{equation} \label{masrels}
\begin{split}
{\mathfrak m}_0&=1, \\ 
{\mathfrak m}'_0&={\mathfrak m}''_0=0, \\
\mm_j^{\pm}&=\mm_j^{'\pm}=\mm_j^{''\pm}=0, \qquad \qquad j=1,2,...
\end{split}
\end{equation}
Nevertheless, it is useful to keep and bookkeep these coefficients in the formulas, because on the one hand %of two reasons:
%\newline
the derivation may be valid for other models, where the Thermodynamic Bethe Ansatz (TBA) 
 \cite{Zamolodchikov:1989cf}  descrition is used to 
describe the finite size effects and where the measure is not simply one at the infinity along the real axis. 
%\newline
On the other hand, when doing numerical computations, implementing the $\pm i \, 0$ prescription
and working along the real axis is impossible. 
In this case the representation ${\mathfrak m}(x)={\cal L}_+(x+i \, 0)+{\cal L}_-(x-i \, 0)$ should be used,
 such that the integrals along $x+i \, 0$ should be shifted to $x+i \, \eta$ and the integrals along $x-i \, 0$ 
should be shifted to $x-i \, \eta,$ with $\eta$ being an appropriately small positive number.
In this case instead of ${\mathfrak m},$ the formulas will contain ${\cal L}_\pm$ 
which has non-zero expansion coefficients as $x \to \pm\infty.$ Assuming the non-zeroness of the coefficients in (\ref{masympts}) 
and keeping them during the calculations in the real axis formulation, it will be easy to figure 
out how to shift the integrals correctly away from the real axis. We will discuss this technical issue in the 
next section. 

Now, we are in the position to write down the final result for the next to leading order correction. 
It is composed of 4 terms.
\begin{equation} \label{omegacoeff2}
\begin{split}
\tilde{\omega}_{2k-1,1-2j|2}^{(\alpha=0)}= \tilde{\omega}^{as}_{0|2}+\delta \tilde{\omega}^{as}_{0|2}+
\delta \tilde{\omega}^{as}_{\delta|2}+\delta \tilde{\omega}^{corr}_{|2},
\end{split}
\end{equation}
where the label "$as$" means, that the corresponding correction comes from the asymptotic solution, and the label 
"$corr$" means, that it comes from the contributions of the  kink correction functions: $\delta Z_\pm(x)$. 
See appendix \ref{appA} for the details of the derivation in the special case of $\tom_{1,-1}(\alpha).$ 

Let's start with the seemingly most complicated "correction" term, since there is no subtlety in its integrands at infinity.

\begin{equation} \label{omt2}
\begin{split}
\delta \tilde{\omega}^{corr}_{|2}&=\int\limits_{-\infty}^\infty dx \, {\cal G}_{-(2j-1)}(x) \, \delta Z_-(x) \, 
\delta {\cal G}_{+(2k-1)}^{(-)}(x) \, {\mathfrak m}'[Z_-(x)]+\\
&+\int\limits_{-\infty}^\infty dx \, {\cal G}_{+(2k-1)}(x) \, \delta Z_+(x) \, 
\delta {\cal G}_{-(2j-1)}^{(+)}(x) \, {\mathfrak m}'[Z_+(x)]+ \\
&+c_1^{(-(2j-1))}\, \int\limits_{-\infty}^\infty dx \, e^{-\tfrac{2 \, x}{p+1}}\,{\cal G}_{+(2k-1)}(x) \, \delta Z_+(x) \, {\mathfrak m}'[Z_+(x)]+ \\
&+c_1^{(+(2k-1))}\, \int\limits_{-\infty}^\infty dx \, e^{\tfrac{2 \, x}{p+1}}\,{\cal G}_{-(2j-1)}(x) \, \delta Z_-(x) \, {\mathfrak m}'[Z_-(x)].
\end{split}
\end{equation}
The next term is the one, which comes from the corrections to the "leading asymptotic 
solution"\footnote{For a more detailed explanation: see appendix \ref{appA}. }:
\begin{equation} \label{omasd}
\begin{split}
\delta \tilde{\omega}^{as}_{\delta|2}&=c_1^{(-(2j-1))}\, \int\limits_{-\infty}^\infty dx \, 
e^{-\tfrac{2 \, x}{p+1}}\, \left({\mathfrak m}_+(x)-{\mathfrak m}_0 \right) \, \delta {\cal G}_{+(2k-1)}^{(+)}(x)+ \\
&+\tilde{c}_1\, \int\limits_{-\infty}^\infty dx \, 
e^{\tfrac{2 \, x}{p+1}}\, {\cal G}_{-(2j-1)}(x) \, {\mathfrak m}'[Z_-(x)] \, \delta {\cal G}_{+(2k-1)}^{(-)}(x).
\end{split}
\end{equation}
The next term is the simplest one, and comes from the "leading asymptotic solution":
\begin{equation} \label{omas0}
\begin{split}
\tilde{\omega}^{as}_{0|2}=\CP\limits_{\ell \to 0} \int\limits_{-\log\tfrac{2}{\ell}}^\infty dx \,\, e^{\left(2k-1-\tfrac{4}{p+1} \right)\, x} \, \left( c_2^{(-(2j-1))} \, {\mathfrak m}_+(x)+c_1 \, c_1^{(-(2j-1))} \, {\mathfrak m}'[Z_+(x)] \right),
\end{split}
\end{equation}
where we introduced the symbolic operator: $\CP\limits_{\ell \to 0},$ which means, that the constant part in the 
$\ell \to 0$ series expansion should be computed. The only $\ell$ dependence in the rhs. of (\ref{omas0}) is 
in the lower integration limit. Analyzing the large $x$ dependence of the integrand it is easy to see, that for 
$p>3$ the integral in (\ref{omas0}) is convergent for any $k,j \geq 1,$ if the lower limit tends to minus infinity.  
Thus for $p>3,$ the constant part prescription corresponds to taking the integral from $-\infty$ to $\infty.$ 
We note, that in case of $p<3$ for $k$ large enough the integral is convergent on $\mathbb{R}$, and the constant part corresponds to integrating along the whole real axis. 
%The correct treatment of (\ref{soma}) in the regime $1<p<3$ can be found for $\tom_{1,-1}(\alpha)$ in appendix \ref{appA}.
The way to compute the constant part of (\ref{omas0}) in the regime $1<p<3$ is demonstrated through 
the example of $\tom_{1,-1}(\al)$ in \ref{A21} subsection of appendix \ref{appA}. 

The contribution still to be discussed from (\ref{omegacoeff2}), is composed of two terms: 
\begin{equation} \label{omas2}
\begin{split}
\delta \tilde{\omega}^{as}_{0|2}=\delta \tilde{\omega}^{as}_{\Sigma|2}+\delta \tilde{\omega}^{as}_{\int|2},
\end{split}
\end{equation}
where $\delta \tilde{\omega}^{as}_{\Sigma|2}$ is a kind of "surface term" coming from taking the primitive function at 
infinity, while  $\delta \tilde{\omega}^{as}_{\int|2}$ is given in a form of integrals. 
The surface term takes the form as follows:
\begin{equation} \label{surf}
\begin{split}
\delta \tilde{\omega}^{as}_{\Sigma|2}&=\frac{p+1}{4} \, c_1^{(-(2j-1))}\, \left( c_2^{(+(2k-1))} \, \mm_1^+ +
 c_1^{(+(2k-1))} \, \mm_2^+  \right)+\\
&+\frac{p+1}{4} \, c_1^{(-(2j-1))}\, \left( \tilde{c}_1 \, c_2^{(+(2k-1))} \, {\mathfrak m}'_0 +
\tilde{c}_2 \, c_1^{(+(2k-1))} \, {\mathfrak m}'_0+
\frac{\tilde{c}_1^2 \, c_1^{(+(2k-1))}}{2} \, {\mathfrak m}''_0  \right).
\end{split}
\end{equation}
Taking into account the relations (\ref{masrels}) %remark made after (\ref{masympts})
, this contribution in the sine-Gordon model is zero:
\begin{equation} \label{surf=0}
\begin{split}
\delta \tilde{\omega}^{as}_{\Sigma|2}=0.
\end{split}
\end{equation}
We presented (\ref{surf}), because we think that it may be useful and applicable to other TBA models, 
where the measure doesn't take such a special form and has non-zero asymptotic coefficients in the 
plateau regime.

Finally, we present the  integral term, which is the most difficult to treat numerically.
\begin{equation} \label{omasint}
\begin{split}
\delta \tilde{\omega}^{as}_{\int|2}&=\CP\limits_{\ell \to 0} \! \! \! \int\limits_{-\log\tfrac{2}{\ell}}^{\infty} 
\! \! \! dx \,
\left[ {\cal G}_{+(2k-1)}(x)-e^{(2k-1)x} \right] \, e^{-\tfrac{4 \, x}{p+1}}  \\ 
&\times \bigg( c_2^{(-(2j-1))} \, \left( {\mathfrak m}_+(x)- {\mathfrak m}_0 \right) +
c_1 \, c_1^{(-(2j-1))} \, \left( {\mathfrak m}'[Z_+(x)]-{\mathfrak m}'[z_0] \right)
\bigg)+ \\
&+\tilde{c}_1 \, c_1^{(+(2k-1))} \CP\limits_{\ell \to 0} \! \! \! \int\limits_{-\infty}^{\log\tfrac{2}{\ell}}
\! \! \! dx \, {\cal G}_{-(2j-1)}(x) \, e^{\tfrac{4 \, x}{p+1}} \, {\mathfrak m}'[Z_-(x)].
\end{split}
\end{equation}

When splitting the measure ${\mathfrak m} (x)\to {\cal L}_+(x+i \, 0)+{\cal L}_-(x-i \, 0)$, the individual 
integrals along the lines $x\pm i \, 0$ in (\ref{omasint}) seem to be divergent, 
when $\ell \to 0,$ but in reality their sums become constants. In the next section we discuss, how to  
compute the constant part as  a sum of integrals running along straight lines 
with non-infinitesimal distance from the real axis.

To close this section, we just summarize that, the next to leading order result 
in the UV series representation for $\tom_{2k-1,1-2j}(\al=0)$ is given by the 
formulas (\ref{omegacoeff2})-(\ref{omasint}).

\section{Changing the integration contours from $x \pm i \, 0$ to  $x\pm i \, \eta$}
\label{7}

One can see, that our results contain integrals running along the lines $\mathbb{R}\pm i  \, 0.$ 
Though, this formulation is mathematically correct, it is not appropriate for numerical computations. 
This is why, in this section we show how to rephrase our integrals running along the lines 
$\mathbb{R}\pm i  \, 0$ to integrals running along the shifted lines $\mathbb{R}\pm i  \, \eta,$ 
with $\eta$ being an appropriately small positive parameter.

The rephrasing consists of 3 subsequent steps:
\newline
1. The integrals running along the real axis and containing the measure ${\mathfrak m}(x)$ are 
written as a sum of two integrals. One is running along ${\mathbb R}+i \, 0$ and the other runs on  
${\mathbb R}-i \, 0.$ This corresponds to the ${\mathfrak m} (x)= {\cal L}_+(x+i \, 0)+{\cal L}_-(x-i \, 0)$ 
representation of the measure.
\newline 
2. We separate the divergent and constant terms in the $\ell \to 0$ limit, such that the constant term is 
represented as an integral running on the full real axis ${\mathbb R}$.
\newline
3. In the integrals running from $-\infty$ to $\infty,$ we perform the necessary contour deformations:
$x \pm  i \, 0 \to x \pm i \, \eta  .$

Now, we demonstrate the method on formal prototype integrals, and then we specify the functions for the special case 
corresponding to (\ref{omasint}). 

First, we rewrite (\ref{omasint}) as a sum of two integrals:
\begin{equation} \label{621}
\begin{split}
\delta \tilde{\omega}^{as}_{\int|2}=\delta \tilde{\omega}^{as}_{\int|2/1}
+\delta \tilde{\omega}^{as}_{\int|2/2},
\end{split}
\end{equation}
where $\delta \tilde{\omega}^{as}_{\int|2/1}$ denotes the first integral on the rhs. in (\ref{omasint}), 
while $\delta \tilde{\omega}^{as}_{\int|2/2}$ denotes the second one. 

Let's start with the type of second integral in (\ref{omasint}):

\begin{equation} \label{f2}
\begin{split}
I_2=\int\limits_{-\infty}^{\log\tfrac{2}{\ell}} \! \! dx \, \left( f_+(x)+f_-(x) \right),
\end{split}
\end{equation}
where $f_\pm(x)$ correspond to the integrands running on ${\mathbb R}\pm i \, 0$ respectively. 
They are integrable at $-\infty,$ but have the asymptotics at $+\infty$ as follows:
\begin{equation} \label{f2as}
\begin{split}
f_\pm(x)=A_\pm\, e^{\tfrac{2 \, x}{p+1}}+B_\pm+O(e^{\tfrac{-2 \, x}{p+1}}).
\end{split}
\end{equation}
One can define an appropriate compensating function, that allows one to subtract and analytically integrate
 the divergent terms at $+\infty:$
\begin{equation} \label{sepf2}
\begin{split}
S_{B_\pm}(x)=B_{\pm}\, \frac{1+\tanh\left( \tfrac{6 \, x}{p+1}  \right)}{2}.
\end{split}
\end{equation}
A simple and straightforward calculation leads to the result as follows:

\begin{equation} \label{f2res}
\begin{split}
&\int\limits_{-\infty}^{\log\tfrac{2}{\ell}} \! \! dx \,  f_\pm(x)=
\frac{A_\pm (p+1)}{2}  \left(\frac{2}{\ell} \right)^{\tfrac{2}{p+1}}\!\!\!+\!B_\pm \, \log\frac{2}{\ell}
\!+\!\!\!\int\limits_{-\infty}^{\infty} \!\! dx 
 \left[f_\pm(x)\!\! -\!\! A_\pm \, e^{\tfrac{2 \, x}{p+1}}\!\!-\!\!S_{B_\pm}(x) \right]\!\!+...=\\
&=\frac{A_\pm (p+1)}{2} \left(\frac{2}{\ell} \right)^{\tfrac{2}{p+1}}\!\!\!+\!B_\pm \, \log\frac{2}{\ell}
\!+\!\!\! \int\limits_{-\infty}^{\infty} \!\! dx 
 \left[f_\pm(x\pm i \, \eta )-\!\! A_\pm \, e^{\tfrac{2 \, (x\pm i \, \eta )}{p+1}}\!\!-\!\!S_{B_\pm}(x\pm i \, \eta ) \right]\!\!+\!...,
\end{split}
\end{equation}
where the dots stand for terms being at least of order $\ell^{\tfrac{2}{p+1}}.$

Now, one can specify the result of (\ref{f2res}) to the second integral in (\ref{omasint}). 
The necessary choices of functions and parameters are as follows:
\begin{equation} \label{omas2/2spec}
\begin{split}
f_\pm(x)&={\cal G}_{-(2j-1)}(x) \, e^{\tfrac{4 \, x}{p+1}} \, {\cal L}'_\pm[Z_-(x)], \\
A_\pm&=c_1^{(-(2j-1))} \, {\cal L}_\pm'[z_0], \qquad \qquad \qquad
B_\pm=c_1^{(-(2j-1))} \, \tilde{{\cal L}}'_{\pm,1}+c_2^{(-(2j-1))} \, {\cal L}_\pm'[z_0], \\
\end{split}
\end{equation}
where we used the coefficients entering the asymptotic series expansions on the plateau:
\begin{equation} \label{calLplatser}
\begin{split}
{\cal L}_\pm[Z_-(x\pm i \,0)]&\stackrel{x \to \infty}{=}{\cal L}_\pm[z_0]
+\tilde{{\cal L}}_{\pm,1}\, e^{-\tfrac{2 \, (x \pm i \, 0)}{p+1}}
+\tilde{{\cal L}}_{\pm,2}\, e^{-\tfrac{4 \, (x \pm i \, 0)}{p+1}}+...\\
{\cal L}'_\pm[Z_-(x\pm i \,0)]&\stackrel{x \to \infty}{=}{\cal L}'_\pm[z_0]
+\tilde{{\cal L}}'_{\pm,1}\, e^{-\tfrac{2 \, (x \pm i \, 0)}{p+1}}
+\tilde{{\cal L}}'_{\pm,2}\, e^{-\tfrac{4 \, (x \pm i \, 0)}{p+1}}+...\\ \\
{\cal L}_\pm[Z_+(x\pm i \,0)]&\stackrel{x \to -\infty}{=}{\cal L}_\pm[z_0]
+\hat{{\cal L}}_{\pm,1}\, e^{\tfrac{2 \, (x \pm i \, 0)}{p+1}}
+\hat{{\cal L}}_{\pm,2}\, e^{\tfrac{4 \, (x \pm i \, 0)}{p+1}}+...\\
{\cal L}'_\pm[Z_+(x\pm i \,0)]&\stackrel{x \to -\infty}{=}{\cal L}'_\pm[z_0]
+\hat{{\cal L}}'_{\pm,1}\, e^{\tfrac{2 \, (x \pm i \, 0)}{p+1}}
+\hat{{\cal L}}'_{\pm,2}\, e^{\tfrac{4 \, (x \pm i \, 0)}{p+1}}+...\\
\end{split}
\end{equation}
As a consequence of (\ref{masrels}), 
%the remarks after (\ref{masympts})
 the coefficients satisfy the following relations:
\begin{equation} \label{calLrel}
\begin{split}
&{\cal L}_+[z_0]+{\cal L}_-[z_0]={\mathfrak m}[z_0]=1, \\
&{\cal L}'_+[z_0]+{\cal L}'_-[z_0]={\mathfrak m}'[z_0]=0, \\
&\tilde{{\cal L}}_{+,j}+\tilde{{\cal L}}_{-,j}=\tilde{{\cal L}}'_{+,j}+\tilde{{\cal L}}'_{-,j}=0, \qquad \qquad j=1,2,...\\
&\hat{{\cal L}}_{+,j}+\hat{{\cal L}}_{-,j}=\hat{{\cal L}}'_{+,j}+\hat{{\cal L}}'_{-,j}=0, \qquad \qquad j=1,2,...\\
\end{split}
\end{equation}
As a consequence of these relations, $A_\pm$ and $B_\pm$ defined in (\ref{omas2/2spec}) also satisfy 
simple sum rules:
\begin{equation} \label{ABrel}
\begin{split}
A_++A_-=0, \qquad \qquad B_++B_-=0,
\end{split}
\end{equation}
which ensures, that the divergent in $\ell \to 0$ terms predicted by (\ref{f2res}) cancel out, and only 
an integral taken along the whole real axis remains upto non-vanishing in $\ell \to 0$ orders.
%in the 2nd term in (\ref{omasint}) cancel and only the integral terms remain. 

Thus, the constant part of the 2nd integral in (\ref{omasint}) will take the form: 
\begin{equation} \label{omasint2/2}
\begin{split}
\delta \tilde{\omega}^{as}_{\int|2/2}=\tilde{c}_1 \, c_1^{(+(2k-1))} \! \! \int\limits_{-\infty}^{\infty} dx \,
\bigg\{ f_+(x+i \, \eta)+f_-(x-i \, \eta)
&-A_+\, \left( e^{\tfrac{2 (x+i \, \eta)}{p+1}}-  
e^{\tfrac{2 (x-i \, \eta)}{p+1}} \right)  \\
&-\left[ S_{B_+}(x+i \, \eta)-S_{B_+}(x-i \, \eta) \right]\bigg\},
\end{split}
\end{equation}
where the functions and constants are specified in (\ref{sepf2}),(\ref{omas2/2spec}).

Now, let's turn to the first integral term in (\ref{omasint}). It is computed in a very similar fashion to the 2nd integral.

%%%%%%%%%%%%%%%%%%%%%%%%%%%%%%%%%%%%%%%%%%%%%%%%%%%%%%%%%%%%%%%%%%%%%%%%%%%%%%%%%%%%%%%%%%%%%%%%%%%%%%%%%%%%%%%%%%%%%%%%%%%%%%

Now, the prototype integral is as follows:
\begin{equation} \label{f1}
\begin{split}
I_1=\int\limits_{-\log\tfrac{2}{\ell}}^{\infty} \! \! dx \, \left( \hat{f}_+(x)+\hat{f}_-(x) \right),
\end{split}
\end{equation}
where $\hat{f}_\pm(x)$ correspond to the integrands running along ${\mathbb R}\pm i \, 0$ respectively. 
They are integrable at $\infty,$ but have constant asymptotics at $-\infty$:
\begin{equation} \label{f1as}
\begin{split}
\hat{f}_\pm(x)=\hat{B}_\pm+O(e^{2x/(p+1)}).
\end{split}
\end{equation}
One can define again an appropriate compensating function, that allows to subtract and analytically integrate
 the divergent terms at $-\infty:$
\begin{equation} \label{sepf1}
\begin{split}
\hat{S}_{\hat{B}_\pm}(x)=\hat{B}_{\pm}\, \frac{1-\tanh\left( \tfrac{6 \, x}{p+1}  \right)}{2}.
\end{split}
\end{equation}
A simple and straightforward calculation leads to the result as follows:

\begin{equation} \label{f1res}
\begin{split}
&\int\limits_{-\log\tfrac{2}{\ell}}^{\infty} \! \! dx \,  \hat{f}_\pm(x)=
\!\hat{B}_\pm \, \log\frac{2}{\ell}
\!+\!\!\!\int\limits_{-\infty}^{\infty} \!\! dx  \left[\hat{f}_\pm(x) \!
 -\!\hat{S}_{\hat{B}_\pm}(x) \right]\!\!+O(\ell^{\tfrac{2}{p+1}})=\\
&=\!\hat{B}_\pm \, \log\frac{2}{\ell}
\!+\!\!\! \int\limits_{-\infty}^{\infty} \!\! dx 
 \left[\hat{f}_\pm(x\pm i \, \eta ) \! -\! \hat{S}_{\hat{B}_\pm}(x\pm i \, \eta ) \right]\!\!++O(\ell^{\tfrac{2}{p+1}}).
\end{split}
\end{equation}

Now, one can specify the result of (\ref{f1res}) to the first integral in (\ref{omasint}) by the following specification 
of functions and constants:
\begin{equation} \label{omas2/1spec}
\begin{split}
\hat{f}_\pm(x)&=\left[ {\cal G}_{+(2k-1)}(x)-e^{(2k-1)x} \right] \, e^{\tfrac{-4 \, x}{p+1}} \, 
\bigg( 
c_2^{(-(2j-1))} \, \left( {\cal L}_\pm[Z_+(x)]-{\cal L}_\pm[z_0] \right)+\\
&+c_1 \, c_1^{(-(2j-1))} \left( {\cal L}'_\pm[Z_+(x)]-{\cal L}'_\pm[z_0] \right)\, 
\bigg), \\
\hat{B}_\pm&=c_1^{(+(2k-1))} \, \left( c_2^{(-(2j-1))} \, \hat{{\cal L}}_{\pm,1}  +c_1 \, c_1^{(-(2j-1))} \, \hat{{\cal L}}'_{\pm,1}  \right),
\end{split}
\end{equation}
where we used the coefficients entering the asymptotic series expansions on the plateau:
 (\ref{calLplatser}).
As a consequence of (\ref{calLrel}), $\hat{B}_+ +\hat{B}_-=0.$ Thus, upto non-vanishing in 
$\ell \to 0$ orders, only the integral terms  from (\ref{f1res}) will contribute to (\ref{f1}), if 
the functions are specified to the ones given in (\ref{omas2/1spec}).

Putting everything together, one obtains, that the first integral in (\ref{omasint}) takes the form  as follows, when formulated along the $\pm i \, \eta$ shifted contours: 
 \begin{equation} \label{omasint2/1}
\begin{split}
\delta \tilde{\omega}^{as}_{\int|2/1}= \! \! \int\limits_{-\infty}^{\infty} dx \,
\bigg\{ \hat{f}_+(x+i \, \eta)+\hat{f}_-(x-i \, \eta)
&-\left[ \hat{S}_{\hat{B}_+}(x+i \, \eta)-\hat{S}_{\hat{B}_+}(x-i \, \eta) \right]\bigg\},
\end{split}
\end{equation}
where the functions and constants are specified in (\ref{sepf1}),(\ref{omas2/1spec}).

%%%%%%%%%%%%%%%%%%%%%%%%%%%%%%%%%%%%%%%%%%%%%%%%%%%%%%%%%%%%%%%%%%%%%%%%%%%%%%%%%%%%%%%%%%%%%%%%%%%%%%%%%%%%%%%%%%%%%%%%%%%%%

The full result for the contribution $\delta \tilde{\omega}^{as}_{|2}$ is as follows:

\begin{equation} \label{omtasfull}
\begin{split}
\delta \tilde{\omega}^{as}_{0|2}=\delta \tilde{\omega}^{as}_{\int|2}=\delta \tilde{\omega}^{as}_{\int|2/1}+\delta \tilde{\omega}^{as}_{\int|2/2},
\end{split}
\end{equation}
with $\delta \tilde{\omega}^{as}_{\int|2/1}$ and $\delta \tilde{\omega}^{as}_{\int|2/2}$ given in (\ref{omasint2/1}) and (\ref{omasint2/2}) , respectively.

The above listed next to leading order results were tested and verified numerically with 8-9 digits of precision 
for the first few $\om$s. In section \ref{8}, we will present a few more details on the numerical work.

\section{The leading order UV coefficient for $\om_{j,k}(\al)$ with subscripts having the same sign} \label{7b}

So far we have discussed the UV behaviour of $\om_{2k-1,1-2j}(\al),$ when $j,k=1,2...$ 
Namely, we focused on the cases, when the signs of the two subscripts of $\om$ are different.

For completeness, in this section we would like to  discuss briefly the case, when the two subscripts have the same sign. 
We will concentrate on only the leading order term. First of all, we recall, that the formulas (\ref{omtom}), (\ref{omegatilde}) 
and (\ref{G1m2j}) are valid in the $j<0$ regime as well. Now, we will discuss the case when $k \geq 1$ and $j \leq -1.$ 
If one starts to analyse (\ref{omegatilde}) in this subscript regime, it turns out, that in the UV limit the dominant 
contributions come from $x \in (0,\infty)$ regime, which is dominated by the "+"-kink contributions. 
Thus, the determination of the leading order term is straightforward. Only an $x \to x'=x-\ltl$ change of variables should be performed 
in (\ref{omegatilde}) and (\ref{G1m2j}), which after a trivial  $1-2j \to 2j-1$ relabeling,  immediately leads to the simple leading order result as follows:
\begin{equation} \label{tomsame}
\begin{split}
\tom_{2k-1,2j-1}(\al)\simeq \!\!  \left( \frac{2}{\ell}\right)^{2(k+j-1)} \!\!\! \intinf \!\! dy \, e^{(2k-1)x} \, \mm_+(x) \, 
\scG_{2j-1}^{(\al),(+)}(x)\!\!+..., 
\quad \text{for} \quad j,k=1,2,...
\end{split}
\end{equation}
where $\scG_{2j-1}^{(\al),(+)}(x)$ satisfies the equation: 
\begin{equation} \label{Gpp}
\begin{split}
\scG_{2j-1}^{(\al),(+)}(x)-\int\limits_{-\infty}^\infty dy \, G_{\alpha}(x-y) \, {\mathfrak m}_+(y) \, {\cal G}^{(\alpha),(+)}_{2j-1}(y)=e^{(2j-1)x}, \quad j=1,2...
\end{split}
\end{equation} 
Due to the transposition symmetry (\ref{cMalTR}) of the integral operator of the linear problem  (\ref{G1m2j}), the following transposition 
property holds for $\om_{2k-1,2j-1}(\al):$
\begin{equation} \label{tromsame}
\begin{split}
\om_{2k-1,2j-1}(\al)=\om_{2j-1,2k-1}(-\al),  \quad \text{for} \quad j,k\in\mathbb{Z}.
\end{split}
\end{equation}

So far we discussed, the case when both subscripts of $\om$ were positive. For the case when both subscripts are negative, a very similar train of thoughts
lead to the leading order UV behaviour as follows:
\begin{equation} \label{tomsame1}
\begin{split}
\tom_{1-2k,1-2j}(\al)\simeq \!\!  \left( \frac{2}{\ell}\right)^{2(k+j-1)} \!\!\! \intinf \!\! dy \, e^{-(2k-1)x} \, \mm_-(x) \, 
\scG_{1-2j}^{(\al),(-)}(x)\!\!+..., 
\quad \text{for} \quad j,k=1,2,...
\end{split}
\end{equation}
where $\scG_{1-2j}^{(\al),(-)}(x)$ satisfies the equation: 
\begin{equation} \label{Gmm}
\begin{split}
\scG_{1-2j}^{(\al),(-)}(x)-\int\limits_{-\infty}^\infty dy \, G_{\alpha}(x-y) \, {\mathfrak m}_-(y) \, {\cal G}^{(\alpha),(-)}_{1-2j}(y)=e^{-(2j-1)x}, \quad j=1,2...
\end{split}
\end{equation}
It is easy to see, that the final leading order result is much simpler in these cases, as it is in the case when the two subscripts of $\om$ have  opposite signs. 
This is because in this case, the left and right moving kink-functions do not interact at leading order. Thus only one type of kink dominates the leading order formula.

\section{Numerical checks in a nutshell} \label{8}

In this section, we would like to  discuss shortly some subtle points of the numerical method, and 
at the end we summarize the numerical checks performed.

Let's start with the numerical method. The primary task in the numerical investigations is to 
solve linear- and nonlinear- integral equations numerically. The typical equations are (\ref{kinksol}), 
(\ref{kinkmalpha0}) and (\ref{kinkpalpha0}). Each equation is solved in terms of the most 
commonly used iterative method. All of these equations can be naturally arranged into the form as follows:
$\text{"unknown function"}=\text{"source function"}+\text{"integral terms containing the unknown"}.$ 
At each step of the iteration, the "unknown function" is determined from the right hand side of this 
formal equation evaluated at the value of the unknown corresponding to the previous iteration step. 
The initial value for the "unknown" is simply the source function of the equation. 

When doing this iteration method numerically, two important issues should be taken care of. 
The first one is the numerical integration method to be used. The second one is the choice of cutoffs 
and the treatment of subtle integrals at the infinities. 
In our numerical method, we use the $x-$space representation of all functions entering the equations. 
This means, that we store the necessary  values of the functions at discretized points of the $x-$space in a 
reduced range $[-\Lambda_-,\Lambda_+].$  All integrals in our formulas run from $-\infty$ to 
$\infty,$ and in all cases the integrands tend to zero at least exponentially at the infinities. 
This means, that the cutoff parameters $\Lambda_\pm$ can be chosen such, that the integration 
outside of their range is negligible within the required precision. In our experience, the 
simplest and best choice is to use equidistant quadrature for each functions and to apply the 
rectangle rule for evaluating the integrals. For the typical integrals arising in our problem,
this simple method works much better, than any  
higher order polynomial interpolation based equidistant quadrature formula.

The next very important issue is the treatment of the functions at the infinities. 
In all cases, the   behaviour of the integrands at the infinities, can be derived only 
from analytical considerations.

No matter how large one chooses the cutoff values $\Lambda_\pm,$ the numerical solutions  
of the equations significantly loose precision when the cutoff values $\Lambda_\pm$ are approached. 
Nevertheless, the accurate knowledge of the unknown functions close to the 
cutoffs is very important for the trustable and precise numerical evaluation of our final 
UV formulas. This is because in some cases, the integrand is decaying at the infinity, but 
in a  way, that an exponentially diverging function is multiplied with two exponentially 
decaying ones. In this case,  the evaluation of the product of the  
3 functions should be accurately evaluated even close to the cutoffs. This requires 
the very accurate knowledge of each functions at the cutoffs. To evaluate such integrals 
precisely, one should proceed as follows. The numerical solution for the unknown looses precision 
as it approaches the cutoffs, but in a region, where $x$ is far from the cutoffs $\Lambda_\pm,$ but
$|x|$ is still large, the solution can be very accurate. 
In these intermediate regimes several coefficients of 
the large $x$ series representations (\ref{kinkplat}), (\ref{scGmasy0}), (\ref{scGpasy0}) of the 
unknown functions can be fitted accurately. 

Knowing these coefficients, the series representations can be used to evaluate accurately 
the unknown functions in regimes being close to the cutoff values $\Lambda_\pm$, as well. 
This makes possible the accurate evaluation of all integrals entering the UV formula.

We close this section with a short summary of our numerical checks. Our numerical tests were performed 
around the $p=5$ value. The quantities we computed, were the $\om$s being necessary to compute the 
expectation values of $\Phi_{2 \, \tfrac{1-\nu}{\nu}}$ and $\Phi_{4 \, \tfrac{1-\nu}{\nu}}$. 

Namely, we performed the numerical study of the following four $\om$-functions: \newline
$\tilde{\omega}_{1,-1}, \tilde{\omega}_{1,-3}, \tilde{\omega}_{3,-1}, \tilde{\omega}_{3,-3}.$ 

At the $\al=0$ point, first we computed all the four $\om$s numerically with a $\sim 35$ digits of precision 
at 50 values of the volume parameter $\ell$ in the range $[10^{-1},10^{-6}]$. Then, assuming the leading order 
behaviour given by (\ref{omexp}) and that the corrections go as positive integer powers in $\ell^{\tfrac{4}{p+1}},$ 
we fitted the first two coefficients of this UV series for the four $\om$s under consideration. 

Then, using our UV formulas, we computed the first two coefficient of these UV series from the 
integrable description of the UV CFT limit of the theory. The coefficients obtained in the two different ways 
agreed within the numerical errors of our method. Namely, the first coefficients showed 16-digits, and the 
second coefficients of the series showed 8-digits of precision agreement. 

In the special case of $\om_{1,-1}(0),$ the leading order result coming from our UV formulas 
could be compared to the prediction of the Liouville CFT (\ref{VEVcft}). This comparison led to 
25-digits of precision agreement between the analytical prediction and the numerical result. 
The same test was performed for the $\al \neq 0$ case, and agreement with the same precision was 
experienced. 

The second order results at $\al=0$ were also tested analytically against the prediction of the Liouville CFT (\ref{VEVcft}). This was done through 
computing the UV limit of $\langle \Phi_{4 \, \tfrac{1-\nu}{\nu}} \rangle$ in two different ways. 

First, one can compute this quantity, by inserting into (\ref{primVEV})  with $m=2$, 
 the numerically upto second order evaluated $\om$s. The other way of determining the UV limit of 
$\langle \Phi_{4 \, \tfrac{1-\nu}{\nu}} \rangle$, is to evaluate (\ref{VEVcft}) at $m=2.$ 
The two different ways of computation led to 8 digits of precision agreement, 
which was within the range of the numerical errors of our method.
 
We provided with some tables of numerical data in appendix \ref{appC}.

\section{Discussion} \label{9}

In this short closing section, we would like to discuss, what our results can be used for.

Our formula for the leading order term in the UV expansion (\ref{UVcoeff1}), allows one to determine,  
purely from integrability data, the UV limit of the following ratio of expectation values of the primaries:
${\langle  \Phi_{\alpha+2 \, \tfrac{1-\nu}{\nu}}(0) \rangle}/{\langle  \Phi_{\alpha}(0) \rangle},$ 
provided the parameters of the model are in the range as follows $1<p<\infty$ and $0 \leq \al < 2 .$

Then, the second order formula at $\al=0,$ with the help of (\ref{primVEV}), allows one to 
compute the UV limit of the expectation values of the primaries: 
$\Phi_{2 \, \tfrac{1-\nu}{\nu}}$ and $\Phi_{4 \, \tfrac{1-\nu}{\nu}}$ 
in terms of data coming from the integrable formulation of the UV CFT limit of the sine-Gordon model.

In general, to compute $\langle \Phi_{2 \,m \, \tfrac{1-\nu}{\nu}} \rangle,$  requires an $m(m-1)/2+1$th order 
computation around our  
asymptotic solutions, which for $m \geq 3,$ becomes a tantalizing, though in principle doable work. 

In this paper we restricted ourselves upto the 2nd order contributions, because of two reasons. 
First, we wanted to gain some experience on, 
how the complexity of the expressions of the coefficients in the UV series 
expands, as the order of the computation increases. 

Second, even these 2nd order results allows one to investigate interesting expectation values 
or 3-point couplings.  

 Namely, the application of our results to twisted NLIE cases with $p \in \mathbb{N}$  
\cite{Fioravanti:1996rz,Feverati:1999sr}, 
allows one to give an integrable description to several diagonal 3-point couplings of primary fields in 
the  unitary Virasoro minimal models $Vir(p,p+1).$   
At the very special twist values corresponding to these minimal models, the analytical structure of the solutions gets 
significantly more complicated from the point of view of the methods presented in this paper. 
Thus, the investigation of these models is still an interesting challenge, which is postponed to a future publication.

There are two more interesting models, where the fermionic description of states is available in the literature. 
The $N=1$ supersymmetric (SUSY) sine-Gordon model \cite{Babenko:2019tvv} and the sinh-Gordon model \cite{Negro:2013wga,Negro:2013rxa,Bajnok:2019yik}. 
Due to the periodicity of the functions entering the integrable description of the spectral problem, our method can be almost literally 
applied to the $N=1$ SUSY sine-Gordon model. In the sinh-Gordon model there is no such obvious periodicity present in the formulation of the 
spectral problem. Nevertheless, we hope, that after some appropriate modification, the mathematical techniques worked out in this 
paper, can be applied to give an integrability based description of the UV limit of 
expectation values in this 
%the sinh-Gordon
model, too.

\vspace{1cm}
{\tt Acknowledgments}

\noindent 
The author would like to thank to Zoltan Bajnok, Romuald Janik for useful discussions.
This research was supported  by the NKFIH grant K134946.

\appendix

\section{The derivation of the first two coefficients in the $\ell \to 0$ series representation of $\tilde{\omega}_{1,-1}(\alpha)$ } \label{appA}

In this appendix, we present the logic and the main steps of the derivation of the formulas for the first two coefficients in 
the $\ell \to 0$ series representation of $\tilde{\omega}_{1,-1}(\alpha).$ At the end of the appendix, we 
note, how 
the formulas, should be trivially modified, when the UV formula for the general 
$\tilde{\omega}_{2k-1,1-2j}(\alpha)$ quantities are considered. The logic and the main steps of the 
derivation is the same as in the special case of $\tilde{\omega}_{1,-1}(\alpha),$ apart from different 
scaling of certain terms with respect to $\ell.$

The definition of $\tom_{1,-1}(\al)$ is given by an integral expression containing the solution of a linear 
integral equation related to the solution of the NLIE (\ref{DDV}) of the finite volume problem. Here, we will give two 
simply related equivalent formulations: 

\begin{equation} \label{om1m1def}
\begin{split}
\tom_{1,-1}(\al)=\intinf \! dx \, e^x\, \mm(x) \, \scG_{-1}^{(\al)}(x)
=\intinf \! dx \, e^x \, \scG_{-1,m}^{(\al)}(x),
\end{split}
\end{equation}
where $\mm(x)\equiv\mm[Z(x)]$ is the integration measure (\ref{m}) related to the solution of 
the NLIE (\ref{DDV}) and 
$\scG_{-1}^{(\al)}(x)$ and $\scG_{-1,m}^{(\al)}(x)$ are two functions simply related to each other by:
\begin{equation} \label{scGvsGm}
\begin{split}
\scG_{-1,m}^{(\al)}(x)=\mm(x)\, \scG_{-1}^{(\al)}(x).
\end{split}
\end{equation}
They satisfy the linear equations as follows\footnote{This equation is the $j=1$ special case of 
(\ref{G1m2j}). }:
\begin{eqnarray} \label{scGeqs}
\scG_{-1}^{(\al)}(x)-\intinf \! dy \, G_\al(x-y)\, \mm(y) \, \scG_{-1}^{(\al)}(y)&=&e^{-x}, \\
\frac{\scG_{-1,m}^{(\al)}(x)}{\mm(x)}-\intinf \! dy \, G_\al(x-y)\,  \scG_{-1}^{(\al)}(y)&=&e^{-x},
\label{scGeqsm}
\end{eqnarray}
where $G_\al(x)$ is given in (\ref{Galpha}).
In the papers \cite{Jimbo:2010jv,Hegedus:2019rju}, originally $\om_{1,-1}(\al)$ is defined in terms of $\scG_{-1}^{(\al)}(x).$
Nevertheless, in our subsequent computations and considerations the formulation 
in terms of  $\scG_{-1,m}^{(\al)}(x)$ 
proves to be useful. The reason for that is, that the integral operator of (\ref{scGeqsm}) has useful 
symmetry properties under transposition. If (\ref{scGeqsm}) is formulated in an operatorial way:
\begin{equation} \label{Opeq}
\begin{split}
\intinf \! dy \, {\cal M}_\al(x,y) \, \scG_{-1,m}^{(\al)}(y)=e^{-x}, 
\end{split}
\end{equation}
where
\begin{equation} \label{cMal}
\begin{split}
{\cal M}_\al(x,y)=\frac{\delta(x-y)}{\mm(x)}-G_\al(x-y),
\end{split}
\end{equation}
denotes the kernel of the corresponding integral operator\footnote{Here, $\delta$ stands for Dirac delta.}, 
then the transposed kernel differs from the original one only by an $\al \to -\al$ replacement: 
\begin{equation} \label{cMalTR}
\begin{split}
{\cal M}_\al(y,x)={\cal M}_{-\al}(x,y).
\end{split}
\end{equation}
This means that at $\al=0$ it is a symmetric operator. These transposition and symmetry properties will 
play an important role in our further considerations\footnote{For example, the definition (\ref{om1m1def}) 
together with this property implies the identity: $\om_{1,-1}(\al)=\om_{-1,1}(-\al)$}. 

Before going deeper into the details of the derivation, we would like to emphasize, that the integration 
measure $\mm(x)$ is a function of the solution of the NLIE (\ref{DDV}) of the finite volume problem. If the solution 
of the NLIE is denoted by $Z(x)$ as usual, then the explicit form of the measure is given by the formula 
(\ref{m}): 
\begin{equation} \label{mZx}
\begin{split}
\mm(x)\equiv\mm[Z(x)]=\scL_+[Z(x+i\, 0)]+\scL_-[Z(x-i\, 0)], \quad \mbox{with} \quad 
\scL_\pm[Z(x)]=\frac{e^{\pm \, i\,Z(x)}}{1+e^{\pm \, i\,Z(x)}}.
\end{split}
\end{equation}
It follows, that the measure can be considered as a function of an arbitrary function. 
Based on this viewpoint, we define several useful measure functions corresponding to some important 
choices of this arbitrary function. In our subsequent computations four measure functions will play 
an important role. 
\begin{itemize}
\item The measure of the original problem defined from the exact solution of the NLIE: $\mm(x)=\mm[Z(x)],$

\item The measure functions defined from the $Z_\pm(x)$ kink NLIE solutions (\ref{Zpmdef}): 
\newline $\mm_\pm(x)=\mm[Z_\pm(x)],$

\item The measure corresponding to the asymptotic solution of the NLIE (\ref{Zasdef}): 
\newline $\mm_{as}(x)=\mm[Z_{as}(x)],$ where the asymptotic solution is defined %made
 out of the kink solutions and the plateau value as follows:  $Z_{as}(x)=Z_+(x-\log\tfrac{2}{\ell})+Z_-(x+\log\tfrac{2}{\ell})-z_0.$

\end{itemize}

It is important to note, that $\mm(x)$ and $\mm_{as}(x)$ are $\ell$-dependent functions. They 
exponential to exponential rapidly tend to zero at both infinities. This makes, integral expressions 
containing their products with any exponentially blowing functions, always convergent on the whole real axis.

On the other hand the kink measures $\mm_\pm(x)$ are independent of $\ell,$ and decay "exponential to exponential" rapidly to zero only  at one of the infinities\footnote{I.e. $+\infty$ or $-\infty$.}. In the other infinity, they tend to a constant 
value corresponding to the plateau nature of the $\ell \to 0$ limit of the mathematical problem. The 
typical integrals in the UV computations, which contain these measures, will diverge at the 
infinity corresponding to the plateau regime. Thus refined considerations are necessary to treat their  
integrals, when $\ell \to 0$. 

With the help of these measures, one can define the important kink linear problems, whose solutions 
enter the final formulas of the UV expansion. These are as follows\footnote{These equations agree with those 
defined in (\ref{kinkmalpha}) and (\ref{kinkpalpha}).}: 
\begin{equation} \label{scGkinkpmj}
\begin{split}
\scG_{\pm j}^{(\al),(\pm)}(x)-\intinf \! dy \, G_\al(x-y)\, \mm_\pm(y) \, \scG_{\pm j}^{(\al),(\pm)}(x)=e^{\pm j\, x}, 
\qquad j=1,2,...
\end{split}
\end{equation}
where in the notation for the solution function $\scG_{\pm j}^{(\al),(\pm)}(x),$ the superscript 
refers to the kink regime given by the measure $\mm_\pm(x)$ and the subscript $\pm j$ refers to the 
source term $e^{\pm j\, x}.$

The main assumptions and the strategy of the $\ell \to 0$ UV expansion of $\tom_{1,-1}(\al)$ are as follows. 
We assume, that the asymptotic solution is a very good approximation of the solution to the problem 
in the UV limit. It is supposed to be so good, that the leading $\ell \to 0$ contributions are 
completely encoded into this function. Thus, we assume, that the deviation of the exact solution 
from the asymptotic one, is small in the UV regime and can be treated as a minor correction.

Thus, our first step in the UV expansion, is to expand (\ref{om1m1def}) around the asymptotic solution. 
Both the leading asymptotic term and the corrections to it, have nontrivial $\ell$ dependence, 
but based on our assumptions, each term in the expansion becomes smaller and smaller in the 
$\ell \to 0$ limit, as the order of the expansion increases. 

In the second step, we further expand these leading and correction terms, in $\ell$ 
 and stop in the expansion at the desired order. 
In the sequel, we will formulate these steps in precise mathematical formulas. 

The corrections to the leading asymptotic term comes from the corrections of the measure:
$\mm(x)=\mm_{as}(x)+\delta \mm(x),$ which is a consequence of the small in the UV limit deviation between  
$Z(x)$ and $Z_{as}(x).$ Namely at linear order, $\delta \mm(x) \simeq \mm'[Z_{as}(x)]\, \delta Z(x)$ with 
$\delta Z(x)=Z(x)-Z_{as}(x).$

We start with the definition of the asymptotic solution of the equations (\ref{scGeqs}) and (\ref{scGeqsm}). 
They are defined as solutions of the asymptotic equations, as follows:
\begin{eqnarray} \label{scGeqsAS}
\scG_{-1}^{(\al),(as)}(x)&-&\intinf \! dy \, G_\al(x-y)\, \mm_{as}(y) \, \scG_{-1}^{(\al),(as)}(y)=e^{-x}, \\
\frac{\scG_{-1,m}^{(\al),(as)}(x)}{\mm_{as}(x)}&-&\intinf \! dy \, G_\al(x-y)\,  \scG_{-1,m}^{(\al),(as)}(y)=e^{-x}, 
\label{scGeqsmAS} \\
\scG_{-1,m}^{(\al),(as)}(x)&=& \mm_{as}(x) \, \scG_{-1}^{(\al),(as)}(x).
\end{eqnarray}
The functions $\scG_{-1}^{(\al),(as)}(x)$ and $\scG_{-1,m}^{(\al),(as)}(x)$ are called the asymptotic solutions
of the linear problems (\ref{scGeqs}) and (\ref{scGeqsm}), respectively. They contain the leading 
$\ell \to 0$ solution of the problem and so they form the background around which, 
the original problem is expanded in the UV limit. Nevertheless, it is obvious, that these functions have 
nontrivial $\ell$ dependence, which requires further careful treatment to get the UV series expansion in $\ell$. 

In principle, one should solve (\ref{scGeqsAS}) or (\ref{scGeqsmAS}) to get the asymptotic solution. 
Nevertheless, these asymptotic equations are still very complicated and cannot be solved exactly. 
The only thing we can do, is to expand their solutions around the "leading asymptotic solutions", which are 
in principle known functions. They capture the leading in small $\ell$ behaviour of these solutions.

They can be found by a very simple train of thoughts. The source term in both (\ref{scGeqsAS}) and 
(\ref{scGeqsmAS}) is $e^{-x}.$ The dominant contribution of this function comes from the $(-\infty,0)$ regime,  
where the $\mm_{as}(x) \to \mm_-(x+\log\tfrac{2}{\ell})$ replacement can be done.
 %in (\ref{scGeqsAS}) and (\ref{scGeqsmAS}). 

Performing this replacement in (\ref{scGeqsAS}), making the $x \to x - \ltl$ change of variables and 
comparing the result to the kink equation (\ref{scGkinkpmj}), one can conclude, that the leading 
asymptotic solutions of (\ref{scGeqsAS}) and (\ref{scGeqsmAS}), can be expressed in terms of 
the kink solutions of (\ref{scGkinkpmj}) as follows: 
\begin{equation} \label{asscG}
\begin{split}
\scG_{-1}^{(\al),(as)}(x)&\simeq \frac{2}{\ell} \, \scG_{-1}^{(\al),(-)}(x+\log \tfrac{2}{\ell}),  \\
\scG_{-1,m}^{(\al),(as)}(x)&\simeq \frac{2}{\ell} \mm_{as}(x) \, \scG_{-1}^{(\al),(-)}(x+\log \tfrac{2}{\ell}).
\end{split}
\end{equation}
Since, we will use extensively these functions, we introduce a notation for them, by putting a $0$ subscript 
on the  asymptotic solutions. Namely:
\begin{equation} \label{asscG0}
\begin{split}
\scG_{-1,0}^{(\al),(as)}(x)&= \frac{2}{\ell} \, \scG_{-1}^{(\al),(-)}(x+\log \tfrac{2}{\ell}),  \\
\scG_{-1,m,0}^{(\al),(as)}(x)&= \frac{2}{\ell} \mm_{as}(x) \, \scG_{-1}^{(\al),(-)}(x+\log \tfrac{2}{\ell}).
\end{split}
\end{equation}
Now, we are in the position to formulate the expansion of the original linear problem (\ref{scGeqsm}). 
As a first step, we expand it around the leading asymptotic solution $\scG_{-1,m,0}^{(\al),(as)}(x).$ 
We identify two basic kinds of corrections to this background: 
\begin{equation} \label{dscG}
\begin{split}
\scG_{-1,m}^{(\al)}(x)=\scG_{-1,m,0}^{(\al),(as)}(x)+\delta \scG_{-1,m}^{(\al),(as)}(x)+\delta \scG_{-1,m}^{(\al),\delta m}(x)+\dots.
\end{split}
\end{equation}
The first correction $\delta \scG_{-1,m}^{(\al),(as)}(x)$ comes purely from the asymptotic equation 
(\ref{scGeqsmAS}), while the second one $\de \scG_{-1,m}^{(\al),\delta m}(x),$ comes from the corrections  of the 
measure to the asymptotic equation. By the computations detailed in the rest of this appendix, it can be shown 
that upto the first two leading terms in the $\ell \to 0$ expansion, no further corrections are needed to be 
taken into account. Namely, $\sim \delta \mm(x)^2$ terms would give higher than 2nd order terms in the UV 
expansion.

It is not hard to derive the linear problems satisfied by the functions being on the right hand side of (\ref{dscG}).
Now, we present them in the subsequent lines. 
The leading asymptotic solution satisfies the following equation:
\begin{equation} \label{scGmas0}
\begin{split}
\frac{\scG_{-1,m,0}^{(\al),(as)}(x)}{\mm_{as}(x)}-\intinf \! dy \, G_\al(x-y)\, 
\frac{\mm_-(y+\log\tfrac{2}{\ell})}{\mm_{as}(y)} \, \scG_{-1,m,0}^{(\al),(as)}(y)=e^{-x}.
\end{split}
\end{equation}
The asymptotic correction: $\delta \scG_{-1,m}^{(\al),(as)}(x)=\scG_{-1,m}^{(\al),(as)}(x)- \scG_{-1,m,0}^{(\al),(as)}(x),$ is the solution of the linear problem as follows:
\begin{equation} \label{dscGmas}
\begin{split}
\frac{\de \scG_{-1,m}^{(\al),(as)}(x)}{\mm_{as}(x)}-\intinf \! dy \, G_\al(x-y)\,  
\de \scG_{-1,m}^{(\al),(as)}(y)=\de RH^{(as)}(x), \\
\de RH^{(as)}(x)=\frac{2}{\ell} \intinf \! dy \, G_\al(x-y) \, \scG_{-1}^{(\al),(-)}(y+\log\tfrac{2}{\ell}) \, 
\left[ \mm_{as}(y)-\mm_-(y+\log\tfrac{2}{\ell}) \right].
\end{split}
\end{equation}
Finally, $\de \scG_{-1,m}^{(\al),\delta m}(x)$ fulfills a linear equation, the source term of which is proportional 
to the correction of the measure $\delta \mm(x)=\mm(x)-\mm_{as}(x):$
\begin{equation} \label{dscGdm}
\begin{split}
\frac{\de \scG_{-1,m}^{(0),\de m}(x)}{\mm_{as}(x)}-\intinf \! dy \, G(x-y)\,  
\de \scG_{-1,m}^{(0),\de m}(y)=\frac{\de \mm(x)}{\mm_{as}^2(x)} \, \scG_{-1,m}^{(0),(as)}(x),
\end{split}
\end{equation}
where we presented the equation at the only relevant for us $\al=0$ 
point\footnote{The reason for this is that 
this function will contribute only to the 2nd order term in $\tom_{1,-1}(\al),$ and we are interested in  
this term only, when $\al=0.$}.

Based on the representation (\ref{dscG}), in the UV limit $\tom_{1,-1}(\al)$ can be given as a sum of 3 terms:
\begin{equation} \label{tom1m1terms}
\begin{split}
\tom_{-1,1}(\al)\! &=\!\!\!\! \intinf \!\! dx \, e^x \, \scG_{-1,m}^{(\al)}(x)=\!\!\!\!
 \intinf \!\! dx \, e^x \,\left( \scG_{-1,m,0}^{(\al),(as)}(x)+\de 
\scG_{-1,m}^{(\al),(as)}(x)+\de \scG_{-1,m}^{(\al),\de m}(x) \right)+....
\end{split}
\end{equation}
The 3 terms in the rhs. of (\ref{tom1m1terms}) are denoted by $\tom_0^{as}(\al), $ $\de \tom^{as}(\al), $ and 
$\de \tom^{\de m}(\al),$ respectively: 
\begin{equation} \label{A20}
\begin{split}
\tom_{-1,1}(\al)\! =\tom_0^{as}(\al)+\de \tom^{as}(\al)+
\de \tom^{\de m}(\al)+....
\end{split}
\end{equation}
The UV expansion of each term will be presented separately, but first we bring them into such forms, 
which are more appropriate for the UV expansion. 

The first and simplest term does not require any special treatment. 
It is governed by only the leading asymptotic solution:
\begin{equation} \label{tomas0}
\begin{split}
\tom_0^{as}(\al)\! =\!\!\!\! \intinf \!\! dx \, e^x \, \scG_{-1,m,0}^{(\al),(as)}(x)\!=\frac{2}{\ell}
\!\! \intinf \!\! dx \, e^x \, \mm_{as}(x) \, \scG_{-1}^{(\al),(-)}(x+\log\tfrac{2}{\ell}).
\end{split}
\end{equation}
The next term is:
\begin{equation} \label{dtomasdef}
\begin{split}
\de \tom^{as}(\al)=\! =\!\!\!\! \intinf \!\! dx \, e^x \, \de \scG_{-1,m}^{(\al),(as)}(x).
\end{split}
\end{equation}
To rewrite it in a form being appropriate for the UV expansion, we represent the solution of (\ref{dscGmas}) 
in a formal way using the asymptotic counterpart of the linear operator (\ref{cMal}) of the problem\footnote{
Here, we mean on asymptotic counterpart, the operator: ${\cal M}^{(as)}_\al(x,y)=\frac{\de(x-y)}{\mm_{as}(x)}-G_\al(x-y),$ which has the same transposition properties as  
those of ${\cal M}_\al(x,y).$ See: (\ref{cMalTR}).}:
\begin{equation} \label{dscGmasformal}
\begin{split}
\de \scG_{-1,m}^{(\al),(as)}(x)=\!\! \intinf \! dy \,{\cal M}_\al^{(as),-1}(x,y) \, \de RH^{(as)}(y).
\end{split}
\end{equation}
Inserting (\ref{dscGmasformal}) into (\ref{dtomasdef}) and exploiting the transposition property (\ref{cMalTR}),  
one obtains the formula as follows:
\begin{equation} \label{dtomasfin}
\begin{split}
\de \tom^{(as)}(\al)=\frac{2}{\ell} \! \intinf \! dx \, \scG_{-1}^{(\al),(-)}(x+\log\tfrac{2}{\ell}) \, 
\left[ \mm_{as}(x)-\mm_-(x+\log\tfrac{2}{\ell}) \right] \,
\left[ \frac{\scG_{+1,m}^{(-\al),(as)}(x)}{\mm_{as}(x)}-e^x \right],
\end{split}
\end{equation}
where the action of the inverse transpose of ${\cal M}_\al^{(as)}(x,y)$ on $e^x$ was defined by the function 
\begin{equation} \label{dscGp1maldef}
\begin{split}
\scG_{+1,m}^{(-\al),(as)}(x)=\intinf \! dy \, {\cal M}_\al^{(as),-1}(y,x) \, e^y.
\end{split}
\end{equation}
As a consequence, it satisfies  a linear equation, that we also exploited when deriving (\ref{dtomasfin}):
\begin{equation} \label{dscGp1mal}
\begin{split}
\intinf dy \, G_{-\al}(x-y) \scG_{+1,m}^{(-\al),(as)}(y)=\frac{\scG_{+1,m}^{(-\al),(as)}(x)}{\mm_{as}(x)}-e^x.
\end{split}
\end{equation}
To treat (\ref{dscGp1maldef}) in the UV limit, we define again the corresponding leading asymptotic solution:
\begin{equation} \label{scGlas}
\begin{split}
\scG_{+1,m,0}^{(-\al),(as)}(x)=\frac{2}{\ell} \, \mm_{as}(x) \, \scG_{+1}^{(-\al),(+)}(x-\log\tfrac{2}{\ell}).
\end{split}
\end{equation}
This definition came from a similar train of thoughts, which led to (\ref{asscG}).

The proper solution of (\ref{dscGp1mal}) is sought of an expansion around this background:
\begin{equation} \label{scGpm0def}
\begin{split}
\scG_{+1,m}^{(-\al),(as)}(x)=\scG_{+1,m,0}^{(-\al),(as)}(x)+\de \scG_{+1,m}^{(-\al),(as)}(x).
\end{split}
\end{equation}
With the help of (\ref{dscGp1mal}), one can derive the linear equations satisfied by  
the leading order background $\scG_{+1,m,0}^{(-\al),(as)}(x)$ 
and the correction $\de \scG_{+1,m}^{(-\al),(as)}(x).$ 

The linear equation for the leading order background takes the form:
\begin{equation} \label{scGpas0}
\begin{split}
\frac{\scG_{+1,m,0}^{(-\al),(as)}(x)}{\mm_{as}(x)}-\intinf \! dy \, G_{-\al}(x-y)\, 
\frac{\mm_+(y-\log\tfrac{2}{\ell})}{\mm_{as}(y)} \, \scG_{+1,m,0}^{(-\al),(as)}(y)=e^{x}.
\end{split}
\end{equation}
The asymptotic correction: $\delta \scG_{+1,m}^{(-\al),(as)}(x)=\scG_{+1,m}^{(-\al),(as)}(x)- \scG_{+1,m,0}^{(-\al),(as)}(x),$ 
is the solution of the linear problem as follows:
\begin{equation} \label{dscGpas}
\begin{split}
\frac{\de \scG_{+1,m}^{(-\al),(as)}(x)}{\mm_{as}(x)}-\intinf \! dy \, G_{-\al}(x-y)\,  
\de \scG_{+1,m}^{(-\al),(as)}(y)=\de RH^{(-\al)}(x), \\
\de RH^{(-\al)}(x)=\frac{2}{\ell} \intinf \! dy \, G_{-\al}(x-y) \, \scG_{+1}^{(-\al),(+)}(y-\log\tfrac{2}{\ell}) \, 
\left[ \mm_{as}(y)-\mm_+(y-\log\tfrac{2}{\ell}) \right].
\end{split}
\end{equation}
The formulas (\ref{dscGp1mal})-(\ref{dscGpas}) will play an important role during the concrete 
computation of the coefficients in the $\ell \to 0$ series representation of $\de \tom^{as}(\al).$ 

The last term in (\ref{A20}) at $\al=0$:
\begin{equation} \label{dtomdmdef}
\begin{split}
\de \tom^{\de m}(0)=\! =\!\!\!\! \intinf \!\! dx \, e^x \, \de \scG_{-1,m}^{(0),\de m}(x),
\end{split}
\end{equation}
where $\de \scG_{-1,m}^{(0),\de m}(x)$ is the solution of (\ref{dscGdm}). 
It is useful again to use the following formal operatorial representation 
of the solution: 
\begin{equation} \label{dscGdmop}
\begin{split}
\de \scG_{-1,m}^{(0),\de m}(x)=\intinf \! dy \, {\cal M}_0^{(as),-1}(x,y) \, \frac{\de \mm(y)}{\mm_{as}^2(y)} \, \scG_{-1,m}^{(0),(as)}(y).  
\end{split}
\end{equation}
Inserting this representation into (\ref{dtomdmdef}), 
and pushing the action of ${\cal M}_0^{(as),-1}(x,y),$ 
to the $e^x$ term in (\ref{dtomdmdef}), one obtains the formula as follows: 
\begin{equation} \label{dtomdmfin}
\begin{split}
\de \tom^{\de m}(0)=\! =\!\!\!\! \intinf \!\! dx  \,  \scG_{-1,m}^{(0),(as)}(x) \, \frac{\de \mm(y)}{\mm_{as}^2(y)} \, \scG_{+1,m}^{(0),(as)}(x),
\end{split}
\end{equation}
where the functions $\scG_{\pm 1,m}^{(0),(as)}(x)$ satisfy the equations:
\begin{equation} \label{scGpm1meq}
\begin{split}
\frac{\scG_{\pm 1,m}^{(0),(as)}(x)}{\mm_{as}(x)}-\intinf \! dy \, G(x-y) \, \scG_{\pm 1,m}^{(0),(as)}(y)=e^{\pm x}.
\end{split}
\end{equation}
Here $\de \mm(x) \simeq \mm'[Z_{as}(x)]\, \de Z(x),$ where $\de Z(x)$ denotes the $O(\ell^{\tfrac{4}{p+1}})$ corrections to $Z_{as}(x).$ (See, (\ref{ZUVansatz}), (\ref{dZ1}).) Its mathematical treatment will 
be given at a later point of this appendix.

To compute $\de \tom^{\de m}(0)$ upto 2nd order, we will need to use the by now usual representation: 
$\delta \scG_{\pm 1,m}^{(0),(as)}(x)=\scG_{\pm 1,m,0}^{(0),(as)}(x)+\delta \scG_{\pm 1,m}^{(0),(as)}(x).$

{\bf To summarize:} the formulas (\ref{tomas0}), (\ref{dtomasfin}) and (\ref{dtomdmfin})  constitute the 
starting point for the further steps of the UV expansion.

Now, we are in the position to start the $\ell \to  0$ expansion of $\tom_{1,-1}(\al).$ We will do that separately for each of the 3 terms: 
$\tom_0^{(as)}(\al), \,\, \de \tom^{(as)}(\al)$ and $\de \tom^{\de m}(\al).$ The leading in small $\ell$ contribution will be computed in the repulsive regime ($1<p$), 
when the  parameter $\al$ is in the range  
$0\leq\al<2,$ while the next to leading order term will be determined only at the $\al=0$ point. 
We will discuss the UV treatment of the simplest quantity $\tom_0^{(as)}(\al)$ in detail. 
The same way of thinking leads to the $\ell \to 0$ series expansions of the other two terms.  

\subsection{The basic strategy of computing the coefficients of the UV series }

Now, we just write down the basic strategy of computing the first two terms in the $\ell \to 0$ series representation of $\tom_{1,-1}(\al).$ 
First, one has to recognize, that each of the 3 terms of $\tom_{1,-1}(\al)$ in  (\ref{A20}), %(\ref{tom1m1terms}), 
 has nontrivial $\ell$ dependence which should be extracted to reach our goal. Each term in $\tom_{1,-1}(\al)$ is  an integral expression of 
products of left- and right-mover combinations of some $\ell$-independent kink-functions.\footnote{Here, we call an $\ell-$independent function $f,$ 
left- or right-mover, if its argument is shifted by $+\log\tfrac{2}{\ell}$ or $-\log\tfrac{2}{\ell},$ respectively. Namely: $f(x\pm\log\tfrac{2}{\ell}).$}
This means, that in the relevant for us expressions, the important $\ell$ dependence is present in some $\ell-$dependent shifts 
in the arguments of the constituting $\ell-$independent functions.

To disentangle this $\ell-$dependency structure, we proceed as follows. Each $\intinf$ integral is cut into two pieces, to be a sum of an $\int\limits_{0}^\infty$
and an $\int\limits_{-\infty}^0$ integrals. The integral on $(0,\infty)$  is transformed to an $\int\limits_{-\log\tfrac{2}{\ell}}^\infty$ integral by an 
$x \to x+\log\tfrac{2}{\ell}$ change of variables. The advantage of this transformation is, that on the one hand, 
the arguments of the constituent right-mover "+"-kink functions are liberated from their $\ell-$dependence, and on the other hand, the 
constituent "-"-kink functions will be in their plateau regime, where they can be expanded in small $\ell.$ This expansion of the "-"-kink 
functions brings some definite $\ell$ powers in front of the integral expressions, but the integrand becomes $\ell-$independent. Thus, the 
$\ell-$dependence of the integral under consideration comes from a pure power expression of $\ell$ coming from the "-"-kink expansion and 
from the lower bound of the integral multiplying it. Since the integrand is $\ell-$independent, 
the $\ell-$dependence coming from the lower bound,  
can be easily determined from the $x \sim -\infty$ series representation of the integrand. According to our assumptions, this is in principle known, 
since this corresponds to the expansion of the integrand in its plateau regime. With the help of the method described in the previous lines, the small $\ell$ 
series representation of any of the $0$ to $\infty$ integrals appearing in  (\ref{A20}) %(\ref{tom1m1terms}) 
can be expanded systematically when $\ell \to 0.$ 

The $\ell \to 0$ treatment of the integrals running from $-\infty$ to $0$, is completely analogous. 
So, the small $\ell$ dependence of each of the 3 terms in (\ref{A20}) %(\ref{tom1m1terms})
 can be determined in terms of this method. 
To get the method clearer, now we demonstrate its application in more details to the simplest quantity $\tom_0^{(as)}(\al).$

\subsection{The computation of the $\ell \to 0$ series of  $\tom_0^{(as)}(\al)$}

We proceed as described in the previous subsection.

Starting from (\ref{tomas0}), cutting the integral into 2 pieces and shifting the two integrals appropriately, $\tom_0^{(as)}(\al)$ can be written as a sum 
of two terms:
\begin{equation} \label{tomas0sum}
\begin{split}
\tom_0^{(as)}(\al)=\tom_0^{(as),+}(\al)+\tom_0^{(as),-}(\al),
\end{split}
\end{equation}
where 
\begin{eqnarray}\label{tomas0p}
\tom_0^{(as),+}(\al)&=& \left(\frac{2}{\ell}\right)^2 \, \int\limits_{-\log\tfrac{2}{\ell}}^\infty \! dx \, e^x \, \mm_{as}(x+\log\tfrac{2}{\ell}) \,
\scG_{-1}^{(\al),(-)}(x+2 \log\tfrac{2}{\ell}),\\
\tom_0^{(as),-}(\al)&=&\int\limits_{-\infty}^{\log\tfrac{2}{\ell}} \! dx \, e^x \, \mm_{as}(x-\log\tfrac{2}{\ell}) \,
\scG_{-1}^{(\al),(-)}(x).
\label{tomas0m}
\end{eqnarray}
First, consider $\tom_0^{(as),+}(\al).$
In order to expand it at $\ell \to 0,$ one needs to know a few series representations: 
\begin{equation} \label{serscGm1al}
\begin{split}
\scG_{-1}^{(\al),(-)}(x) &\overset{x \to \infty}{=} e^{\tfrac{\al-2}{p+1} x}\, \sum\limits_{k=0}^\infty c_{k+1}^{(-1,\al)} \, e^{-\tfrac{2 k}{p+1}x}, \\
Z_-(x) &\overset{x \to \infty}{=}z_0+ c_1 \, e^{-\frac{2 x}{p+1}}+c_2 \, e^{-\frac{4 x}{p+1}}+... 
\end{split}
\end{equation}
As a consequence, the following small $\ell$ series representations are valid in the
\newline $-\ltl<x<\infty$ range of the integration:
\begin{equation} \label{smallell1}
\begin{split}
\scG_{-1}^{(\al),(-)}(x+2 \, \ltl) &\simeq \left(\frac{2}{\ell}\right)^{\tfrac{2 \al-4}{p+1}} \, c_1^{(-1,\al)} \, e^{\tfrac{\al-2}{p+1}x} +..., \\
Z_-(x+2 \, \ltl)-z_0 &\simeq \left(\frac{\ell}{2}\right)^{\tfrac{4}{p+1}} \, c_1 \,  e^{-\frac{2 x}{p+1}}+\left(\frac{\ell}{2}\right)^{\tfrac{8}{p+1}} \, c_2 \, e^{-\frac{4 x}{p+1}}+... \\
\mm_{as}(x+\ltl) &= \mm[Z_+(x)+Z_-(x+ 2 \, \ltl)-z_0]\simeq\mm_+(x)+O(\ell^{\tfrac{4}{p+1}}).
\end{split}
\end{equation}
Inserting (\ref{smallell1}) into  (\ref{tomas0p}), at leading order one ends up with:
\begin{equation} \label{tomas0pfin0}
\begin{split}
\tom_0^{(as),+}(\al)&\simeq \left(\frac{2}{\ell}\right)^{2+\tfrac{2 \al}{p+1}-\tfrac{4}{p+1}} \, c_1^{(-1,\al)} \intli \!\! dx \, e^x 
\, \mm_+(x) \, \, e^{\tfrac{\al-2}{p+1} x}.
\end{split}
\end{equation}
The last step is to analyse the integral when $\ell \to 0,$ which corresponds to sending the lower bound to $-\infty.$
It is easy to show, that the integral is convergent at $+\infty,$ since here $\mm_+(x)$ tends to zero  exponential to exponential rapidly.  
At $-\infty$ the integrand behaves as $e^{\left(1+\tfrac{\al-2}{p+1}\right) x},$ which means, 
that in our relevant regime ($1<p<\infty, 0<\al<2$) 
at leading order in $\ell,$ the lower bound of the integral can be sent to $-\infty.$ Thus, the leading order in small $\ell$ expression 
for $\tom_0^{(as),+}(\al)$ takes the form:
\begin{equation} \label{tomas0pfin1}
\begin{split}
\tom_0^{(as),+}(\al)&\simeq \left(\frac{2}{\ell}\right)^{2+\tfrac{2 \al}{p+1}-\tfrac{4}{p+1}} \, c_1^{(-1,\al)} \intinf \!\! dx \, e^x 
\, \mm_+(x) \,\, e^{\tfrac{\al-2}{p+1} x}.
\end{split}
\end{equation}
The other term can be treated similarly. One ends up with:
\begin{equation} \label{tomas0mfin}
\begin{split}
\tom_0^{(as),-}(\al)&\simeq\intil \!\! dx \, e^x \, \mm_-(x) \, \scG_{-1}^{(\al),(-)}(x).
\end{split}
\end{equation}
The analysis of the behaviour of the integrand at $\pm \infty$ reveals, that it is convergent at $-\infty$ due to the fast decay of $\mm_-(x).$ 

At $+\infty$ the integrand behaves as $e^{\left(1+\tfrac{\al-2}{p+1}\right) x},$ which means, 
that in our relevant regime ($1<p<\infty, \, 0<\al<2$) it is divergent. Integrating it upto $\ltl,$  
one obtains the leading in small $\ell$ behaviour as follows:
$$ \tom_0^{(as),-}(\al) \sim \left(\frac{2}{\ell}\right)^{1+\tfrac{\al-2}{p+1}}.$$
It is easy to show, that in the regime of our interest\footnote{I.e. $1<p<\infty,\, 0<\al<2.$}, it is negligible 
with respect to the leading contribution (\ref{tomas0pfin1}) of $\tom_0^{(as),+}(\al).$ 

Thus, the leading contribution to $\tom_0^{(as)}(\al)$ comes form $\tom_0^{(as),+}(\al).$  
It has the form in $\ell,$ which agrees with field theoretical predictions.

To summarize, the leading in small $\ell$ contribution to $\tom_0^{(as)}(\al)$ takes the form:
\begin{equation} \label{tomas01fin}
\begin{split}
\tom_0^{(as)}(\al){\bigg |}_1&= \left(\frac{2}{\ell}\right)^{2+\tfrac{2 \al}{p+1}-\tfrac{4}{p+1}} \, c_1^{(-1,\al)} \intinf \!\! dx \, e^x 
\, \mm_+(x) \,\, e^{\tfrac{\al-2}{p+1} x}.
\end{split}
\end{equation}

We close this subsection with some comments concerning the small $\ell$ behaviour coming from (\ref{tomas0mfin}). 
The leading small $\ell$ behaviour of $\tom_0^{(as),-}(\al=0)$ does not fit into the pattern 
$\ell^{-2+\frac{4 n}{p+1}},$ with $n \in \mathbb{N},$ predicted by field theory (See, (\ref{om1m10ser})). 
It is not hard to show, that this proves to be true for the corrections, too. Namely, the typical 
small $\ell$ behaviour of the terms in the UV expansion of $\tom_0^{(as),-}(0)$ takes the form:
$\sim \ell^{-1+\tfrac{2+4 \,n}{p+1}},$ with $n \in \mathbb{N}.$ Since we do not expect such a 
behaviour in the UV limit, we simply do not pay attention to these terms and compute only those, whose 
small $\ell$ behaviour is in agreement with field theoretical predictions. 
Here, and in the sequel, we assume, that these non-predicted terms are either zero or 
cancel with contributions coming from  other terms of (\ref{A20}). 

We mention, that our high precision numerical studies, supported these assumptions.

\subsubsection{The 2nd order correction at $\al=0$} \label{A21}

Proceeding completely analogously to the leading order computation, but making the expansion to 
higher order terms in $\ell$, one obtains the following result for the 2nd order contribution at $\al=0$:
\begin{equation} \label{tomas02fin}
\begin{split}
\tom_0^{(as)}(0){\bigg |}_2&= \left(\frac{2}{\ell}\right)^{2-\tfrac{8}{p+1}} \, 
\CP\limits_{\ell \to 0} \intli \! dx \, e^{(1-\tfrac{4}{p+1})\, x} \, \left( c_2^-\, \mm_+(x)+c_1 \, c_1^-\, \mm'[Z_+(x)] \right),
\end{split}
\end{equation}
where we introduced the short notations $c_j^\pm$ for $c_j^{(\pm1,{\al=0})}, \quad j=1,2,....$
In (\ref{tomas02fin}) 
the operation $\CP\limits_{\ell \to 0}$ on the expression following it, 
means that the constant term in the $\ell \to 0$ 
series representation  should be computed. The concrete integral 
representation of the result of this prescription can be $p$-dependent, 
as we will see that in the subsequent lines. 

To get a concrete expression for this constant part, the $\ell-$dependence of the integral in (\ref{tomas02fin}) 
should be analyzed. The integral is obviously convergent at $+\infty,$ but depending on the value of $p,$ can be 
even diverging at $-\infty.$ Thus, the behaviour of the integrand at $-\infty$ will be crucial. 
It is easy to see, that the integrand in (\ref{tomas02fin}) behaves as 
$\sim  e^{(1-\tfrac{4}{p+1})\, x} \, \left( c_2^-\, \mm[z_0]+c_1 \, c_1^-\, \mm'[z_0] \right)$
when $x \to -\infty.$
It follows, that for $p>3,$ the integral is convergent at $-\infty.$ So, the constant part is given by 
simply sending the lower bound of the integral to $-\infty:$ 
\begin{equation} \label{tomas02finpg3}
\begin{split}
\tom_0^{(as)}(0){\bigg |}_2\!&\!=  \!\left(\frac{2}{\ell}\right)^{2-\tfrac{8}{p+1}}  
\!  \! \intinf \!\!\! dx \, e^{(1-\tfrac{4}{p+1})\, x}  \left( c_2^-\, \mm_+(x)+c_1 \, c_1^-\, \mm'[Z_+(x)] \right), \,\,\, \mbox{for} \,\, p>3.
\end{split}
\end{equation}
On the other hand, when $1<p<3,$ the integral is divergent at $-\infty.$ Since $-\infty$ corresponds to the 
plateau regime, from the series representations (\ref{masympts}),  %(\ref{serscGm1al}),
 it can be easily shown, that the next to leading in $x$ correction to the integrand at $-\infty$, 
 is of order $\sim  e^{(1-\tfrac{2}{p+1})\, x},$ which is integrable, when $1<p<3.$ Since the leading 
at $-\infty$ term $\sim  e^{(1-\tfrac{4}{p+1})\, x}$ is integrable at $+\infty,$ when $1<p<3,$ the integral in 
 (\ref{tomas02fin}) can be rephrased as follows:
\begin{equation} \label{p13int}
\begin{split}
I^{(as)}_0=&\intli \! dx \, e^{(1-\tfrac{4}{p+1})\, x} \, \left( c_2^-\, \mm_+(x)+c_1 \, c_1^-\, \mm'[Z_+(x)] \right)= \\
&\intli \! dx \, e^{(1-\tfrac{4}{p+1})\, x} \, \bigg( c_2^-\, (\mm_+(x)-\mm[z_0])+c_1 \, c_1^-\, (\mm'[Z_+(x)]-\mm[z_0]) \bigg)+ \\
&\intli \! dx \, e^{(1-\tfrac{4}{p+1})\, x} \, \left( c_2^-\, \mm[z_0]+c_1 \, c_1^-\, \mm'[z_0] \right).
\end{split}
\end{equation}
Now, it is easy to compute the constant part, since the first integral is convergent on the whole real axis, 
while the 2nd one can be computed exactly. Thus, one obtains the following result, when $\ell \to 0:$
\begin{equation} \label{p13intfin}
\begin{split}
I^{(as)}_0=&\intinf \! dx \, e^{(1-\tfrac{4}{p+1})\, x} \, \left( c_2^-\, (\mm_+(x)-\mm[z_0])+c_1 \, c_1^-\, (\mm'[Z_+(x)]-\mm[z_0]) \right)+ \\
&+ \, \frac{p+1}{p-3} \left( \frac{\ell}{2}\right)^{1-\tfrac{4}{p+1}} \left( c_2^-\, \mm[z_0]+c_1 \, c_1^-\, \mm'[z_0] \right)+{\cal O}(\ell),
\end{split}
\end{equation}
where ${\cal O}(\ell)$ denotes terms tending to zero in the $\ell \to 0$ limit.

It is worth mentioning, that in (\ref{p13intfin}) beyond the expected constant term, 
there is a divergent one at $\ell \to 0$, as well. 
The extra term gives, an $\sim \ell^{1-\tfrac{4}{p+1}}$ contribution to $\tom_0^{(as)}(\al).$ From field 
theoretical considerations, we do not expect such a term, thus we do not pay attention to it. We assume, that 
after proper evaluation, that correction is either
 zero\footnote{In the derivation, we did not assume any relation 
between, the plateau expansion coefficients entering (\ref{p13intfin}). If they are fine tuned as:
$c_2^-\, \mm[z_0]+c_1 \, c_1^-\, \mm'[z_0]=0,$ than the extra contribution cancels. 
Unfortunately, in our method such a cancellation, can be checked only numerically. }
 or in the full quantity $\tom_{1,-1}(\al)$, it cancels  with similar contributions coming from other terms. 

To summarize, when $1<p<3,$ the 2nd order correction to $\tom_0^{(as)}(0)$  takes the form: 
\begin{equation} \label{tomas02finpg13}
\begin{split}
\tom_0^{(as)}(0){\bigg |}_2 \! &= \!\!\left(\frac{2}{\ell}\right)^{2-\tfrac{8}{p+1}} \!\!\! \intinf \!\!\! dx \, e^{(1-\tfrac{4}{p+1})\, x}  \left( c_2^-\, (\mm_+(x)\!-\!\mm[z_0])+c_1 \, c_1^-\, 
(\mm'[Z_+(x)]\!-\!\mm'[z_0]) \right).
\end{split}
\end{equation}

In the sequel, when evaluating the contributions of the other terms, one can determine the constant parts in a 
very similar way as we did in the previous lines. In general, we will present our formulas using the constant part 
prescription, since the concrete constant part formula may depend on the actual value of $p.$ 

Now, we continue with the UV expansion of $\delta \tom^{(as)}(\al).$

\subsection{The computation of the $\ell \to 0$ series of  $\delta \tom^{(as)}(\al)$}

Our starting formula is (\ref{dtomasfin}). First, we write it as a sum of two terms. 
\begin{equation} \label{dtomassum}
\begin{split}
\delta \tom^{(as)}(\al)=\delta \tom_0^{(as)}(\al)+\delta \tom_\de^{(as)}(\al).
\end{split}
\end{equation}

One of them contains only the leading order asymptotic solutions:
\begin{equation} \label{dtomas00}
\begin{split}
\de \tom_0^{(as)}(\al)&=\frac{2}{\ell} \! \intinf \! dx \, \scG_{-1}^{(\al),(-)}(x+\log\tfrac{2}{\ell}) \, 
\left[ \mm_{as}(x)-\mm_-(x+\log\tfrac{2}{\ell}) \right] \,\times \\
&\times \left[\frac{2}{\ell} \, \scG_{+1}^{(-\al),(+)}(x-\ltl)-e^x \right].
\end{split}
\end{equation}
and the other one is proportional to the asymptotic correction term:
\begin{equation} \label{dtomasdG}
\begin{split}
\de \tom_\de^{(as)}(\al)=\frac{2}{\ell} \! \intinf \! dx \, \scG_{-1}^{(\al),(-)}(x+\log\tfrac{2}{\ell}) \, 
\left[ \mm_{as}(x)-\mm_-(x+\log\tfrac{2}{\ell}) \right] \,
 \frac{\de \scG_{+1,m}^{(-\al),(as)}(x)}{\mm_{as}(x)}.
\end{split}
\end{equation}
It is not hard to see, that (\ref{dtomasdG}) starts to contribute only in the 2nd order $(\sim 
\ell^{-2+\tfrac{8}{p+1}}).$ Thus, to get the leading order contribution, 
only (\ref{dtomas00}) should be analyzed. As explained in the previous section, we rewrite it as a sum of two 
integrals:
\begin{equation} \label{dtomas0}
\begin{split}
\de \tom_0^{(as)}(\al)=\de \tom_0^{(as),+}(\al)+\de \tom_0^{(as),-}(\al),
\end{split}
\end{equation}
where
\begin{equation} \label{dtomas0p}
\begin{split}
\de \tom_0^{(as),+}(\al)&=\left(\frac{2}{\ell}\right)^2 \! \intli \! dx \,
 \scG_{-1}^{(\al),(-)}(x+2\, \log\tfrac{2}{\ell}) \, 
\left[ \mm_{as}(x+\ltl)-\mm_-(x+2 \, \log\tfrac{2}{\ell}) \right] \times \\
&\times \left[ \scG_{+1}^{(-\al),(+)}(x)-e^x \right],
\end{split}
\end{equation}
and
\begin{equation} \label{dtomas0m}
\begin{split}
\de \tom_0^{(as),-}(\al)\!&=\!\frac{2}{\ell} \!\!\! \intil \!\! dx \,
 \scG_{-1}^{(\al),(-)}(x)  \left[ \mm_{as}(x-\ltl)\!-\!\mm_-(x) \right] \!
 \left[\frac{2}{\ell} \, \scG_{+1}^{(-\al),(+)}(x-2 \, \ltl)\!-\!\frac{\ell}{2} e^x \right].
\end{split}
\end{equation}
 Let's start with $\de \tom_0^{(as),+}(\al).$ Using the expansions collected in (\ref{smallell1}), 
one immediately obtains the leading order result:
\begin{equation} \label{dtomas0pres}
\begin{split}
\de \tom_0^{(as),+}(\al)&=\left(\frac{2}{\ell}\right)^{2+\tfrac{2 \al -4}{p+1}} 
\, c_1^{(-1,\al)}\! \intinf \! dx \,\, e^{\tfrac{\al-2}{p+1}x} \, 
\left( \mm_{+}(x)-\mm[z_0] \right)\, \left( \scG_{+1}^{(-\al),(+)}(x)-e^x \right),
\end{split}
\end{equation}
where due to the convergence of the integral, the lower bound could be sent to $-\infty.$ 

The other term $\de \tom_0^{(as),-}(\al)$ is negligible at the order of our interest, 
which we sketchily show in the following lines. In the $-\infty<x<\ltl$ range of integration, 
the following series representations can be used:  
\begin{equation} \label{smallellp1}
\begin{split}
Z_+(x-2 \, \ltl)-z_0&=\left( \frac{\ell}{2}\right)^{\tfrac{4}{p+1}} \, \tilde{c}_1 \,\, e^{\tfrac{2 x}{p+1}}
+O(\ell^{\tfrac{8}{p+1}}) \\
\scG_{+1}^{(-\al),(+)}(x- 2\, \ltl) &\simeq \, \left( \frac{\ell}{2}\right)^{\tfrac{4- 2\al}{p+1}} c_1^{(+1,-\al)} \, e^{\tfrac{2-\al}{p+1} \, x}+..., \\
\mm_{as}(x-\ltl)-\mm_-(x)&=\left( \frac{\ell}{2}\right)^{\tfrac{4}{p+1}} \, \tilde{c}_1 \,\, e^{\tfrac{2 x}{p+1}} \, \mm'[Z_-(x)]++O(\ell^{\tfrac{8}{p+1}}),
\end{split}
\end{equation}
which were derived with the help of the  plateau expansions, as follows:
\begin{equation} \label{A550}
\begin{split}
Z_+(x)-z_0&\overset{x \to -\infty}{=}\tilde{c}_1 \, e^{\tfrac{2 x}{p+1}}+... \\
\scG_{+1}^{(-\al),(+)}(x) &\overset{x \to -\infty}{=} c_1^{(+1,-\al)} \, \, e^{\tfrac{2-\al}{p+1} x}+... 
\end{split}
\end{equation}
Inserting the leading in small $\ell$ terms form (\ref{smallellp1}) into (\ref{dtomas0m}), 
the contribution can be written as a sum of two terms:
\begin{equation} \label{dtomas0msum}
\begin{split}
\de \tom_0^{(as),-}(\al)=\de \tom_\scG^{(as),-}(\al)+\de \tom_e^{(as),-}(\al),
\end{split}
\end{equation}
where
\begin{equation} \label{dtomeG}
\begin{split}
\de \tom_e^{(as),-}(\al)&=-\left( \frac{2}{\ell}\right)^{\tfrac{4}{p+1}} \, \tilde{c}_1 \, 
\intil \! dx \, \scG_{-1}^{(\al),(-)}(x) \,\, e^{(1+\tfrac{2}{p+1}) x} \, \mm'[Z_-(x)], \\
\de \tom_\scG^{(as),-}(\al)&=\left( \frac{2}{\ell}\right)^{2+\tfrac{2 \al-8}{p+1}} \, 
\tilde{c}_1 \,\, c_1^{(+1,-\al)} \!\! \intil \!\! dx \, \scG_{-1}^{(\al),(-)}(x) \,\, e^{\tfrac{4-\al}{p+1} x} \, \mm'[Z_-(x)].
\end{split}
\end{equation}
Both integrals running from $-\infty$ to $\ltl$ in (\ref{dtomeG}), are divergent in the $\ell \to 0$ limit. 
Nevertheless, the leading divergent terms can be easily computed from the series representations of 
the integrands at $+\infty.$ Doing so, the following estimates for the magnitudes in $\ell$ are obtained:
\begin{equation} \label{dtomas0mmagn}
\begin{split}
\de \tom_e^{(as),-}(\al)&\sim \left( \frac{2}{\ell}\right)^{1+\tfrac{\al-4}{p+1}}, \\
\de \tom_\scG^{(as),-}(\al)&\sim \left( \frac{2}{\ell}\right)^{2+\tfrac{2\, \al-6}{p+1}}.
\end{split}
\end{equation}
It is easy to see, that both contributions are negligible with respect to that of $\de\tom_{0}^{(as),+}.$ 
Thus the leading order contribution of $\de \tom^{(as)}$ is given by (\ref{dtomas0pres}):
\begin{equation} \label{dtomas01}
\begin{split}
\de \tom^{(as)}(\al){\bigg{|}_1}&=\left(\frac{2}{\ell}\right)^{2+\tfrac{2 \al -4}{p+1}} 
\, c_1^{(-1,\al)}\! \intinf \! dx \,\, e^{\tfrac{\al-2}{p+1}x} \, 
\left( \mm_{+}(x)-\mm[z_0] \right)\, \left[ \scG_{+1}^{(-\al),(+)}(x)-e^x \right].
\end{split}
\end{equation}
 To close this part, we mention again, that the estimated scalings in (\ref{dtomas0mmagn}) 
considered at the  $\al=0$ point, 
 do not fit into the field theory predicted pattern %$\ell^{-2+\tfrac{4 \,n-2 \al}{p+1}}$ 
$\ell^{-2+\tfrac{4 \,n}{p+1}}$ with $n \in \mathbb{N}.$ 
Thus, we do not pay attention to them. We assume, that such  corrections cancel out, 
when all contributions are taken into account. 
In the framework of our derivation, the cancellation cannot be seen analytically, since 
a priori we do not know the values of the integrals and the expansion coefficients 
appearing in our formulas. 
Only knowing them, it is possible to prove the disappearance of these "unwanted" terms.  
In our numerical work, by fitting in $\ell$ the numerical values of $\tom_{1,-1}(\al)$ 
in the UV regime, %for small $\ell,$
 we could see, that the actual $\ell-$dependence is consistent with the field theoretical prediction. 

\subsubsection{ The 2nd order corrections at $\al=0$ }

We saw, that the leading order UV contribution to $\de \tom^{(as)}(\al)$ came from  $\de \tom_0^{(as)}(\al).$ 
Beyond the leading order, it also contributes to the 2nd order, as well.
Using the standard technique explained in the previous subsections, its 2nd order contribution can be computed, 
as well. 
It takes the form as follows:
\begin{equation} \label{dtomas0fin}
\begin{split}
\delta \tilde{\omega}_0^{as}(0){\big{|}_2}&=\left(\frac{2}{\ell} \right)^{2-\tfrac{8}{p+1}}
\bigg\{ \CP\limits_{\ell \to 0} \! \! \! \int\limits_{-\log\tfrac{2}{\ell}}^{\infty} 
\! \! \! dx \,
\left[ \scG_{+1}^{(0),(+)}(x)-e^{x} \right] \, e^{-\tfrac{4 \, x}{p+1}}  \\ 
&\times \bigg( c_2^{-} \, \left( {\mathfrak m}_+(x)- {\mathfrak m}_0 \right) +
c_1 \, c_1^{-} \, \left( {\mathfrak m}'[Z_+(x)]-{\mathfrak m}'[z_0] \right)
\bigg)+ \\
&+\tilde{c}_1 \, c_1^{+} \CP\limits_{\ell \to 0} \! \! \! \int\limits_{-\infty}^{\log\tfrac{2}{\ell}}
\! \! \! dx \, \scG_{-1}^{(0),(-)}(x) \, e^{\tfrac{4 \, x}{p+1}} \, {\mathfrak m}'[Z_-(x)]+ \\
&+\frac{p+1}{4} \, c_1^- \, \left( c_2^+ \, \mm_1^+ +c_1^+ \, \mm_2^+ + \tilde{c}_1 \, c_2^+ \, \mm'[z_0]+
c_1^+ \, \tilde{c}_2 \, \mm'[z_0]+\frac{c_1^+ \, \tilde{c}_1^2}{2} \, \mm''[z_0] 
\right)
\bigg\}. 
\end{split}
\end{equation}
Here, we mention, that last line of (\ref{dtomas0fin}) is zero, if one takes into account the actual 
values of the plateau constants coming from the measure (See, (\ref{calLrel})). 

As we mentioned in the previous part, $\de \tom_\de^{(as)}(\al)$ in (\ref{dtomasdG}) 
starts to contribute only to the 2nd order. To compute this contribution, we rephrase (\ref{dtomasdG}) a bit, 
by introducing a new function:
 \begin{equation} \label{dscGpk}
\begin{split}
\de \scG_{+}^{(-\al),(as)}(x) = \frac{\de \scG_{+1,m}^{(-\al),(as)}(x)}{\mm_{as}(x)},
\end{split}
\end{equation}
which, following from (\ref{dscGpas}), satisfies the linear equation as follows:
\begin{equation} \label{dscGpaseq}
\begin{split}
\de \scG_{+}^{(-\al),(as)}(x)-\intinf \! dy \, G_{-\al}(x-y)\,  \mm_{as}(y) \,
\de \scG_{+}^{(-\al),(as)}(y)=\de RH^{(-\al)}(x), 
\end{split}
\end{equation}
with the source term $\de RH^{(-\al)}$ given in (\ref{dscGpas}). The solution at leading order in $\ell,$ 
can be sought similarly to that of the NLIE (\ref{ZUVansatz}). Namely, it is composed of a left- and a right-mover kink function
connected at their common plateau value:
\begin{equation} \label{dscGkpas}
\begin{split}
\de \scG_{+}^{(0),(as)}(x)=\left( \frac{2}{\ell} \right)^{1-\tfrac{4}{p+1}} 
\left( \de \scG_{++}(x-\ltl)+\de \scG_{+-}(x+\ltl)-\de \scG_{+0} \right)+O(\ell^{-1}),
\end{split}
\end{equation}
where the functions $\scG_{+\pm}(x)$ are the $\ell-$independent kink functions, 
while $\de \scG_{+0}$ denotes their common plateau value.
Inserting this Ansatz into (\ref{dscGpaseq}), one obtains the corresponding linear problems for the kink
functions:
\begin{equation} \label{dscGpkaseq}
\begin{split}
\de \scG_{+ \pm}(x)-\intinf \! dy \, G(x-y)\, \mm_\pm(y) \,
\de \scG_{+ \pm}(y)=(G \star \de F_\pm)(x), 
\end{split}
\end{equation}
where functions $\de F_\pm(x)$ on the right hand side are given by:
\begin{equation}  \label{dFpm}
\begin{split}
\de F_+(x)&=c_1 \, e^{-\tfrac{2 x}{p+1}} \, \scG_+(x) \, \mm'[Z_+(x)], \\
\de F_-(x)&=c_1^+ \, e^{\tfrac{2 x}{p+1}} \,  \left(\mm_-(x)-\mm[z_0]\right),
\end{split}
\end{equation}
and the $\star$ stands for convolution: $(G \star \de F_\pm)(x)=\intinf \!\! dy \, G(x-y) \, \de F_{\pm}(y).$ 
The analysis of the equations reveals, that the solutions tend to a constant in their plateau regime, which we denoted by $\de \scG_{+0}:$
\begin{equation}  \label{dGp0}
\begin{split}
\de \scG_{+ \pm}(\mp \infty)=\de \scG_{+0}.
\end{split}
\end{equation}
Furthermore, it can be shown, that $\de \scG_{+0}=0$ and the corrections to this plateau value go 
as $e^{\pm \tfrac{2 x}{p+1}}$ for $\de \scG_{+ \pm}(x),$ respectively.

Using (\ref{dscGpk}), inserting (\ref{dscGkpas}) into (\ref{dtomasdG}) and carrying out our standard UV analysis for each term, one obtains the following result 
for the 2nd order contribution of $\de \tom_{\de}^{(as)}(0):$
\begin{equation}  \label{dtomasdefin}
\begin{split}
\de \tom_{\de}^{(as)}(0){\bigg{|}_2} &=\left( \frac{2}{\ell} \right)^{2-\tfrac{8}{p+1}} \bigg\{ 
c_1^- \, 
\CP\limits_{\ell \to 0}
\! \intli \!\! dx \, e^{-\tfrac{2 x}{p+1}} \, \left(\mm_+(x)-\mm[z_0]\right) \, \de \scG_{++}(x)+ \\
&+\tilde{c}_1 \, 
\CP\limits_{\ell \to 0} 
\! \intil \!\! dx \, \scG_-(x) \, e^{\tfrac{2 x}{p+1}} \, \mm'[Z_-(x)] \, \de \scG_{+-}(x)
\bigg\}.
\end{split}
\end{equation}
We note, that due to $\de \scG_{+0}=0,$ the $\CP$ description can be deleted and all integrals 
in (\ref{dtomasdefin}) can be taken from $-\infty$ to $\infty.$

To summarize: the 2nd order correction coming from the term $\de \tom^{(as)}(0)$ is the sum of $\de \tom_{0}^{(as)}(0){\bigg{|}_2}$ and $\de \tom_{\de}^{(as)}(0){\bigg{|}_2}$ 
given in (\ref{dtomas0fin}) and (\ref{dtomasdefin}). 

Thus, our next, but last task is to compute the 2nd order contribution coming from $\de \tom^{\de m}(0).$

\subsection{The 2nd order UV correction from $\de \tom^{\de m}(0)$}

As we mentioned earlier, $\de \tom^{\de m}(0)$ does not contribute to $\tom_{1,-1}(\al)$ at leading  order {\small{($\sim \ell^{-2+\tfrac{4}{p+1}}$)}}. 
It starts to contribute at the 2nd order ($\sim \ell^{-2+\tfrac{8}{p+1}}$). In this subsection we elaborate this contribution.

Our starting formula is (\ref{dtomdmfin}). In order to compute the relevant for us UV contributions, we need to define two more correction 
functions. One of them is just the "-"-analog of $\de \scG_{+}^{(-\al),(as)}(x),$ with the definition as follows:
 \begin{equation} \label{dscGmk}
\begin{split}
\de \scG_{-}^{(\al),(as)}(x) = \frac{\de \scG_{-1,m}^{(\al),(as)}(x)}{\mm_{as}(x)}.
\end{split}
\end{equation}
It is the solution of the linear equation as follows:
\begin{equation} \label{dscGmaseq}
\begin{split}
\de \scG_{-}^{(\al),(as)}(x)-\intinf \! dy \, G_{\al}(x-y)\,  \mm_{as}(y) \,
\de \scG_{-}^{(\al),(as)}(y)=\de RH^{(as)}(x), 
\end{split}
\end{equation}
with the source term $\de RH^{(as)}$ given in (\ref{dscGmas}). The solution at leading order in $\ell,$ 
can be sought similarly to that of the NLIE. Namely, it is composed of a left- and a right-mover kink function
connected at their common plateau value\footnote{From here, we consider the problem at $\al=0.$}:
\begin{equation} \label{dscGkmas}
\begin{split}
\de \scG_{-}^{(0),(as)}(x)=\left( \frac{2}{\ell} \right)^{1-\tfrac{4}{p+1}} 
\left( \de \scG_{-+}(x-\ltl)+\de \scG_{--}(x+\ltl)-\de \scG_{-0} \right)+O(\ell^{-1}),
\end{split}
\end{equation}
the functions $\scG_{-\pm}(x)$ are the $\ell-$independent kink functions, while $\de \scG_{-0}$ is their common plateau value.
Inserting this Ansatz into (\ref{dscGmaseq}), one obtains the corresponding linear problems for the kinks:
\begin{equation} \label{dscGmkaseq}
\begin{split}
\de \scG_{- \pm}(x)-\intinf \! dy \, G(x-y)\, \mm_\pm(y) \,
\de \scG_{- \pm}(y)=(G \star \de {\cal F}_\pm)(x), 
\end{split}
\end{equation}
where the functions $\de {\cal F}_\pm(x)$ on the right hand side are given by:
\begin{equation}  \label{dcFpm}
\begin{split}
\de {\cal F}_+(x)&=c_1^- \, e^{-\tfrac{2 x}{p+1}} \, \left(\mm_+(x)-\mm[z_0]\right), \\
\de {\cal F}_-(x)&=\tilde{c}_1 \, e^{\tfrac{2 x}{p+1}} \, \scG_-(x) \, \mm'[Z_-(x)],
\end{split}
\end{equation}
and the $\star$ stands for convolution.
%: $(G \star \de F_\pm)(x)=\intinf \!\! dy \, G(x-y) \, \de F_{\pm}(y).$ 
The analysis of the equations reveals, that the solutions tend to a constant in their plateau regime, which we denoted by $\de \scG_{-0}:$
\begin{equation}  \label{dGm0}
\begin{split}
\de \scG_{- \pm}(\mp \infty)=\de \scG_{-0}.
\end{split}
\end{equation}
Furthermore, it can be shown, that $\de \scG_{-0}=0,$ 
and the corrections to this plateau value go as 
$e^{\pm \tfrac{2 x}{p+1}}$ for $\de \scG_{- \pm}(x),$ respectively. 

The other important correction function to compute the UV contributions coming from $\de \tom^{\de m}(0),$ is the 
correction to the asymptotic solution of the NLIE. It is defined by:
\begin{equation} \label{dZdef}
\begin{split}
\de Z(x)=Z(x)-Z_{as}(x).
\end{split}
\end{equation}
Inserting this form into the NLIE (\ref{DDV}), and assuming that $\de Z(x)$ is small in the UV limit, 
with the help of the equations (\ref{kinksol}) and (\ref{plateq}), 
a linear equation can 
be derived for its leading in small $\ell$ part\footnote{Namely, possible $\sim \de Z^2$ terms are neglected.}. It takes the form:  
\begin{equation}  \label{dZeq}
\begin{split}
\de Z^{(1)}(x)-\intinf \! dy G(x-y) \, \mm_{as}(y) \, \de Z^{(1)}(y)=(G \star \de f_Z)(x),
\end{split}
\end{equation}
where $\star$ stands for convolution and the source $\de f_Z(x)$ is given by:
\begin{equation}  \label{dfZ}
\begin{split}
\de f_Z(x)&=\frac{1}{i} \! \bigg\{ \! \bigg( \! L_+[Z_{as}(\! x+ \!i \,0)]\!+\! L_+[z_0]\!-\!
L_+[Z_{+}(x\!+\!i \, 0\!-\! \ltl)] \! - \!L_+[Z_{-}(x\!+\!i \, 0\!+\! \ltl)] \! \bigg)\!+ \\
&+\bigg( \! L_-[Z_{as}(\!x\!-\!i \, 0)]\!+\! L_-[z_0]\!-\!
L_-[Z_{+}(\!x\!-\!i \, 0\!-\!\ltl)]  \!-\! L_-[Z_{-}(x\!-\!i \, 0 \!+\!\ltl)] \bigg)
\bigg\}
\end{split}
\end{equation}
where
\begin{equation}  \label{Lpm}
\begin{split}
L_\pm[Z]=\log(1+e^{\pm i\, Z}).
\end{split}
\end{equation}

As it was stated in (\ref{ZUVansatz}) in section \ref{3}, 
the structure of the solution to this equation, is assumed to be similar to that of the asymptotic solution.
 Namely, it is composed of a left- and a right-mover kink function
connected at their common plateau value\footnote{
We note, that: $\de Z_0=\de Z_0^{(1)}$ and $\de Z_\pm(x)=\de Z^{(1)}_\pm(x)$ in (\ref{ZUVansatz}).}:
\begin{equation} \label{dZk}
\begin{split}
\de Z^{(1)}(x)=\left( \frac{\ell}{2} \right)^{\tfrac{4}{p+1}} 
\left( \de Z_{+}(x-\ltl)+\de Z_{-}(x+\ltl)-\de Z_{0} \right)+O(\ell^{\tfrac{8}{p+1}}),
\end{split}
\end{equation}
where the functions $\de Z_{\pm}(x)$ are the $\ell-$independent kink correction functions, while $\de Z_{0}$ is their common plateau value.
Inserting this Ansatz into (\ref{dZeq}), one obtains the corresponding linear problems for the corrections to the kinks:
\begin{equation} \label{dscGpkaseq}
\begin{split}
\de Z_{\pm}(x)-\intinf \! dy \, G(x-y)\, \mm_\pm(y) \,
\de Z_{\pm}(y)=(G \star \de f_\pm)(x), 
\end{split}
\end{equation}
where functions $\de f_\pm(x)$ on the right hand side are given by:
\begin{equation}  \label{dfpm}
\begin{split}
\de f_+(x)&=c_1 \, e^{-\tfrac{2 x}{p+1}} \, (\mm_+(x)-\mm[z_0]), \\
\de f_-(x)&=\tilde{c}_1 \, e^{\tfrac{2 x}{p+1}}  \, (\mm_-(x)-\mm[z_0]),
\end{split}
\end{equation}
and the $\star$ stands for convolution.
%$(G \star \de F_\pm)(x)=\intinf \!\! dy \, G(x-y) \, \de F_{\pm}(y).$ 
The analysis of the equations reveals, that the solutions tend to a constant in their plateau regime, which we denoted by $\de Z_{0}:$
\begin{equation}  \label{dZ0}
\begin{split}
\de Z_{\pm}(\mp \infty)=\de Z_{0}.
\end{split}
\end{equation}
Furthermore, it can be shown, that $\de Z_{0}=0$ 
and the corrections to this plateau value go as $e^{\pm \tfrac{2 x}{p+1}}$ for $\de Z_{\pm}(x),$ respectively.

Now, we are in the position to start to extract the leading in $\ell \to 0$ terms in $\de \tom^{\de m}(0).$
To do that, we insert $\de \mm(x) \simeq \mm'[Z_{as}(x)]\, \de Z(x),$ and  the following representations for the necessary asymptotic solutions into (\ref{dtomdmfin}):
\begin{equation}  \label{Gasform}
\begin{split}
\scG_{\pm1,m}^{(0),(as)}(x)=\scG_{\pm1,m,0}^{(0),(as)}(x)+\mm_{as}(x) \, \de \scG_{\pm}^{(0),(as)}(x).
\end{split}
\end{equation}
As a result, one ends up with:
\begin{equation}  \label{dtomdm1}
\begin{split}
\de \tom^{\de m}(0)&=\intinf \!\! dx \, \left[ \frac{\scG_{-1,m,0}^{(0),(as)}(x)}{\mm_{as}(x)}+\de \scG_-^{(0),(as)}(x) \right] \,
  \mm'[Z_{as}(x)]\, \de Z(x) %\times 
  \\
&\times \left[ \frac{\scG_{+1,m,0}^{(0),(as)}(x)}{\mm_{as}(x)}+\de \scG_+^{(0),(as)}(x) \right].
\end{split}
\end{equation}
Expanding the integrand in (\ref{dtomdm1}), one  obtains 4 terms, the UV contributions of which will be discussed separately. Concretely:
\begin{equation}  \label{dtomdm1s}
\begin{split}
\de \tom^{\de m}(0)&= \sum\limits_{k=1}^4 \de \tom_k^{\de m}(0).
\end{split}
\end{equation}
where 
\begin{equation} \label{dtomdm4}
\begin{split}
\de \tom_1^{\de m}(0)&=\left(\frac{2}{\ell}\right)^2  \intinf  \!\! dx \, \scG_{-1}^{(0),(-)}(x+\ltl) \, \mm'[Z_{as}(x)]\, \de Z(x) \,   \scG_{+1}^{(0),(+)}(x-\ltl),   \\
\de \tom_2^{\de m}(0)&=\frac{2}{\ell}  \intinf  \!\! dx \, \de \scG_-^{(0),(as)}(x) \, \mm'[Z_{as}(x)]\, \de Z(x) \,   \scG_{+1}^{(0),(+)}(x-\ltl),   \\
\de \tom_3^{\de m}(0)&=\frac{2}{\ell}  \intinf  \!\! dx \,  \scG_{-1}^{(0),(-)}(x+\ltl)  \, \mm'[Z_{as}(x)]\, \de Z(x) \, \de \scG_+^{(0),(as)}(x), \\
\de \tom_4^{\de m}(0)&=\intinf  \!\! dx \, \de \scG_-^{(0),(as)}(x)  \, \mm'[Z_{as}(x)]\, \de Z(x) \, \de \scG_+^{(0),(as)}(x). 
\end{split}
\end{equation}
We note, that to obtain (\ref{dtomdm4}) from(\ref{dtomdm1}), we used (\ref{asscG0}) and (\ref{scGlas}).

Inserting (\ref{dscGkpas}), (\ref{dscGkmas}) and (\ref{dZk}) into (\ref{dtomdm4}) and carrying out the method 
described in the previous subsections, one obtains the following 2nd order contributions from them. 
%$  \scG_{+1}^{(0),(+)}(x)  $ 

\begin{equation}  \label{dtomdm1c}
\begin{split}
\de \tom_1^{\de m}(0) &\simeq \left(\frac{2}{\ell} \right)^{2-\tfrac{8}{p+1}} \bigg\{ c_1^- \, \CP\limits_{\ell \to 0}  \! \intli \!\! dx \, e^{-\tfrac{2 x}{p+1}} \, 
\scG_{+1}^{(0),(+)}(x) \, \de Z_+(x) \, \mm'[Z_+(x)]+ \\
&+c_1^+ \, \CP\limits_{\ell \to 0}  \! \intil \!\! dx \, e^{\tfrac{2 x}{p+1}} \, \scG_{-1}^{(0),(-)}(x) \, \de Z_-(x) \, \mm'[Z_-(x)]
\bigg\},  \\
\de \tom_2^{\de m}(0) &\simeq \left(\frac{2}{\ell} \right)^{2-\tfrac{8}{p+1}} \! \intinf \!\! dx \, 
\scG_{+1}^{(0),(+)}(x) \, \de Z_+(x) \, \de \scG_{-+}(x)\, \mm'[Z_+(x)], \\
\de \tom_3^{\de m}(0) &\simeq \left(\frac{2}{\ell} \right)^{2-\tfrac{8}{p+1}} \! \intinf \!\! dx \, 
\scG_{-1}^{(0),(-)}(x) \, \de Z_-(x) \, \de \scG_{+-}(x)\, \mm'[Z_-(x)], \\
\de \tom_4^{\de m}(0) &\sim \ell^{-2+\tfrac{10}{p+1}}, \qquad \mbox{negligible upto next to leading order.}
\end{split}
\end{equation}
Here, we note that due to $\de Z_0=0,$ the $\CP$ prescription is superfluous in $\de \tom_1^{\de m}(0),$ 
and the integrals can be taken from $-\infty$ to $\infty.$ 
%=\\de \scG_{+0}=\de \scG_{-=}=0,$ 

Thus, the 2nd order contribution to $\de \tom^{\de m}(0)$ is the sum of the previous expressions. Collecting  the leading and the 2nd order contributions 
from $\tom^{(as)}(\al)$ in (\ref{tomas01fin}) and  (\ref{tomas02fin}), from $\de \tom^{(as)}(\al)$ given in (\ref{dtomas01}) and (\ref{dtomas0fin}), (\ref{dtomasdefin})
and from $\de \tom^{\de m}(0)$ given in (\ref{dtomdm1c}), 
one obtains the leading and next to leading terms\footnote{Next to leading terms are computed only at $\al=0.$} in the $\ell \to 0$ series expansion of $\tom_{1,-1}(\al).$

\subsection{Comments concerning the derivation of the UV formula for the more general $\tom_{2k-1,1-2j}(\al)$ 
set of quantities }

In the previous appendix, we described, the derivation of the first two terms in the 
UV series representation of $\tom_{1,-1}(\al).$ 
In the main text of the paper, we presented the UV formula for a more general set of quantities: $\tom_{2k-1,1-2j}(\al),$ with $j,k=1,2,...$
Without going deeply into the lengthy computations again, one can recognize, that for these quantities, 
the derivation and the bookkeeping of the correction terms 
is almost literally the same as in the case of $\tom_{1,-1}(\al),$ apart from some simple modification of the powers of $\frac{2}{\ell}$ in front of certain 
terms and apart from using kink functions with different subscripts $-1 \to 1-2j, +1 \to 2k-1$. This subscript change comes from the  different definition of $\tom_{2k-1,1-2j}(\al).$ 
In this case, the source terms of the typical linear problems will change: $e^x \to e^{(2k-1)x},$ and $e^{-x} \to e^{(1-2j) x}.$ This change in the source terms 
implies a different scaling in $\ell,$ since during the UV computations the source terms are shifted with $\pm \ltl.$  
As a consequence of these shifts, instead of a factor $\frac{2}{\ell}$, the factor $\left(\frac{2}{\ell}\right)^{2k-1}$ or $\left(\frac{2}{\ell}\right)^{2j-1}$ 
will arise, during the computations. 
Otherwise, the logic and the mathematical steps of eliminating the different terms in the UV limit are the same as described for 
$\tom_{1,-1}(\al).$

Now, we just collect the main replacement rules, which allows one to obtain the 2nd order UV formula for  
$\tom_{2k-1,1-2j}(\al),$ from that of $\tom_{1,-1}(\al).$

\begin{itemize}

\item{ The $\ell-$dependent prefactors change: $\left(\frac{2}{\ell}\right)^{2-\tfrac{4}{p+1}} \to \left(\frac{2}{\ell}\right)^{2(j+k-1)-\tfrac{4}{p+1}}$ 

and $\left(\frac{2}{\ell}\right)^{2-\tfrac{8}{p+1}} \to \left(\frac{2}{\ell}\right)^{2(j+k-1)-\tfrac{8}{p+1}}.$}

\item{ The kink functions are also changed: $\scG_{-1}^{(\al),(-)}(x) \to \scG_{-(2j-1)}^{(\al),(-)}(x)$ and \newline 
$\scG_{+1}^{(-\al),(+)}(x) \to \scG_{+(2k-1)}^{(-\al),(+)}(x).$ }

\item{As a consequence of this change in the kink functions, the leading order asymptotic solutions 
will also scale differently. In the analog of (\ref{asscG0}) a $\frac{2}{\ell} \to \left( \frac{2}{\ell}\right)^{2j-1},$ and in the analog of (\ref{scGlas}) a  $\frac{2}{\ell} \to \left( \frac{2}{\ell}\right)^{2k-1},$ changes in the "scaling prefactors" should be performed.
}

\item{The plateau expansion coefficients entering the UV formula ($c_n^{\pm},c_n^{(\pm1,\al)},$ etc.), 
should be interpreted as the analogous expansion coefficients of the new kink functions. Namely, 
%\newline
$c_1^{(-1,\al)}$ and $c_1^-=c_{1}^{(-1,0)}$ etc. should be considered as coefficients in the plateau expansion (\ref{scGmasy}) of $\scG_{-(2j-1)}^{(\al),(-)}(x),$ while
$c_1^{(+1,\al)}$ and $c_1^+=c_{1}^{(+1,0)}$ etc. should be considered as coefficients in the plateau expansion (\ref{scGpasy}) of $\scG_{+(2k-1)}^{(\al),(+)}(x).$  
These lead to the replacement rules: %$c_1^-=c_1^{(-1,\alpha)} \to c_1^{(-(2j-1),\al)}$ and $c_1^+=c_1^{(+1,-\alpha)} \to c_1^{(+(2k-1),\al)}.$
$c_1^{(-1,\alpha)} \to c_1^{(-(2j-1),\al)},$ 
$c_1^-=c_1^{(-1,0)} \to c_1^{(-(2j-1))}$ and  $c_1^+=c_1^{(+1,-\alpha)} \to c_1^{(+(2k-1),\al)}, $ $c_1^+=c_1^{(+1,0)} \to c_1^{(+(2k-1))}.$
}

\item{ The source terms in the equations of the kink functions should be replaced as: $e^{-x} \to e^{(1-2 j)x},$ and $e^x \to e^{(2k-1)x}.$ }

\item{Finally, in the final result everywhere the: $\scG_{+1}^{(-\al),(+)}(x)-e^x \to \scG_{+(2k-1)}^{(-\al),(+)}(x)-e^{(2k-1)x}$ 
replacement should be done.}

\end{itemize}

These "generalizating" replacements in the UV result of $\tom_{1,-1}(\al)$ of the previous sections, lead to the result for $\tom_{2k-1,1-2j}(\al)$ 
presented in the main text of the paper.

\section{The large argument expansion of the kernel $G_\al(x)$} \label{appB}

The knowledge of the large argument series expansion of the kernel $G_\al(x)$ (\ref{Galpha}), is useful for two reasons. 
First, it offers an alternative useful series representation of the kernel for its numerical computation 
in a regime, where the accurate numerical evaluation of the original Fourier-representation (\ref{Galpha}) 
becomes challenging due to the highly oscillating nature of the integrand. 

Second, such a series representation of $G_\al(x),$ allows one to compute the qualitative large argument 
behaviours of the kink functions from the linear equations they satisfy. Namely, the second series 
representations in (\ref{scGmasy}) and (\ref{scGpasy}) can be derived from the equations 
(\ref{kinkmalpha}) and (\ref{kinkpalpha}) respectively, 
with the help of the large argument series of $G_\al(x).$

The large argument series representation of $G_\al(x)$ can be computed from its Fourier-integral 
representation (\ref{Galpha}), using the residue theorem. 
We present the form of the series at $+\infty.$ It takes the form:
\begin{equation} \label{Gsoral}
\begin{split}
G_\al(x)&=\sum\limits_{k=0}^\infty \hat{C}_k^{(\al)} \, e^{-(1+2k) x} +
\sum\limits_{k=1}^\infty \tilde{C}_k^{(\al)} \, e^{-\tfrac{2k-\al}{p} x}, 
\qquad 0\leq \al <2,
\\
\mbox{with} \qquad
\hat{C}_k^{(\al)}&=-\frac{1}{\pi} \cot\left[ (1+2 k) \tfrac{\pi \, p}{2}+\al \right],  \qquad
\tilde{C}_k^{(\al)}=\frac{1}{p \, \pi} \tan\left[ \tfrac{\pi}{2 \, p} (2k-\al) \right].
\end{split}
\end{equation}
The series representation at $-\infty,$ can be obtained from (\ref{Gsoral}) by 
using the reflection symmetry identity implied by the Fourier-representation (\ref{Galpha}):
\begin{equation}  \label{Grefl}
\begin{split}
G_\al(-x)=G_{-\al}(x).
\end{split}
\end{equation}

We close this short appendix by the remark, that the large argument series representation  
(\ref{Gsoral}) allows one to obtain simple exact analytical formulas for $G_\al(x)$ 
in certain special choices of the coupling constant. For example, when $p \in \mathbb{N},$ 
by summing the series in (\ref{Gsoral}), simple analytical formulas for $G_\al(x)$ can be found.  
 
Only for demonstrational reasons, here we present the simple analitycal form for $G(x)$ and $G_\alpha(x)$ at the 
$p=5$ point. The function $G(x)$ takes the form:
\begin{equation}  \label{Gp5}
\begin{split}
G(x)&=\frac{d}{dx} \chi(x), \qquad \text{where}: \\
\chi(x)&=-\frac{i}{2 \pi} \, \log 
\frac{\cosh\left(\tfrac{2x+3 \, i \, \pi}{10} \right)\, \cosh\left(\tfrac{2x-i \, \pi}{10} \right)}{\cosh\left(\tfrac{2x-3 \,i \, \pi}{10} \right) \, \cosh\left(\tfrac{2x+i \, \pi}{10} \right)}, 
\end{split}
\end{equation}
which implies the following form for the soliton-soliton S-matrix element at this special value of the coupling constant:
\begin{equation}  \label{S5}
\begin{split}
S(x)&=e^{2 \pi \, i \, \chi(x)}=\frac{\cosh\left(\tfrac{2x+3 \, i \, \pi}{10} \right)\, \cosh\left(\tfrac{2x-i \, \pi}{10} \right)}{\cosh\left(\tfrac{2x-3 
\,i \, \pi}{10} \right) \, \cosh\left(\tfrac{2x+i \, \pi}{10} \right)}.
\end{split}
\end{equation}
This is a much simpler expression, than the form given in \cite{Zamolodchikov:1978xm}, which contains  
infinite products of $\Gamma-$functions. We also note, that these special points lie in the repulsive regime and should not be confused with the 
well-known reflectionless points ($1/p \in \mathbb{N}$), which lie in the attractive regime,  
and where the S-matrix similarly simplifies to products of hyperbolic functions \cite{Zamolodchikov:1977py}.

To close this appendix, we also give the analytical form for the kernel $G_\al(x)$ at $p=5,$ which takes the form as follows: 
\begin{equation}  \label{Galp5}
\begin{split}
G_\al(x)&=\frac{\tan (\tfrac{\pi \, \al }{2})}{2 \, \pi \, \sinh x}-\frac{e^{\left(\tfrac{\al}{5}-1\right)x}}{10 \, \pi \sinh x} \, 
\sum\limits_{k=1}^4 e^{\tfrac{2k}{5}x} \, \tan \left( \frac{\pi}{10}(2 \, k+\al)\right), \qquad 0<\al<2.
\end{split}
\end{equation}

%%%%%%%%%%%%%%%%%%%%%%%%%%%%%%%%%%%%%%%%%%%%%%%%%%%%%%%%%%%%%%%%%%%%%%%%%%%%%%%%%%%%%%%%%%%%%%%%%%%%%%%%%%%%%%%%%%%%%%%%%%%%%
%%%%%%%%%%%%%%%%%%%%%%%%%%%%%%%%%%%%%%%%%%%%%%%

\section{Some tables of numerical data} \label{appC}

In this appendix, we provide with some concrete numerical data at the $p=5$ value of the coupling constant. 

In the sequel, we just add a few comments to facilitate the interpretation of the content of these tables.

\begin{itemize} 

\item{
In tables \ref{text1} and \ref{text2} the scaled groundstate energies and $\om$ elements are 
given by the definitions:
\begin{equation}
\begin{split}
E_0^{\text{scaled}}(\ell)&=\ell\, E_0(\ell), \\
\om_{2k-1,1-2j}^{\text{scaled}}(0|\ell)&=\frac{1}{2 \, \pi \,i}\, \ell^{2(k+j-1)-\tfrac{4}{p+1}}\,\tom_{2k-1,1-2 j}(0), \qquad j,k \in \mathbb{N}.
\end{split}
\end{equation}
}

\item{
In tables \ref{text3}-\ref{text6}, we gave the first few coefficients in the large argument series expansion 
of some $"-"$-kink type functions. The analogous coefficients of the $"+"$-kink type functions can be easily determined from them, using the simple reflection relations being valid at $\al_z=0$:
\begin{equation}
\begin{split}
Z_+(x)&=-Z_-(-x), \qquad \scG_{+(2j-1)}^{(-\al)(+)}(x)=\scG_{-(2j-1)}^{(\al)(-)}(-x), \qquad j\geq 1.
\end{split}
\end{equation}  
}

\item{
One can recognize, that the tables \ref{text5} and \ref{text6}, do not contain the $b$-type coefficients 
entering (\ref{scGmasy}).  The reason for that, is that they are all zero in the $\al=0, \, \,p=5$ point, 
because at this value of the coupling constant  each coefficient $\hat{C}^{(0)}_k$  becomes zero in 
(\ref{Gsoral}).
}

\item{
Finally, we provide the analytical formula, which allows one to test the numerical estimates for  
the leading coefficient in the UV series representation (\ref{UVexpansion}) of $\tom_{1,-1}(\al)$ 
 (\ref{omegatilde}).
The comparison of (\ref{primVEV}) to (\ref{VEVcft}) leads to the following formula:
\begin{equation}
\begin{split}
\tom_{1,-1|1}^{(\al)}=\pi\, \nu \, 2^{-2+4 \, \nu-2\, \nu\, \al} \, 
\frac{(2 \, \pi)^{2\,(\Delta_{\al+2 (1-\nu)/\nu}-\Delta_\al)} \, {\cal V}(\al,0) }
{\Pi(\nu)^{2-2\, \nu +2 \, \al \, \nu} \, C_1(\al)},
\end{split}
\end{equation} 
where the definition of the constituting functions are given in (\ref{Dalfa}), (\ref{C1}), (\ref{muM}) and 
(\ref{prim2CFT}).
}

\end{itemize}

\begin{table}[h]
\begin{center}
\begin{tabular}{|c|c|c|}
\hline
$\ell$ & $ E_0^{\text{scaled}}(\ell)$   & $ \om_{1,-1}^{\text{scaled}}(0|\ell)$  \tabularnewline
 \hline
 $10^{-1}$  & $-0.509118046201947953856845548671$ & $-0.0700350183615299127182564862836$    \\
 \hline
$10^{-2}$  & $-0.521752815793292914037183625983$ & $-0.0309091889458714362530643260619$    \\
 \hline
$10^{-3}$  & $-0.523250032152706409743954909889$ & $-0.0240856216233222182216321884171$    \\
 \hline
$10^{-4}$  & $-0.523525716645838246935956941589$ & $-0.0227794902633984171366865719750$    \\
 \hline
$10^{-5}$  & $-0.523583128972127245620052105468$ & $-0.0225096883140453029421020333958$    \\
 \hline
\end{tabular}\label{table_E0_om1m1}
\bigskip
\caption{Numerical data for the scaled energy and $\om_{1,-1}(0)$ at $\al=\al_z=0$ and $p=5.$
}
\label{text1}
\end{center}
\end{table}
\normalsize

\begin{table}[h]
\begin{center}
\begin{tabular}{|c|c|c|}
\hline
$\ell$ & $ \om_{3,-1}^{\text{scaled}}(0|\ell)$   & $ \om_{3,-3}^{\text{scaled}}(0|\ell)$  \tabularnewline
 \hline
 $10^{-1}$  & $0.260532706117901469152318223857$ & $-1.19347271532320375565434527039$    \\
 \hline
$10^{-2}$  & $0.166179778849068486546832734456$ & $-0.989449022915246710045399824164$    \\
 \hline
$10^{-3}$  & $0.149047671194975286250021848064$ & $-0.943651632832537560388849145642$    \\
 \hline
$10^{-4}$  & $0.145496356228911664605692627914$ & $-0.933740670328518852006310913534$    \\
 \hline
$10^{-5}$  & $0.144737637849769879343670291690$ & $-0.931603754402221822888144591531$    \\
 \hline
\end{tabular}\label{table_om_3m1_3m3}
\bigskip
\caption{Numerical data for the scaled versions of $\om_{3,-1}(0)=\om_{1,-3}(0)$ and $\om_{3,-3}(0)$ at $\al=\al_z=0$ and $p=5.$}
\label{text2}
\end{center}
\end{table}
\normalsize

%%%%%%%%%%%%%%%%%%%%%%%%%%%%%%%%%%%%

\begin{table}[h]
\begin{center}
\begin{tabular}{|c|c|c|c|}
\hline
$c_{\text{plateau}}$ & $ Z_-(x)$   \tabularnewline
 \hline
 $c_1$  & $-0.215096113203562341763036402395$  \\
 \hline
$c_2$  & $-0.352028885033093452354544480456$  \\
 \hline
$c_3$  & $-0.333333333333333333333333333333$   \\
 \hline
$c_4$  & $-0.210210394810533434744446287053$  \\
 \hline
\end{tabular}\label{table_Zm_cjk}
\bigskip
\caption{The numerical values of the first four coefficients of $Z_-(x)$ in the series (\ref{kinkplat}) at $\al_z=0$ and $p=5.$}
\label{text3}
\end{center}
\end{table}
\normalsize

%%%%%%%%%%%%%%%%%%%%%%%%%%%%%%%%%%%%

\begin{table}[h]
\begin{center}
\begin{tabular}{|c|c|c|c|}
\hline
$c_{\text{plateau}}$  & $ \scG_{-1}^{(0)(-)}(x)$ & $\scG_{-3}^{(0)(-)}(x)$ \tabularnewline
 \hline
 $c_1$  &  $0.07169870440118744725434570550$ &  $-0.1154653832068193979886878318$ \\
 \hline
$c_2$  &  $0.2346859233553956348834547228$  & $-0.1643891946493538739353415988$ \\
 \hline
$c_3$  &  $0.3333333333333333333333333333$ & $0$  \\
 \hline
$c_4$  &  $0.2802805264133130184291060880$  & $0.1288839794551031895820186132$ \\
 \hline
\end{tabular}\label{table_scG_cjk}
\bigskip
\caption{The numerical values of the first four coefficients of $ \scG_{-1}^{(0)(-)}(x)$ and $ \scG_{-3}^{(0)(-)}(x)$ in the series (\ref{scGmasy}) at $\al=\al_z=0$ and $p=5.$}
\label{text4}
\end{center}
\end{table}
\normalsize

%%%%%%%%%%%%%%%%%%%%%%%%%%%%%%%%%%%%%%%%%%%%%%%%%%%%%%%%%%%%%%%%%%%%%%%%%%%%%%%%%%%%%%%%%

%%%%%%%%%%%%%%%%%%%%%%%%%%%%%%%%%%%%

\begin{table}[h]
\begin{center}
\begin{tabular}{|c|c|c|c|}
\hline
$a$ & $ Z_-(x)+e^{-x}$   \tabularnewline
 \hline
 $a_1$  & $-0.101871105117020148643657276208$  \\
 \hline
$a_2$  & $-0.288637775812982232679774725985$  \\
 \hline
$a_3$  & $0.330733810276246723130170092328$   \\
 \hline
\end{tabular}\label{table_Zm_aj}
\bigskip
\caption{The numerical values of the first three coefficients of the series %\newline
$Z_-(x)+e^{-x}\stackrel{x \to -\infty}{=} \sum\limits_{k=1}^\infty a_k \, e^{\tfrac{2 \,k}{p}\, x}+\sum\limits_{k=1}^\infty b_k \, e^{(2 \,k-1)\, x}$  at $\al_z=0$ and $p=5.$ We note, that $b_k=0$ 
for any values of the $k,$ when $p=5.$} 
\label{text5}
\end{center}
\end{table}
\normalsize

%%%%%%%%%%%%%%%%%%%%%%%%%%%%%%%%%%%%

\begin{table}[h]
\begin{center}
\begin{tabular}{|c|c|c|c|}
\hline
$a$  & $ \scG_{-1}^{(0)(-)}(x)$ & $\scG_{-3}^{(0)(-)}(x)$ \tabularnewline
 \hline
 $a_1$  &  $-0.0407484420468080594574629104831$ &  $0.176022191341968338904393436795$ \\
 \hline
$a_2$  &  $-0.288637775812982232679774725985$  & $1.80118092958846187205645190248$ \\
 \hline
$a_3$  &  $0.330733810276246723130170092328$ & $-2.83528817118671599575889889429$  \\
 \hline
\end{tabular}\label{table_scG_ajk}
\bigskip
\caption{The numerical values of the first three coefficients of $ \scG_{-1}^{(0)(-)}(x)$ and $ \scG_{-3}^{(0)(-)}(x)$ in the series (\ref{scGmasy}) at $\al=\al_z=0$ and $p=5.$}
\label{text6}
\end{center}
\end{table}
\normalsize

%%% mod3.tex  vege %%%%%

\begin{table}[h]
\begin{center}
\begin{tabular}{|c|c|c|c|}
\hline
$(k,j)$  & $ \tom_{2k-1,1-2j|1}^{(0)}$ & $\tom_{2k-1,1-2j|2}^{(0)}$ \tabularnewline
 \hline
 $(1,1)$  &  $0.05594523619745321253305345556 \, i$ &  $0.6211105849806047774109704441 \, i$ \\
 \hline
$(1,2)$  &  $-0.09009560479628380986278568022 \, i$  & $-0.4431535146850558715485802866 \, i$ \\
 \hline
$(2,2)$  &  $0.1450922107998478209678591080 \, i$ & $0.3128175595054423275099975234 \, i$  \\
 \hline
\end{tabular}\label{table_tom_0}
\bigskip
\caption{The numerical values of the first two coefficients in the series (\ref{UVexpansion}) 
for a few elements 
of the $\om$ matrix at $\al=\al_z=0$ and $p=5.$ The numerical values  for $ \tom_{2k-1,1-2j|1}^{(0)}$ 
$ \tom_{2k-1,1-2j|2}^{(0)}$ agree with the exact values with 16- and 8- digits of precisions, respectively.}
\label{text7}
\end{center}
\end{table}
\normalsize

%%%%%%%%%%%%%%%%%%%%%%%%%%%%%%%%%%%%

\begin{table}[h]
\begin{center}
\begin{tabular}{|c|c|c|c|}
\hline
$\tom_{1,-1|1}^{(\al)}$  & $ 0.0055071758870437493105284541826 \, i $  \tabularnewline
 \hline
$c^{(-1,\al)}_1$  &  $0.0107368873172055208778869610246$  \\
 \hline
$c^{(-1,\al)}_2$  &  $0.132013741773294529046432357696$  \\
 \hline
$c^{(-1,\al)}_3$  &  $0.288606455083507499100453773739$  \\
 \hline
$a^{(-1,\al)}_0$  &  $-0.0149971066307696298914362950656$  \\
 \hline
$a^{(-2,\al)}_1$  &  $-0.130265441355505842575765697648$  \\
 \hline
$a^{(-3,\al)}_2$  &  $1.49638257654624549444742381634$  \\
 \hline
$b^{(-1,\al)}_1$  &  $-1.41634629055386602312957496165$  \\
 \hline
$b^{(-1,\al)}_2$  &  $10.4035563123513103539247760682$  \\
 \hline
$b^{(-1,\al)}_3$  &  $-219.615596021681549680518140881$  \\
 \hline
\end{tabular}\label{table_alpha}
\bigskip
\caption{The numerical values related to the computation of $\tom_{1,-1|1}^{(\al)}$ at 
$\al=\tfrac{87}{71}, \quad \al_z=0$ and $p=5.$ Beyond the leading order coefficient $\tom_{1,-1|1}^{(\al)}$ of $\tom_{1,-1}(\al),$ 
the coefficients of the function $\scG_{-1}^{(\al)(-)}(x)$ in the series representations (\ref{scGmasy}) are 
also listed.}
\label{text8}
\end{center}
\end{table}
\normalsize

\clearpage

%%%%%%%%%%%%%%%%%%%%%%%%%%%%%%%%%%%%%%%%%%%%%%%%%%%%%%%%%%%%%%%%%%%%%%%%%%%%%%%%%%

%%%%%%%%%%
%%%%%%
%%%%%%%%%%%
%%%%%%%
%%%%%%%
%%%%%%%%

%%%%%%%%%%%%%%%%%%%%%%%%%%%%%%%%%%%%%%%%%%%%%%%%%%%%%%%%%%%%%%%%%%%%%%%%%%%%%%%%%%%%%%%%%%%
%%%%%%%%%%%%%%%%%%%%%%%%%%%%%%%%%%%%%%%%%%%%%%%%%%%%%%%%%%%%%%%%%%%%%%%%%%%%%%%%%%%%%%%%%%%
%%%%%%%%%%%%%%%%%%%%%%%%%%%%%%%%%%%%%%%%%%%%%%%%%%%%%%%%%%%%%%%%%%%%%%%%%%%%%%%%%%%%%%%%%%%

%%%%%%%%%%%%%%%%%%%%%%%%%%%%%%%%%%%%%%%%%%%%%%%%%%%%%%%%%%%%%%%%%%%%%%%%%%%%%%%%%%%%%%%%%%%
%%%%%%%%%%%%%%%%%%%%%%%%%%%%%%%%%%%%%%%%%%%%%%%%%%%%%%%%%%%%%%%%%%%%%%%%%%%%%%%%%%%%%%%%%%%
%%%%%%%%%%%%%%%%%%%%%%%%%%%%%%%%%%%%%%%%%%%%%%%%%%%%%%%%%%%%%%%%%%%%%%%%%%%%%%%%%%%%%%%%%%%

\newpage

\providecommand{\href}[2]{#2}\begingroup\raggedright\endgroup

\end{document}